%% file: syk3d-arxiv-v2.tex
\documentclass[11pt]{article}
\pdfoutput=1
\usepackage{jheppub}

\usepackage[svgnames,table]{xcolor}
\usepackage{graphicx}
\usepackage{amsfonts,amsmath,amssymb,amsthm}
\usepackage{mathrsfs,bm,bbm,esint}
\usepackage[mathscr]{eucal}
\usepackage{physics,tensor}
\usepackage{slashed,soul}
\usepackage{rotating,pdflscape}
\usepackage{enumitem}
\usepackage{multirow,array}
\usepackage{diagbox}
\usepackage[numbers,sort&compress]{natbib}
\usepackage{tikz}
\usetikzlibrary{cd}
\usetikzlibrary{decorations.pathmorphing}
\usetikzlibrary{decorations.pathreplacing}
\usetikzlibrary{decorations.markings}
\tikzset{snake it/.style={decorate, decoration=snake}}
\tikzset{->-/.style={decoration={
  markings,
  mark=at position .5 with {\arrow{>}}},postaction={decorate}}}
\usepackage{subcaption}
\usepackage{float}
\usepackage{afterpage}
\usepackage{cleveref}
\crefname{section}{§\!\!}{§§\!\!}
\Crefname{section}{§}{§§}
\crefname{appendix}{Appendix}{Appendices\!}
\crefname{figure}{Fig.\!}{Figs.\!}
\crefname{table}{Table}{Tables}

\newcommand{\shadeB}{\cellcolor{blue!5}}
\newcommand{\shadeR}{\cellcolor{red!5}}

\newcommand{\thetab}{\bar{\theta}}
\newcommand{\zb}{\bar{z}}
\newcommand{\psib}{\bar{\psi}}
\newcommand{\Phib}{\overline{\Phi}}
\newcommand{\phib}{\bar{\phi}}
\newcommand{\Fb}{\bar{F}}
\newcommand{\Q}{\mathcal{Q}}
\newcommand{\Qb}{\bar{\mathcal{Q}}}
\newcommand{\Gt}{\widetilde{G}}
\newcommand{\Sigt}{\widetilde{\Sigma}}
\newcommand{\Xb}{\overline{X}}
\newcommand{\cope}[1]{\mathcal{C}_{_{#1}}}
\newcommand{\mft}{\text{\tiny MFT}}

\newcommand{\AS}{\mathfrak{X}}

\newcommand{\Ax}{\mathfrak{a}}

\definecolor{rust}{rgb}{0.8,0.2,0.2}

\title{A 3d disordered superconformal fixed point}

\author[a,b]{Chi-Ming Chang}
\author[c]{, Sean Colin-Ellerin}
\author[d]{, Cheng Peng}
\author[c]{, Mukund Rangamani}

\affiliation[a]{Yau Mathematical Sciences Center (YMSC), Tsinghua University, Beijing, 100084, China}

\affiliation[b]{Beijing Institute of Mathematical Sciences and Applications (BIMSA), Beijing, 101408, China}

\affiliation[c]{
Center for Quantum Mathematics and Physics (QMAP),  \\
Department of Physics \& Astronomy, University of California, Davis, CA 95616 USA.}

\affiliation[d]{Kavli Institute for Theoretical Sciences (KITS) and CAS Center for Excellence in Topological Quantum Computation, University of Chinese Academy of Sciences, Beijing 100190, China}

\emailAdd{cmchang@tsinghua.edu.cn, scolinellerin@ucdavis.edu, pengcheng@ucas.ac.cn, mukund@physics.ucdavis.edu}

\abstract{ 
We initiate the study of a three dimensional disordered supersymmetric field theory. Specifically, we consider a $\mathcal{N}=2$ large $N$ Wess-Zumino like model with cubic superpotential  involving couplings drawn from a  Gaussian random ensemble. Taking inspiration from analyses of lower dimensional SYK like models we demonstrate that the theory flows to a  strongly coupled superconformal fixed point in the infra-red. In particular, we obtain leading large $N$ spectral data and operator product coefficients at the critical point.  Moreover, the analytic control accorded by the model allows us to compare our results  against those derived in the conformal bootstrap program and demonstrate consistency with general expectations.
}

\begin{document}
\maketitle


\section{Introduction}
\label{sec:intro}
  
Strongly correlated critical behaviour occurs in a wide class of physical systems and underlies some of the interesting physical phenomena in nature. Over the years we have seen examples of such, not only in  low energy dynamics of many-body systems, but also in the dynamics of black holes in quantum gravity. Tractable  models of strong coupling critical behaviour, while scarce, have the potential to provide valuable insight into the dynamics of a wide class of physical systems. We therefore motivate the study of one such system. 

 The classic example of strong coupling critical behaviour is provided by the Wilson-Fisher fixed point \cite{Wilson:1971dc} occurring in  the simplest field theory, the self-interacting scalar. This example provides the low energy fixed point for the three-dimensional Ising model which has seen incredible precision studies using the conformal bootstrap \cite{ElShowk:2012ht,El-Showk:2014dwa}. Analytic methods for analyzing such systems range from the classic $\epsilon$-expansion \cite{Wilson:1973jj} and the use of large $N$  techniques \cite{Brezin:1972se,Wilson:1973jj}, both of which have the advantage of rendering the analysis of the fixed point amenable within perturbation theory. The large $N$ expansion for these systems may be characterized as being vector-like and is qualitatively different from the planar diagrammatics of matrix-like models (which typically have strong coupling fixed points). The difference can be traced to the presence of nearly conserved higher spin currents of these vector models cf., \cite{Maldacena:2012sf}. In turn this feature has interesting implications for holographic duals of such critical points -- they are given in terms of higher spin theories in an asymptotically AdS spacetime \cite{Klebanov:2002ja}.  Planar field theories typically end up with strongly coupled fixed points with a sparse low lying spectrum with a dual holographic description in terms of classical gravitational dynamics \cite{Maldacena:1997re}; they are  however are hard to analyze directly.

In recent times  a new class of large $N$ models has emerged which provides a happy middle ground between the two classes described above. These models have a different set of diagrams dominating the large $N$ limit, the so-called melonic diagrams, and give rise to strongly coupled fixed points, with a spectrum that is not entirely sparse per se, but one that is nevertheless amenable to direct analysis. The prototype example is the disordered quantum mechanical Sachdev-Ye-Kitaev  (SYK) model \cite{Sachdev:1992fk,Kitaev:2015aa}. As demonstrated in \cite{Maldacena:2016hyu,Kitaev:2017awl} the model is explicitly solvable using large $N$ techniques -- the Schwinger-Dyson equations truncate (owing to the disorder averaging over random couplings). In this model the low energy spectrum is not sparse, as in the matrix models, but the dynamics is largely controlled by a single Goldstone mode that dominates over the rest of the spectrum. Consequently, the system admits a holographic dual in terms of a two dimensional classical gravitational theory, the Jackiw-Teitelboim (JT) gravity  \cite{Maldacena:2016upp,Jensen:2016pah,Kitaev:2017awl}. It additionally exhibits maximal Lyapunov exponent  \cite{Maldacena:2015waa} as measured by the out-of-time order four-point correlation function, a feature it shares with (higher dimensional AdS) black holes, rendering it an invaluable toy model for understanding holography and black hole physics, cf., \cite{Cotler:2016fpe,Maldacena:2018lmt,Saad:2018bqo} for some salient developments. Additionally, the SYK model has played an important role in elucidating local criticality in fermionic systems without quasiparticles \cite{Sachdev:2015efa,Esterlis:2021eth}.

In the past few years various  generalizations of these models have been considered. On the one hand, there are works analyzing quantum mechanical models without disorder using colored  \cite{Witten:2016iux,Bonzom:2011zz}, or uncolored \cite{Klebanov:2016xxf,Carrozza:2015adg} tensor valued degrees of freedom.  The strategy here is to pick Hamiltonians with specific tensor contractions which ensure that only melonic diagrams contribute in the large $N$ limit (the subleading $1/N$ effects are different, and to our knowledge not extensively analyzed). There  have also been attempts to include additional flavour symmetries \cite{Gross:2016kjj,Gu:2019jub} and supersymmetry \cite{Anninos:2016szt,Fu:2016vas} in the quantum mechanical disordered model. 

Beyond quantum mechanics (i.e., $0+1$ dimensional field theories) the class of models that have been investigated have been somewhat limited. In \cite{Murugan:2017eto} two dimensional SYK like models with disorder were analyzed in detail, both with and without supersymmetry. The problem with higher dimensional models is two fold. Firstly, pure fermionic models do not generically have relevant operators (a four-Fermi interaction being marginally irrelevant in two dimensions). Secondly, including  bosonic degrees of freedom is problematic, since disordered Hamiltonians fail to be generically positive definite. Furthermore, even with these issues brought under control one has to account for the generation of relevant operators along the RG flow to low energy, making the analysis more involved.

In \cite{Murugan:2017eto} an interesting class of supersymmetric models was analyzed in detail. In this case the aforementioned problems can be overcome and  one obtains an IR superconformal fixed point. We refer henceforth to the model with $\mathcal{N}=2$ supersymmetry analyzed therein as the MSW model; see \cite{Bulycheva:2018qcp} for further analysis of this model. From the spectral analysis one learns that the low energy collective modes are not sparse enough to admit a classical gravity dual; for one, the Lyapunov exponent is sub-maximal indicative of a classical string description (one which however is as yet unknown). The reason for this can be traced to the fact that the low energy dynamics is not altogether controlled by the energy-momentum tensor but the presence of other  light collective degrees of freedom. There have also been studies of models with lower, $(0,2)$, supersymmetry \cite{Peng:2018zap}, which allow for tuning the interactions so that the Lyapunov exponent can range from the value attained for the MSW model down to zero, when it is an integrable vector model.  In this example one can track the emergence of higher spin symmetries in the integrable limit \cite{Ahn:2018sgn}. Analysis of two dimensional non-disordered models is more intricate and is explored for example in \cite{Chang:2019yug}. 

Thus far there has been no full-fledged analysis of models beyond two dimensions, though a class of tensor models and the disordered SYK model in three dimensions with $\mathcal{N}=1$ supersymmetry was analyzed in \cite{Popov:2019nja},  and a class of tensor models in dimension $1<d<3$ with four supercharges was studied in \cite{Lettera:2020uay}. In this paper we provide a concrete example of a disordered field theory in $2+1$ dimensions that can be solved using large $N$ melonic Schwinger-Dyson equations.  We restrict our attention to supersymmetric models with $\mathcal{N}=2$ supersymmetry to keep the renormalization group flow analysis tractable. In a sense our model is a natural generalization of the MSW model to three dimensions. Parts of our analysis has partial overlap with the earlier work of \cite{Popov:2019nja} though the models we study are different.

Specifically, we explore the low energy dynamics of a disordered $\mathcal{N}=2$ supersymmetric three dimensional field theory. The physical field content comprises of $N$ complex scalars $\phi_i$, complex fermions $\psi_{\alpha,i}$, and auxiliary fields $F_i$, arranged into a suitable chiral multiplet $\Phi_i$ of the superalgebra.  The interactions are governed by a set of cubic couplings that we draw out of a random Gaussian ensemble. Our interest will be in understanding the dynamics of the low energy collective modes deep in the IR in the disorder averaged theory.\footnote{Large $N$ theories with disordered couplings were analyzed in \cite{Aharony:2015aea} while connections to holography were explored in \cite{Adams:2011rj}  and \cite{Hartnoll:2014cua,Hartnoll:2015faa}.}

Our primary motivation for analyzing these models is to understand the dynamics of thermal field theories at the strongly coupled fixed point attained. However, for the present, we will focus on understanding the superconformal theory in its own right and focus on spectral properties and OPE coefficients which one can extract analytically.  The analysis of the vacuum Schwinger-Dyson equations in the model are straightforward (and related to tensor model analysis in \cite{Popov:2019nja}). One can show that the model flows to a supersymmetric critical point, leading to an IR SCFT. While certain properties such as the conformal dimensions of the chiral fields are fixed by the supersymmetry, there are other, non-chiral, aspects of the spectrum that we can explore quantitatively in the model, thanks to the truncation of the aforesaid Schwinger-Dyson equations. One can in fact view the model we discuss in terms of a disordered version of the super-Ising model in three dimensions.  We will exploit this connection and in particular find it useful to compare our results for the IR fixed point with numerical bootstrap results derived for the super-Ising model \cite{Bobev:2015vsa,Bobev:2015jxa}. In addition, using the results from analytic bootstrap \cite{Fitzpatrick:2012yx,Komargodski:2012ek,Costa:2012cb} we are able to obtain various OPE data in both the chiral and non-chiral sector.  

We found it instructive to consider a general set of correlation functions that allows one to also explore the non-singlet part of the OPE data. Most analyses of SYK-like models focus on four-point functions where one averages not only over disorder, but also over the external operator insertions. These  operator averaged correlation functions may be easily understood using the collective field approach. However, they  project one down to the singlet sector under an emergent symmetry in the IR since one restores democracy between the microscopic fields after disorder averaging. For the model we analyze it will be a $U(N)$ symmetry rotating the $N$ chiral multiplets, though as we shall see, there will no flavour currents associated with this that are generated in the IR. The fixed point we obtain is expected to have a large conformal manifold based on general arguments regarding marginal operators \cite{Leigh:1995ep,Chang:2010sg,Green:2010da}. Our analysis will however not shed light on this structure directly as the marginal operators are expected to appear in the triple-twist sector, while our study of four-point functions and bootstrap data only gives insight into double-twist sector.\footnote{We thank Ofer Aharony and Adar Sharon for helpful discussions on this point.} 

The outline of the paper is as follows: we begin in \cref{sec:n2qft} by delineating the class of models we study and  solve for the low-energy fixed point. We then proceed in \cref{sec:ir} to examine the superconformal field theory thus obtained in some detail, obtaining the spectral data for the non-chiral states. In \cref{sec:4pt} and \cref{sec:chiral}  we analyze  the 4-point correlation function to obtain information about the OPE coefficients and the central charges. The discussion in \cref{sec:4pt} pertains to the non-chiral operators in the theory, while \cref{sec:chiral} uses crossing to get information about the chiral sector. We end with a brief discussion in \cref{sec:discuss}. 

Some of the technical details pertinent to our analysis are collected in various appendices: \cref{sec:conventions}  summarizes our supersymmetry conventions. In  \cref{sec:genqd}  we generalize our discussion to $q$-body superpotential in $d$ dimensions for completeness. \cref{sec:cpw}
reviews conformal partial waves necessary for the analysis in \cref{sec:4pt}. In \cref{sec:s3pt} we describe the superconformal three-point function with extended supersymmetry which we use to construct the superconformal partial wave in the supershadow formalism. In \cref{sec:ssops}  we outline the computation of the supershadow coefficients for the superconformal block decomposition and in \cref{sec:det} we present the analysis of the Berezinian for the gauge fixing to compute the measure for the inner product of the superconformal partial waves.

\section{The disordered field theory}
\label{sec:n2qft}

We will  focus on models with $\mathcal{N}=2$ global supersymmetry in three spacetime dimensions.
 To set the stage for the discussion we begin by outlining some basic features and construct the microscopic Lagrangian. We work directly in Euclidean signature as we are interested in vacuum dynamics. We begin in \cref{sec:model} by describing the model of primary interest, and  describe the Schwinger-Dyson equations for the analysis of the RG flow in \cref{sec:ir2pt}.  

\subsection{The 3d model}
\label{sec:model}

The basic field content of an $\mathcal{N}=2$ chiral multiplet in three dimensions comprises of a complex scalar $\phi$, a two component fermion $\psi_\alpha$ and an auxiliary field $F$. These can be succinctly encoded into a single chiral superfield 
\begin{equation}\label{eq:chiral}
\Phi(X)  = \phi(y) + \sqrt{2}\, \theta^\alpha \,\psi_\alpha(y) + \theta^2\, F(y)
\end{equation}	
where $y^\mu \equiv x^\mu - i\, \theta\, \sigma^\mu \thetab$ is the chiral coordinate in superspace $\mathbb{R}^{3|4}$ which we have chosen to coordinatize by $X^\mu =\{x^\mu, \theta_{\sf 1}, \thetab_{\sf 1}, \theta_{\sf 2}, \thetab_{\sf 2}\}$. We delineate some of the details of our supersymmetry conventions in \cref{sec:conventions}.  The anti-chiral superfield is likewise
\begin{equation}\label{eq:achiral}
\Phib(X^\dagger)  = \phib(y^\dagger) + \sqrt{2}\, \thetab^\alpha \,\psib_\alpha(y^\dagger) + \thetab^2\, \Fb(y^\dagger)
\end{equation}	
with $y^{\dagger \mu}  \equiv x^\mu + i\, \theta\, \sigma^\mu \thetab$.  
The theory we  study will have $N$ such chiral and anti-chiral superfields $\Phi_i$, $\Phib^i$, respectively, with $i =1,2,\cdots, N$ in three dimensions. The classical Lagrangian density for this system is a generalized Wess-Zumino model, and comprises of a canonical K\"ahler term and a cubic superpotential with random couplings, viz.,
\begin{equation}\label{eq:syk3dLs}
\mathcal{L} = -\int d^2 \theta \, d^2\thetab\; \Phib^i(y^\dagger)\, \Phi_i(y) - \left[ \int\, d^2\theta\, \frac{1}{3}\, g^{ijk}\, \Phi_i(y) \,\Phi_j(y)\, \Phi_k(y)  + \text{c.c} \right] .
\end{equation}	
The model has a discrete $\mathbb{Z}_3$ global symmetry that phase rotates the chiral superfields $\Phi_i$ by a cubic root of unity.  We give the salient results for an  $q$-body superpotential in arbitrary dimensions in  \cref{sec:genqd}  (though the IR fixed points exist only in $d\leq 3$) for comparison with existing results in lower dimensions \cite{Anninos:2016szt,Murugan:2017eto}. 

The couplings $g^{ijk}$ are taken to be random Gaussian variables, with zero mean and non-vanishing variance  which we normalize suitably to obtain a large $N$ fixed point. They are drawn from a classical ensemble with probability distribution:
\begin{equation}\label{eq:gdist}
\mathcal{P}(g^{ijk}) \propto e^{-N^2\, \frac{g^{ijk} \overline{g}_{ijk}}{3J}} \,, \qquad \expval{ g^{ijk} } = 0 \,, \qquad 
\expval{ g^{ijk} \overline{g}_{pqr}  
} =  \frac{3J}{N^2}\, \delta^i_{(p}\, \delta^k_q\, \delta^k_{r)} \,.
\end{equation}	
The cubic superpotential ensures that the interaction term remains a relevant operator. We can view the theory with  fixed couplings $g^{ijk}$ as generalized Wess-Zumino model or  as a  $\mathcal{N}=2$ super-Ising model.  Indeed, expanding out the superpotential, we find the Lagrangian density 
\begin{equation}\label{eq:syk3d}
\mathcal{L} = -i\, \psib^i\, \slashed{\partial} \psi_i + \partial_\mu \phib^i\, \partial_\mu \phi_i -  \Fb^i \, F_i - g^{ijk} \left(\phi_i \, \phi_j \, F_k - \psi_i\, \psi_j\, \phi_k\right) -  \overline{g}_{ijk} \left(\phib^i \, \phib^j \, \Fb^k - \psib^i\, \psib^j\, \phib^k\right) .
\end{equation}	
Integrating out the auxiliary field $F$ we see that we induce a quartic scalar potential, thus making contact with a Wilson-Fisher like interaction. We note in passing that the undisordered models have been analyzed in the $d=4-\epsilon$ expansion in  \cite{Avdeev:1982jx,Jack:1998iy}.

The  critical point for the undisordered 3d critical Wess-Zumino theory with $N=1$ was analyzed using superconformal bootstrap in \cite{Bobev:2015vsa,Bobev:2015jxa}. We will compare our results to their numerical bootstrap data as well as the general results from analytic bootstrap \cite{Fitzpatrick:2012yx,Komargodski:2012ek,Costa:2012cb} in the course of our discussion. There are related models where one of the superfields is singled out to give a vector-like large $N$ model which will be analyzed elsewhere \cite{Chang:2021vm} (though we make some brief comments in \cref{sec:discuss}).

For the discussion that follows it is useful to record the fact that in the microscopic (UV) theory the scalar fields $\phi_i$ have scaling dimension half, while those of the fermion $\psi_{\alpha,i}$ and the auxiliary field $F_i$ are unity  and $\frac{3}{2}$, respectively.  Let us define the propagators:\footnote{ We will consistently write the two-point functions, self-energies etc.,  with the anti-chiral operator preceding the chiral operator. The fermionic objects in matrix form are boldfaced $\mathbf{G}, \bm{\Sigma}$, but in component form are simply characterized by the  fermionic indices, i.e., $G\indices{_\alpha^\beta}$. The bosonic functions are subscripted by the corresponding field, and bare (free) propagators carry a tilde decoration. Superspace Green's functions will be disambiguated by a mathscript font ($\mathscr{G}$). The argument of the function should make clear whether we are in position space or in the Fourier domain. }
\begin{equation}\label{eq:propG}
\begin{split}
G_\phi(x_{12}) \, \delta^i_j 
&=
	\expval{\phib^i(x_1)\,\phi_j(x_2)},\\
G_\alpha{}^\beta(x_{12})\,\delta^i_j 
&=
	\expval{ \psib_{\alpha}^i(x_1)\,\psi^\beta_{j}(x_2)},\\
G_F(x_{12})\,\delta^i_j 
&=
	\expval{\Fb^i(x_1)\,F_j(x_2)}.
\end{split}
\end{equation}
In the UV the free propagators (denoted by a tilde accent) of the non-interacting theory in momentum space are simply
\begin{equation}\label{eq:G0s}
\Gt_\phi(p) = \frac{1}{p^2},
\qquad 
\Gt\indices{_\alpha^\beta}(p)= - \frac{p^\mu(\sigma^\mu)_\alpha{}^\beta}{p^2},
\qquad 
\Gt_F(p)=-1\,,
 \end{equation}	
consistent with the classical engineering dimensions. The IR fixed point will be dominated by the superpotential and thus leads to non-trivial anomalous dimensions for these operators.

\subsection{The low energy fixed point}
\label{sec:ir2pt}

The effective action for the field theory can be obtained by the path integral for the collective fields after integrating out the bare fields. We define 
\begin{equation}\label{eq:SeffA}
\begin{split}
e^{-S_\text{eff}}=
	\int  \prod d g_{ijk} \,e^{-N^2\, \frac{\abs{g^{ijk}}^2}{3J}} 
	\int [\mathcal{D}\phi_i]\,[\mathcal{D}\psi_{\alpha,i}]\,[\mathcal{D}F_i] \; e^{-\int d^3 x \,\mathcal{L}}\ .
\end{split}
\end{equation}
The collective field effective action is encoded in terms of the two-point functions and self-energies. Integrating out the random couplings which are Gaussian distributed,  the effective action to the leading order in $\frac{1}{N}$ reads \footnote{Strictly speaking we only consider the replica diagonal solution in this analysis. Alternatively, we can promote the coupling $g_{ijk}$ to a slow varying ``heavy" superfield; giving it a vev leads to the same dynamics as we analyze here. The fact that such a field is ``heavy" is automatic in the large $N$ limit since it has 3 indices and any quantum corrections to it is suppressed by powers of $N$ \cite{Peng:2017kro}, therefore this is a valid analysis. We thank Jinwu Ye for raising this question. }
\begin{equation}\label{eq:SeffGS}
\begin{split}
\frac{1}{N}\, S_\text{eff} 
= \int dx dy \Bigg[
&
	-\log \det \left[ i\,\delta(x-y)\, (\sigma^\mu \partial_\mu)^{\alpha\beta}+\Sigma^{\alpha\beta}(x,y)\right] \\
&
	 + \log\det \left[\delta(x-y)\,\partial_{\mu}\partial^{\mu}  +\Sigma_\phi(x,y)\right] +\log\det \left[\delta(x-y)  +\Sigma_F(x,y) \right]\\
&
	+ \Sigma^{\alpha\beta}(x,y)G_{\alpha\beta}(x,y) +\Sigma_\phi(x,y)G_\phi (x,y)
	+\Sigma_F(x,y)G_F (x,y) \\
&
	  -
	J\, \Bigg\{ G_\phi(x,y)^2 \, G_F(x,y)+ 2\, G_\phi(x,y) \, \det(\mathbf{G})
	\Bigg\}\Bigg] \ .
\end{split}
\end{equation}

\begin{figure}[h]
\centering
\begin{tikzpicture}[scale=0.9]
\draw [thick](-2,0)--(-0.5,0);
\draw [thick] (0.5,0) -- (2,0);
\draw [thick, fill=none] (0,0) circle (0.5cm);
\draw (0,0) node{$\scriptstyle{G}$};

\draw (2.5,0) node{$=$};

\draw [thick](3,0)--(6,0);
\draw (6.5,0) node{$+$};

\draw [thick](7,0)--(8,0);
\draw[thick] (8,0) to[out=60,in=210] (9,1);
\draw[thick, fill=none] (9.5,1) circle (0.5cm);
\draw (9.5,1) node{$\scriptstyle{G}$};
\draw[thick]  (10,1) to[out=-30,in=120] (11,0);

\draw[thick] (8,0) to[out=-60,in=150] (9,-1);
\draw[thick, fill=none] (9.5,-1) circle (0.5cm);
\draw (9.5,-1) node{$\scriptstyle{G}$};
\draw[thick]  (10,-1) to[out=30,in=-120] (11,0);

\draw [thick](11,0)--(12,0);
\draw[thick, fill=none] (12.5,0) circle (0.5cm);
\draw (12.5,0) node{$\scriptstyle{G}$};
\draw [thick](13,0)--(14,0);

\end{tikzpicture}
\caption{ The diagrammatic representation of the  Schwinger-Dyson equation for cubic superpotential \eqref{eq:3dSD}.}
\label{fig:SDprop}
\end{figure}
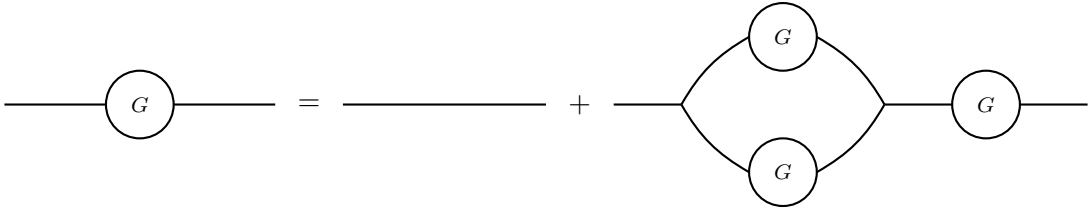

In the large $N$ limit, the two-point functions can be seen to satisfy the Schwinger-Dyson equations
\begin{equation}\label{eq:3dSD}
\begin{split}
G_\phi(x_{12})
&=
	\Gt(x_{12})+\int d^3x_3d^3x_4\; \Gt(x_{13})\, \Sigma_\phi(x_{43}) \, G_\phi(x_{42}),
\\
G\indices{_\alpha^\beta}(x_{12})
&=
	\Gt\indices{_\alpha^\beta}(x_{12})
	+	\int d^3x_3d^3x_4\; \Gt\indices{_\alpha^\gamma}(x_{13}) \, \Sigma^{\delta}{}_\gamma(x_{43})
	\, G\indices{_\delta^\beta}(x_{42}),
\\
G_F(x_{12}) 
&=
	\Gt_F(x_{12})+\int d^3x_3d^3x_4\; \Gt_F(x_{13})\, \Sigma_F(x_{43})\, G_F(x_{42}),
\end{split}
\end{equation}
with the self-energies given by
\begin{equation}\label{eq:3dSigma}
\begin{split}
\Sigma_\phi(x)&= J \left[2\, G_F(x)G_\phi(x) - G\indices{_\alpha^\beta}(x) \,G\indices{^\alpha_\beta}(x) \right],
\\
\Sigma\indices{_\alpha^\beta}(x)&= 2\, J \, G\indices{_\alpha^\beta}(x) \, G_\phi(x) \,, 
\\
\Sigma_F(x)&= J \, G_\phi(x)^2\,.
\end{split}
\end{equation}
The diagrammatic derivation of these equations follows along similar lines to that of the SYK model \cite{Maldacena:2016hyu}. The truncation to the simple set of closed form equations owes to the random couplings that suppresses the higher-point interactions from appearing at the leading order in the $\frac{1}{N}$ expansion, which diagrammatically is illustrated in \cref{fig:SDprop}. 

In momentum space representation the Schwinger-Dyson equations take the form:
\begin{equation}\label{eq:3dSDmom}
\begin{split}
G_\phi(p)
&= \frac{1}{\Gt_\phi(p)^{-1}-\Sigma_\phi(-p)}
	= \frac{1}{ p^2-\Sigt^{\phib\phi}(-p)},
\\
\mathbf{G}(p)
&= 
	\left(\widetilde{\mathbf{G}}(p)^{-1}- \bm{\Sigma}(- p)\right)^{-1}
	=\left(-p^\mu\sigma^\mu-\bm{\Sigma}(- p)\right)^{-1},
\\
G_F(p)
&=
	\frac{1}{\Gt_F^{-1}-\Sigma_F(-p)}
	= \frac{1}{ -1-\Sigma_F(-p)} \,.
\end{split}
\end{equation}	
Since the theory is supersymmetric, the two-point functions and self-energies in a supersymmetric vacuum ought to satisfy the Ward identities 
\begin{equation}\label{eq:susyWard}
\begin{split}
\mathbf{G}(x) 
&=
	-i \,\slashed{\partial} G_\phi(x),\qquad G_F(x)=\partial^2G_\phi(x), \\
\bm{\Sigma}(x) 
&=
	-i \,\slashed{\partial} \Sigma_{F}(x),\qquad \Sigma_\phi(x)=\partial^2\Sigma_F(x).
\end{split}
\end{equation}	

The Schwinger-Dyson equations can readily be solved, and one is explicitly aided by the underlying supersymmetry. It suffices to solve for the scalar propagator, exploiting the Ward identities \eqref{eq:susyWard} to obtain the low energy fixed point.  In momentum space, we obtain the relations:
\begin{equation}\label{eq:IRward}
G_\phi(p) \, \Sigma_\phi(p) = - 1 = -p^2\, G_\phi(p) \, \Sigma_F(p) \,.
\end{equation}	
We pick  a scale invariant anstaz for the propagators
\begin{equation}\label{eq:Gphiansatz}
G^*_\phi(x,y) = \frac{b_\phi}{|x-y|^{2\Delta_\phi}} \,,
\end{equation}	
which leads upon Fourier transforming to (cf., \eqref{eq:propFT})
\begin{equation}
\begin{split}
G^*_\phi(p) 
&= 
	\pi^\frac{3}{2} \, 2^{3-2\Delta_\phi} \, b_\phi \, \frac{\Gamma(\frac{3}{2} - \Delta_\phi)}{\Gamma(\Delta_\phi)} \, 
	|p|^{2\Delta_\phi -3} \,, \\
\Sigma^*_F(p) 
&= 
	J\, \pi^\frac{3}{2} \, 2^{3-4\Delta_\phi} \, b_\phi^2 \, \frac{\Gamma(\frac{3}{2} - 2\Delta_\phi)}{\Gamma(2\Delta_\phi)} \, 
	|p|^{4\Delta_\phi -3} \,,
\end{split}
\end{equation}
The solution is given by demanding consistency with the last Schwinger-Dyson equation  in \eqref{eq:3dSigma} and is given by
\begin{equation}\label{eq:Dbphi}
\Delta_\phi = \frac{2}{3} \,, \qquad b_\phi = \frac{1}{2^\frac{2}{3} \, \sqrt{3}\, \pi\, J^\frac{1}{3}} \,.
\end{equation}	
 The conformal dimensions of the other fields are fixed by supersymmetry to be 
\begin{equation}
\Delta_\psi = \frac{7}{6} \,, \qquad \Delta_F = \frac{5}{3} \,.
\end{equation}	

A quick self-consistency check of the solution can be provided by noting that the self-energy of the scalar field dominates over the bare propagator since  $\Sigma^*_\phi \sim |p|^\frac{5}{3}$ and that the Fourier transformations involved are UV finite, obviating any potential source of supersymmetry breaking along the RG flow.\footnote{ While this behaviour would be unusual it does  occur in some supersymmetric tensor quantum mechanical models \cite{Chang:2018sve} due to strong IR effects in low dimensions. } 

Note that while the collective field action focuses on the singlet correlators it is natural to assume that the low energy SCFT has $N$ chiral multiplets $\Phi_i$ with $\Delta(\Phi_i) = \Delta_\phi = \frac{2}{3}$.\footnote{The $U(N)$ symmetry can also be argued by promoting the random coupling $g^{ijk}$ into a random constant superfield that transforms in the tri-fundamental representation of $U(N)$. Such a symmetry is manifest in a similar analysis \cite{Peng:2017kro}.} Since the spectrum is degenerate there is a $U(N)$ symmetry rotating these superfields into each other. We will make use of this observation below in analyzing higher-point functions.

\section{Spectrum of the IR superconformal fixed point}
\label{sec:ir}

We now turn to analyzing aspects of the fixed point  theory in the infra-red. Since the chiral spectrum is determined by the superconformal symmetry, we will investigate the non-chiral $4$-point function. We find it useful to begin with a general correlator where the external indices are labeled by the $N$ chiral or anti-chiral superfields present in the microscopic theory. This will allow us to understand the spectrum of both the singlet sector and the non-singlet sector under an emergent $U(N)$ rotation of the chiral multiplets $\Phi_i$. 

Our first task will be to obtain the ladder kernel which can be iterated to give a geometric series representation for the 4-point function; this turns out to be most straightforward in superspace. We solve for the eigenspectrum of the ladder kernel to obtain the spectrum of intermediate states in the $\Phib \times \Phi$ OPE demonstrating, in particular, the existence of a stress tensor multiplet.  We explore the asymptotic features of the spectrum and compare against analytic bootstrap and Regge limit predictions. We also take the opportunity to contrast our model with lower dimensional models investigated previously in \cite{Anninos:2016szt,Murugan:2017eto} (and also \cite{Fu:2016vas,Peng:2018zap,Bulycheva:2018qcp}).  We will investigate the four-point function itself in greater detail in \cref{sec:4pt} and \cref{sec:chiral} where we will extract the OPE data and central charges.

\subsection{The  general four-point correlator}
\label{sec:4average}

Let us start by considering the most general un-normalized disorder averaged four-point function  of the chiral and anti-chiral superfields
\begin{equation}\label{eq:4ptikjl}
\mathcal{W}_{kj}^{il}(\Xb_{1},X_{2},X_{3},\Xb_{4}) 
= 
  \expval{\overline{\Phi}^i(\Xb_{1})\,\Phi_k(X_{2}) \, \Phi_j(X_{3})\, \overline{\Phi}^l(\Xb_{4})} \,.
\end{equation}  
where $X_i$ are the super-coordinates on $\mathbb{R}^{3|4}$.

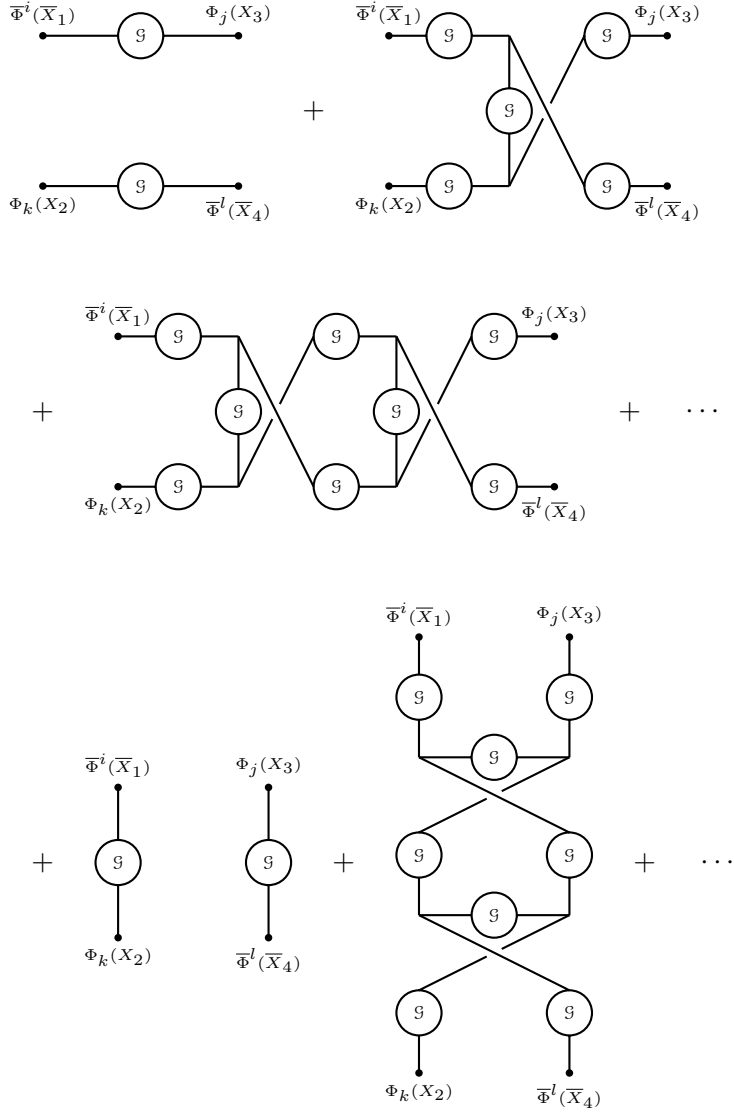
\begin{figure}[tp!]
\centering
\begin{tikzpicture}

\draw[thick, shift={(-2.6,1)}] (0,0) to (1,0) (1.3,0) circle (.3cm) (1.3,0) node{$\scriptscriptstyle{\mathscr{G}}$} (1.6,0) to (2.6,0) (0,0) node[circle,fill,inner sep=1pt]{} (0,0) node[above]{$\scriptscriptstyle{\Phib^i(\Xb_1)}$} (2.6,0) node[circle,fill,inner sep=1pt]{} (2.6,0) node[above]{$\scriptscriptstyle{\Phi_j(X_3)}$};
\draw[thick, shift={(-2.6,-1)}] (0,0) to (1,0) (1.3,0) circle (.3cm) (1.3,0) node{$\scriptscriptstyle{\mathscr{G}}$} (1.6,0) to (2.6,0)
(0,0) node[circle,fill,inner sep=1pt]{} (0,0) node[below]{$\scriptscriptstyle{\Phi_k(X_2)}$} (2.6,0) node[circle,fill,inner sep=1pt]{} (2.6,0) node[below]{$\scriptscriptstyle{\Phib^l(\Xb_4)}$};

\draw (1,0) node{$+$};

\draw[thick, shift={(2,1)}] (0,0) to (.5,0) (.8,0) circle (.3cm) (.8,0) node{$\scriptscriptstyle{\mathscr{G}}$} (1.1,0) to (1.6,0) (0,0) node[circle,fill,inner sep=1pt]{} (0,0) node[above]{$\scriptscriptstyle{\Phib^i(\Xb_1)}$};
\draw[thick, shift={(2,-1)}] (0,0) to (.5,0) (.8,0) circle (.3cm) (.8,0) node{$\scriptscriptstyle{\mathscr{G}}$} (1.1,0) to (1.6,0) (0,0) node[circle,fill,inner sep=1pt]{} (0,0) node[below]{$\scriptscriptstyle{\Phi_k(X_2)}$};

\draw[thick,shift={(3.6,0)}] 
(0,-1) to (0,-.3)
(0,0) circle (.3cm)
(0,0) node{$\scriptscriptstyle{\mathscr{G}}$}
(0,.3) to  (0,1);

\draw[thick, shift={(3.6,1)}] (0,0) to (1,-2) (1.3,-2) circle (.3cm) (1.3,-2) node{$\scriptscriptstyle{\mathscr{G}}$} (1.6,-2) to (2.1,-2) (2.1,-2) node[circle,fill,inner sep=1pt]{} (2.1,-2) node[below]{$\scriptscriptstyle{\Phib^l(\Xb_4)}$};
\draw[thick, shift={(3.6,-1)}] (0,0) to (0.45,0.9) (0.55,1.1) to (1,2)  (1.3,2) circle (.3cm) (1.3,2) node{$\scriptscriptstyle{\mathscr{G}}$} (1.6,2) to (2.1,2) (2.1,2) node[circle,fill,inner sep=1pt]{} (2.1,2) node[above]{$\scriptscriptstyle{\Phi_j(X_3)}$};

\draw (-2.6,-4) node{$+$};

\draw[thick, shift={(-1.6,-3)}] (0,0) to (.5,0) (.8,0) circle (.3cm) (.8,0) node{$\scriptscriptstyle{\mathscr{G}}$} (1.1,0) to (1.6,0) (0,0) node[circle,fill,inner sep=1pt]{} (0,0) node[above]{$\scriptscriptstyle{\Phib^i(\Xb_1)}$};
\draw[thick, shift={(-1.6,-5)}] (0,0) to (.5,0) (.8,0) circle (.3cm) (.8,0) node{$\scriptscriptstyle{\mathscr{G}}$} (1.1,0) to (1.6,0) (0,0) node[circle,fill,inner sep=1pt]{} (0,0) node[below]{$\scriptscriptstyle{\Phi_k(X_2)}$};

\draw[thick,shift={(0,-4)}] 
(0,-1) to (0,-.3)
(0,0) circle (.3cm)
(0,0) node{$\scriptscriptstyle{\mathscr{G}}$}
(0,.3) to  (0,1);

\draw[thick, shift={(.5,-3)}] (-.5,0) to (.5,-2) (.8,0) circle (.3cm) (.8,0) node{$\scriptscriptstyle{\mathscr{G}}$} (1.1,0) to (1.6,0);
\draw[thick, shift={(.5,-5)}] (-.5,0) to (-.05,.9) (.05,1.1) to (.5,2) (.8,0) circle (.3cm) (.8,0) node{$\scriptscriptstyle{\mathscr{G}}$} (1.1,0) to (1.6,0);

\draw[thick,shift={(2.1,-4)}] 
(0,-1) to (0,-.3)
(0,0) circle (.3cm)
(0,0) node{$\scriptscriptstyle{\mathscr{G}}$}
(0,.3) to  (0,1);

\draw[thick, shift={(2.6,-3)}] (-.5,0) to (.5,-2) (.8,0) circle (.3cm) (.8,0) node{$\scriptscriptstyle{\mathscr{G}}$} (1.1,0) to (1.6,0) 
(1.6,0) node[circle,fill,inner sep=1pt]{} (1.6,0) node[above]{$\scriptscriptstyle{\Phi_j(X_3)}$};
\draw[thick, shift={(2.6,-5)}] (-.5,0) to (-.05,.9) (.05,1.1) to (.5,2) (.8,0) circle (.3cm) (.8,0) node{$\scriptscriptstyle{\mathscr{G}}$} (1.1,0) to (1.6,0) 
(1.6,0) node[circle,fill,inner sep=1pt]{} (1.6,0) node[below]{$\scriptscriptstyle{\Phib^l(\Xb_4)}$};

\draw (5.2,-4) node{$+$};
 \draw (6.2,-4) node{$\cdots$};

\draw (-2.6,-10) node{$+$};

\draw[thick, shift={(-1.6,-10)}] (0,-1) to (0,-.3) (0,0) circle (.3cm)
(0,0) node{$\scriptscriptstyle{\mathscr{G}}$}
(0,.3) to  (0,1)
(0,1) node[circle,fill,inner sep=1pt]{} (0,1) node[above]{$\scriptscriptstyle{\Phib^i(\Xb_1)}$} (0,-1) node[circle,fill,inner sep=1pt]{} (0,-1) node[below]{$\scriptscriptstyle{\Phi_k(X_2)}$};

\draw[thick, shift={(.4,-10)}] (0,-1) to (0,-.3) (0,0) circle (.3cm)
(0,0) node{$\scriptscriptstyle{\mathscr{G}}$}
(0,.3) to  (0,1)
(0,1) node[circle,fill,inner sep=1pt]{} (0,1) node[above]{$\scriptscriptstyle{\Phi_j(X_3)}$} (0,-1) node[circle,fill,inner sep=1pt]{} (0,-1) node[below]{$\scriptscriptstyle{\Phib^l(\Xb_4)}$};

 \draw (1.4,-10) node{$+$};

\draw[thick, shift={(2.4,-8)}] (0,-.6) to (0,-.1) (0,.2) circle (.3cm)
(0,.2) node{$\scriptscriptstyle{\mathscr{G}}$}
(0,.5) to  (0,1)
(0,1) node[circle,fill,inner sep=1pt]{} (0,1) node[above]{$\scriptscriptstyle{\Phib^i(\Xb_1)}$};
\draw[thick, shift={(4.4,-8)}] (0,-.6) to (0,-.1) (0,.2) circle (.3cm)
(0,.2) node{$\scriptscriptstyle{\mathscr{G}}$}
(0,.5) to  (0,1)
(0,1) node[circle,fill,inner sep=1pt]{} (0,1) node[above]{$\scriptscriptstyle{\Phi_j(X_3)}$};

\draw[thick,shift={(3.4,-8.6)}] 
(-1,0) to (-.3,0)
(0,0) circle (.3cm)
(0,0) node{$\scriptscriptstyle{\mathscr{G}}$}
(.3,0) to  (1,0);

\draw[thick, shift={(2.4,-10.1)}] (0,-.6) to (0,-.1) (0,.2) circle (.3cm)
(0,.2) node{$\scriptscriptstyle{\mathscr{G}}$}
(0,.5) to  (.9,.95) (1.1,1.05) to  (2,1.5);
\draw[thick, shift={(4.4,-10.1)}] (0,-.6) to (0,-.1) (0,.2) circle (.3cm)
(0,.2) node{$\scriptscriptstyle{\mathscr{G}}$}
(0,.5) to  (-2,1.5);

\draw[thick,shift={(3.4,-10.7)}] 
(-1,0) to (-.3,0)
(0,0) circle (.3cm)
(0,0) node{$\scriptscriptstyle{\mathscr{G}}$}
(.3,0) to  (1,0);

\draw[thick, shift={(2.4,-12.2)}] (0,-.6) to (0,-.1) (0,.2) circle (.3cm)
(0,.2) node{$\scriptscriptstyle{\mathscr{G}}$}
(0,.5) to  (.9,.95) (1.1,1.05) to  (2,1.5)
(0,-.6) node[circle,fill,inner sep=1pt]{} (0,-.6) node[below]{$\scriptscriptstyle{\Phi_k(X_2)}$};
\draw[thick, shift={(4.4,-12.2)}] (0,-.6) to (0,-.1) (0,.2) circle (.3cm)
(0,.2) node{$\scriptscriptstyle{\mathscr{G}}$}
(0,.5) to  (-2,1.5)
(0,-.6) node[circle,fill,inner sep=1pt]{} (0,-.6) node[below]{$\scriptscriptstyle{\Phib^l(\Xb_4)}$};

\draw (5.4,-10) node{$+$};
 \draw (6.4,-10) node{$\cdots$};

\end{tikzpicture}
\caption{The diagrammatic expansion of the four-point function of superfield $\Phi$ and its conjugate \eqref{eq:4ptikjl} in the disorder averaged large $N$ expansion.}
\label{fig:ladder}
\end{figure}

The 4-point function \eqref{eq:4ptikjl} can be evaluated in the large $N$  expansion, thanks to the disorder averaging, by summing over a set of ladder diagrams, which we illustrate diagrammatically in \cref{fig:ladder}.  From the expansion it is easy to see, using \eqref{eq:gdist} for the disorder average, that  
\begin{equation}\label{eq:4ptgeneral}
\begin{split}
\mathcal{W}^{il}_{kj} (\Xb_{1},X_{2},X_{3},\Xb_{4})  &=    ( \mathbf{F}_0)^{il}_{kj}(\Xb_{1},X_{2},X_{3},\Xb_{4})+ \sum_{n=1}^\infty (\mathbf{K}^n \,\mathbf{F}_0)^{il}_{kj} (\Xb_{1},X_{2},X_{3},\Xb_{4})
\\
&\quad+( \mathbf{F}_0)^{il}_{jk}(\Xb_{1},X_{3},X_{2},\Xb_{4})+ \sum_{n=2}^\infty (\mathbf{K}^n \,\mathbf{F}_0)^{il}_{jk}\, (\Xb_{1},X_{3},X_{2},\Xb_{4})  \,.
\end{split}
\end{equation}  
The first line on the r.h.s of \eqref{eq:4ptgeneral} is a sum over the $s$-channel (horizontal) ladder diagrams starting from the zero-rung diagram, and the second line is a sum over the $u$-channel (vertical) ladder diagrams but without one-rung diagram, because the $s$- and $u$-channels have identical one-rung diagram. 
The quantities $(\mathbf{F}_0)^{il}_{kj}$ and $\mathbf{K}^{il}_{kj}$  are can be expressed as 
\begin{equation}\label{eq:bFKdef}
\begin{split}
(\mathbf{F}_0)^{il}_{kj}(\Xb_{1},X_{2},X_{3},\Xb_{4})
&=
  \delta^i_j\delta^l_k\, \mathcal{F}_0(\Xb_{1},X_{2},X_{3},\Xb_{4}) \\
\mathbf{K}^{il}_{kj}(\Xb_1,X_2,X_a, \Xb_b)
&= 
  \frac{N+2}{N^2} \, (\delta^i_k\delta^l_j+\delta^i_j\delta^l_k) \, K(\Xb_1,X_2,X_a, \Xb_b)\,,
\end{split}
\end{equation}
factoring out the contribution from the $U(N)$ tensor structure and super-coordinate dependence.  The latter is encoded in the disconnected 4-point function $\mathcal{F}_0$ and the ladder kernel $K$, which are themselves defined in terms of the super-propagator $\mathscr{G}(\Xb_1,X_2)$, given below in \eqref{eq:Gsuper}, as 
\begin{equation}
\begin{split}
\mathcal{F}_0(\Xb_{1},X_{2},X_{3},\Xb_{4}) 
& =  
  \mathscr{G}(\Xb_1,X_3) \,\mathscr{G}(\Xb_4,X_2)\,,
\\
K(\Xb_{1},X_{2},X_{a},\Xb_{b}) 
  &= 2\,J\, \mathscr{G}(\Xb_1,X_a) \mathscr{G}(\Xb_b,X_a) \, \mathscr{G}(\Xb_b,X_2) \,,
\end{split}
\end{equation}
The action of the kernel is defined by summing over intermediate $U(N)$ labels, and integrating over the intermediate vertex positions, viz.,  
\begin{equation}
\begin{split}
(\mathbf{K} \mathbf{F})^{il}_{kj}(\Xb_{1},X_{2},X_{3},\Xb_{4})
&\equiv 
\sum_{a,b=1}^N\mathbf{K}^{ia}_{kb}  (\mathbf{F})^{bl}_{aj}(\Xb_{1},X_{2},X_{3},\Xb_{4})\,, \\
K\, \mathcal{F}(\Xb_{1},X_{2},X_{3},\Xb_{4}) 
&\equiv 
  \int dX_{a}d\Xb_{b} \, K(\Xb_{1},X_{2},X_{a},\Xb_{b})  \, \mathcal{F}(\Xb_{b},X_{a},X_{3},\Xb_{4})\,.
\end{split}
\end{equation}

To compute the ladder sums note that the  $n^{\rm th}$ power of the kernel $\mathbf{K}^{il}_{kj}$ is 
\begin{equation}
\begin{split}
(\mathbf{K}^n)^{il}_{kj}= \frac{1}{N} \, K^n \delta^i_k\delta^l_j+ \order{N^{-2}}\quad{\rm for}\quad n>1\,,
\end{split}
\end{equation}  
where we have only retained the index label structure to isolate the large $N$ contributions.  This implies that in the large $N$ limit, we have
\begin{equation}\label{eq:4ptgeneralN}
\begin{split}
\mathcal{W}^{il}_{kj} (\Xb_{1},X_{2},X_{3},\Xb_{4})  
&=
  \delta^i_j\delta^l_k \left[
    \mathcal{F}_0(\Xb_{1},X_{2},X_{3},\Xb_{4})+ \frac{1}{N}\, \frac{K}{1-K} \mathcal{F}_0 (\Xb_{1},X_{3},X_{2},\Xb_{4})\right]
\\
&\quad+\delta^i_k\delta^l_j\
  \left[\mathcal{F}_0(\Xb_{1},X_{3},X_{2},\Xb_{4}) +  \frac{1}{N} \, 
    \frac{K}{1-K}\,  \mathcal{F}_0(\Xb_{1},X_{2},X_{3},\Xb_{4}) \right] \\
&\qquad    
  + \order{N^{-2}} \,.
\end{split}
\end{equation}  

In the IR the chiral superfields $\Phi_i$ are  degenerate in their conformal dimension $\Delta_\phi = \frac{2}{3}$, so it is sensible to decompose the correlator \eqref{eq:4ptikjl} using the $U(N)$ symmetry that rotates them. In terms of the projection matrices $P_{\bf 1}$ and $P_{\bf adj}$ that project onto the contributions of operators in the $\overline{\Phi}_i\times \Phi_k$ OPE that transform in the trivial or adjoint representations of the $SU(N)$, one can expand
\begin{equation}\label{eqn:Wikjldecompose}
\mathcal{W}^{il}_{kj}
  = \left(P_\mathbf{1}\right)^{il}_{kj} \;  \mathcal{W}_\mathbf{1} + \left(P_{\bf adj}\right)^{il}_{kj} \; \mathcal{W}_\mathbf{adj} \,.
\end{equation}
The projection matrices themselves are formed by bilinears of the Clebsch-Gordan coefficients as
\begin{equation}
\left(P_{\bf r}\right)^{il}_{kj} \equiv \sum_{a=1}^{{\rm dim}({\bf r})} \left(c_{\bf r}\right)^i_{k,a}\; 
  \left(c_{{\bf r}}\right)^l_{j,a}\quad{\rm for}\quad{\bf r}={\bf 1},{\bf adj}\,.
\end{equation}
and obey the orthogonality condition
\begin{equation}
\sum_{k,l=1}^N \left(P_{\bf r}\right)^{il}_{kj} \; \left(P_{{\bf r}'}\right)^{jn}_{l m} = \delta_{{\bf r}{\bf r}'} \; (P_{\bf r})^{im}_{k m}\,.
\end{equation}
Explicitly, they are given by
\begin{equation}
\left(P_{\bf 1}\right)^{il}_{kj} = \frac{1}{N}\, \delta^i_{k} \delta^l_{j}\,,\quad \left(P_{\bf adj}\right)^{il}_{kj} = \delta^i_{j} \delta^l_{k} - \frac{1}{N} \delta^i_{k} \delta^l_{j}\,.
\end{equation}

As a consequence it follows that the contribution to the singlet and the adjoint sectors can be isolated as 
\begin{equation}\label{eq:Wsinglet}
\begin{split}
\mathcal{W}_{\bf 1}(\Xb_{1},X_{2},X_{3},\Xb_{4})
&=
  N\, \mathcal{F}_0(\Xb_{1},X_{3},X_{2},\Xb_{4})+ \frac{1}{1-K} \, 
    \mathcal{F}_0(\Xb_{1},X_{2},X_{3},\Xb_{4}) \\
& \qquad 
  + \frac{1}{N} \, \frac{K}{1-K}\, \mathcal{F}_0(\Xb_{1},X_{3},X_{2},\Xb_{4})+ \order{N^{-1}}\,, 
\end{split}
\end{equation}
and 
\begin{equation}\label{eq:Wadj} 
\mathcal{W}_{\bf adj}(\Xb_{1},X_{2},X_{3},\Xb_{4})
=
  \mathcal{F}_0(\Xb_{1},X_{2},X_{3},\Xb_{4}) + \frac{1}{N} \, \frac{K}{1-K}\mathcal{F}_0(\Xb_{1},X_{3},X_{2},\Xb_{4})+ \order{N^{-2}}\,,
\end{equation}
respectively. 

It is common in the SYK literature to focus on the singlet contribution to the OPE, which can equivalently be isolated by considering the external operator averaged correlation function: 
\begin{equation}\label{eq:Wsingletav}
\begin{split}
\mathcal{W}(\Xb_{1},X_{2},X_{3},\Xb_{4})  
&\equiv
  \frac{1}{N^2} \sum_{i,j,k,l=1}^N\, \delta^k_i\, \delta^j_l \, \mathcal{W}^{il}_{kj} (\Xb_{1},X_{2},X_{3},\Xb_{4}) \,,  \\
&= 
  \frac{1}{N}\, \mathcal{W}_\mathbf{1}(\Xb_{1},X_{2},X_{3},\Xb_{4})  \,.
\end{split}
\end{equation}  
There is nevertheless a non-singlet part which also can be extracted. We will for the most part focus on the singlet contribution but shall comment on the non-singlet part when we analyze the OPE data in \cref{sec:4pt}.

\subsection{The ladder kernel}
\label{sec:ladder}

Our discussion thus far has focused on the 4-point function itself, but to extract the OPE data it is sensible to normalize this by the two-point function of the theory. We therefore define the normalized correlator:\footnote{ The disorder average is taken independently for each correlation function appearing  in \eqref{eq:Wikjlnormal}.}
\begin{equation}\label{eq:Wikjlnormal}
\widehat{\mathcal{W}}^{il}_{kj} (\Xb_{1},X_{2},X_{3},\Xb_{4})  
\equiv \frac{\mathcal{W}^{il}_{kj} (\Xb_{1},X_{2},X_{3},\Xb_{4}) }{\mathscr{G}(\Xb_1,X_2)\, \mathscr{G}(\Xb_4,X_3)} \,. 
\end{equation}  
We will use the same normalization for the singlet and non-singlet pieces in the decomposition. Note that in particular, with this normalization one has for the singlets 
\begin{equation}\label{eq:4ptfn}
\begin{split}
\widehat{\mathcal{W}}(\Xb_{1},X_{2},X_{3},\Xb_{4}) 
= 
  1+  \frac{1}{N}\, \widehat{\mathcal{F}}(\Xb_1, X_2,X_3,\Xb_4)  + \order{N^{-2}} \,,
\end{split}
\end{equation}  
where we define the connected contribution to the normalized Euclidean four-point function as
\begin{equation}\label{eq:Faverage}
\widehat{\mathcal{F}} (u,v)
=   \frac{1}{1-K}\, \widehat{\mathcal{F}}_0 = 
  \frac{1}{1-K}\, \frac{\mathscr{G}(\Xb_1,X_3)\, \mathscr{G}(\Xb_4,X_2)}{\mathscr{G}(\Xb_1,X_2)\, \mathscr{G}(\Xb_4,X_3)} \,.
\end{equation}  
We have gauge fixed the supercoordinates and written the result in terms of the conformal cross-ratios 
$u,v$. They  can be obtained from the supersymmetric cross-ratios 
$\hat{u} = \frac{z_{12}^2\, z_{43}^2}{z_{13}^2\, z_{42}^2}$ and $\hat{v}= \frac{z_{1234}^2}{z_{13}^2\, z_{42}^2}$ built from the superspace translation invariant \eqref{eq:superz} and another combination $z_{1234}$. The precise form of the latter is  unnecessary as we can use the 8 fermionic generators of the Euclidean superconformal group $OSp(4|2,2)$ to set $\thetab_1= \theta_2 = \theta_3=\thetab_4=0$ and reduce to the standard cross-ratios constructed using $x_{ij}$.

With this preamble we can proceed to analyze the intermediate states appearing in the OPE decomposition of the 4-point function. All we need is the super-propagator, $\mathscr{G}$,  to evaluate \eqref{eq:Faverage}, which can be obtained from the scalar propagator computed hitherto, viz., 
\begin{equation}\label{eq:Gsuper}
\mathscr{G}(\Xb_1,X_2) = G_\phi(z_{12}) \,,
\end{equation}  
using the superspace translational invariant combination: 
\begin{equation}\label{eq:superz}
z_{ij}^\mu  \equiv x_i^\mu-x_j^\mu+i \theta_i \sigma^\mu\thetab_i +i \theta_j \sigma^\mu\thetab_j-2 i \theta_j \sigma^\mu \thetab_i 
= y^{\dagger\mu}_i - y^\mu_j - 2 i \theta_j \sigma^\mu \thetab_i \,.
\end{equation}  
%

We need to obtain the eigenspectrum of the ladder kernel -- diagonalizing it will give us the intermediate states in the $\Phib \times\Phi$ operator product expansion. We can utilize the observation from \cite{Maldacena:2016hyu} that these are given in terms of the (super)conformal three-point functions, $\mathcal{T}_{\tau, \ell}$ where the labels correspond to the spin $\ell$ and the conformal dimension $\Delta$, respectively. The latter can be traded for the twist $\tau = \Delta - \ell$. The eigenfunctions are simplest when one of the operator in the 3-point function is taken to infinity, whence 
\begin{equation}
\mathcal{T}_{\Delta, \ell}(\Xb_4, X_3) = |z_{43}|^{\Delta -\ell -2 \Delta_\phi} \, z_{43, \mu_1} \, \cdots z_{43, \mu_\ell} \; \mathcal{A}^{\mu_1\, \cdots \mu_\ell}\,,
\end{equation}	
where $\mathcal{A}^{\mu_1 \mu_2 \cdots \mu_\ell}$ is a symmetric traceless tensor.  

The eigenvalue equation of the ladder kernel takes the form:
\begin{equation}\label{eq:kernelDl}
k(\Delta, \ell)\, \mathcal{T}_{\Delta , \ell}(\Xb_1,X_2) = \int d^3 x_a\, d^2\theta_a\;  \int  d^3x_b\, d^2\thetab_b \, K(\Xb_1, X_2, X_a, \Xb_b) \, \mathcal{T}_{\Delta ,\ell} (\Xb_b,X_a) \,.
\end{equation}	
Plugging in the expressions for the propagator and the 3-point function, and carrying out the integral in the expression above, we get after some algebra, the sought for eigenvalue:
\begin{equation}\label{eq:ksheqn}
\begin{split}
k(\Delta, \ell)
&= 
	(-1)^{\ell}\, 2^{2-2\Delta_\phi}(2\Delta_\phi-1)
		\frac{\Gamma(\Delta_\phi-1)\, \Gamma(2\,\Delta_\phi)}{\Gamma\left(\frac{\Delta_\phi}{2}\right)^2}
		\, \frac{\Gamma\left(\Delta_\phi-\frac{\Delta-\ell}{2}\right)\Gamma\left(\frac{\Delta+\ell}{2}+\frac{1-\Delta_\phi}{2}\right)}{\Gamma\left(2\Delta_\phi-\frac{\Delta-\ell}{2}\right)\Gamma\left(\frac{\Delta+\ell}{2}+\frac{1+\Delta_\phi}{2}\right)}.
\end{split}	
\end{equation}	

The operators that  appear in the singlet sector of the $\Phib \times \Phi$ OPE\footnote{The OPE decomposition of the operator averaged correlator (see \cref{sec:4ptfn}, Eq.~\eqref{eq:4ptfn_sconfblocks} for its definition) will be denoted without any index decoration on the superfields $\Phi$, $\Phib$. When we discuss the OPE for the general correlator $\mathcal{W}^{il}_{kj}$ we will indicate it with appropriate external operator labels.}  can be read off from the above. Their spectrum is determined by the condition 
\begin{equation}\label{eq:kernev1}
k(\Delta, \ell) =1\,,
\end{equation}	
which owes to the geometric series originating from the ladder summation, cf., \eqref{eq:Faverage}. There will be some additional states in the non-singlet part of the OPE decomposition which we will postpone till  \cref{sec:4ptfn}. For the remainder of this section we will examine the features of the singlet spectrum.

\subsection{Features of the IR spectrum}
\label{sec:IRprops}

There are several observations to be made about the spectrum of operators which we now turn to. We will argue that the spectrum is consistent with known bounds from the conformal bootstrap and confirm our identification of the low-energy dynamics being a non-trivial SCFT by exhibiting the presence of a stress-tensor multiplet. 

Before proceeding with these analyses, we first note that $\Delta=\ell=0$ the eigenvalue equation is proportional to the Schwinger-Dyson equation itself. For a general $q$-body interaction, we expect based on the analysis in \cite{Maldacena:2016hyu,Murugan:2017eto} the coefficient of proportionality to be $1-q$. We indeed verify $k(0,0) =-2$ for our cubic superpotential.

The spectrum contains a supermultiplet with $\Delta=2$ and $\ell=1$ which we identify as the supercurrent multiplet. This multiplet has as its  top component  the spin-2 energy-momentum tensor and the bottom component is the $R$-current. In superfields the multiplet is of the form:
\begin{equation}\label{eq:supercurrent}
\mathcal{R}_\mu = J_{\mu} + \theta \, S_\mu + \thetab\, \bar{S}_\mu  + \theta \sigma^\nu \thetab\; T_{\mu\nu}\,.
\end{equation}	
with $J^\mu$ being the $R$-current. This multiplet can arise in the $\Phib\times \Phi$ operator product, which is the non-chiral part of the spectrum that we are exploring. Its presence confirms our assertion that the low energy dynamics is indeed dominated by a superconformal fixed point.

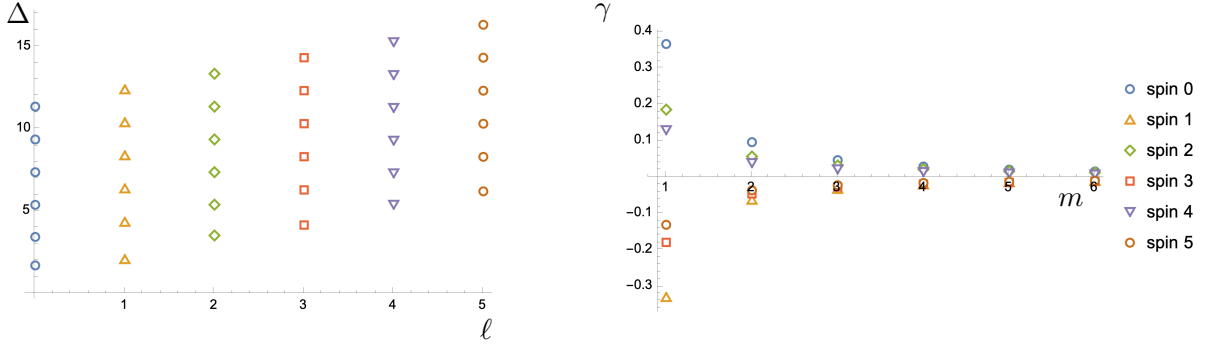
\begin{figure}
\centerline{
\includegraphics[width=2.5in]{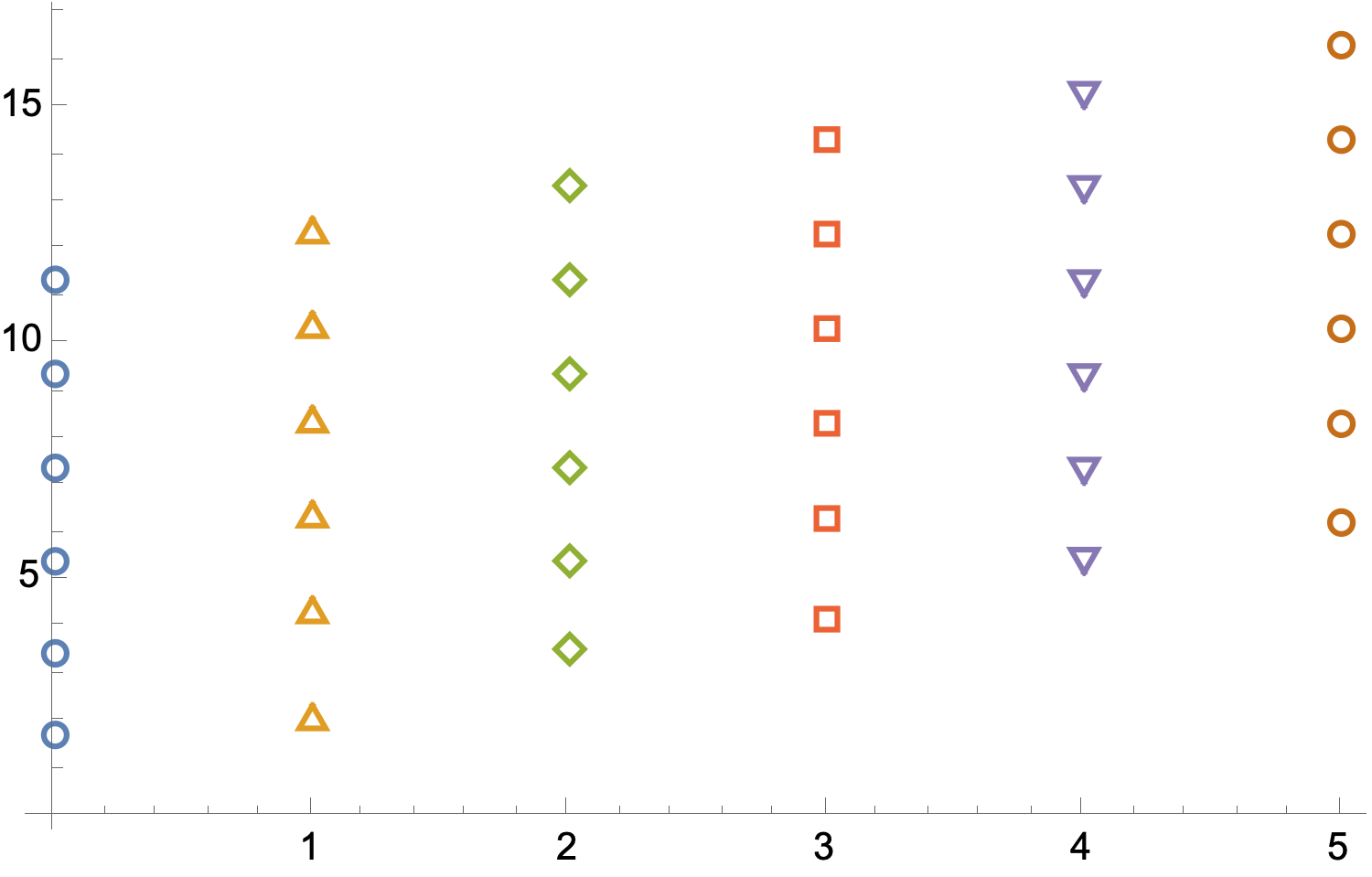} \hspace{1.5cm}
\includegraphics[width=3in]{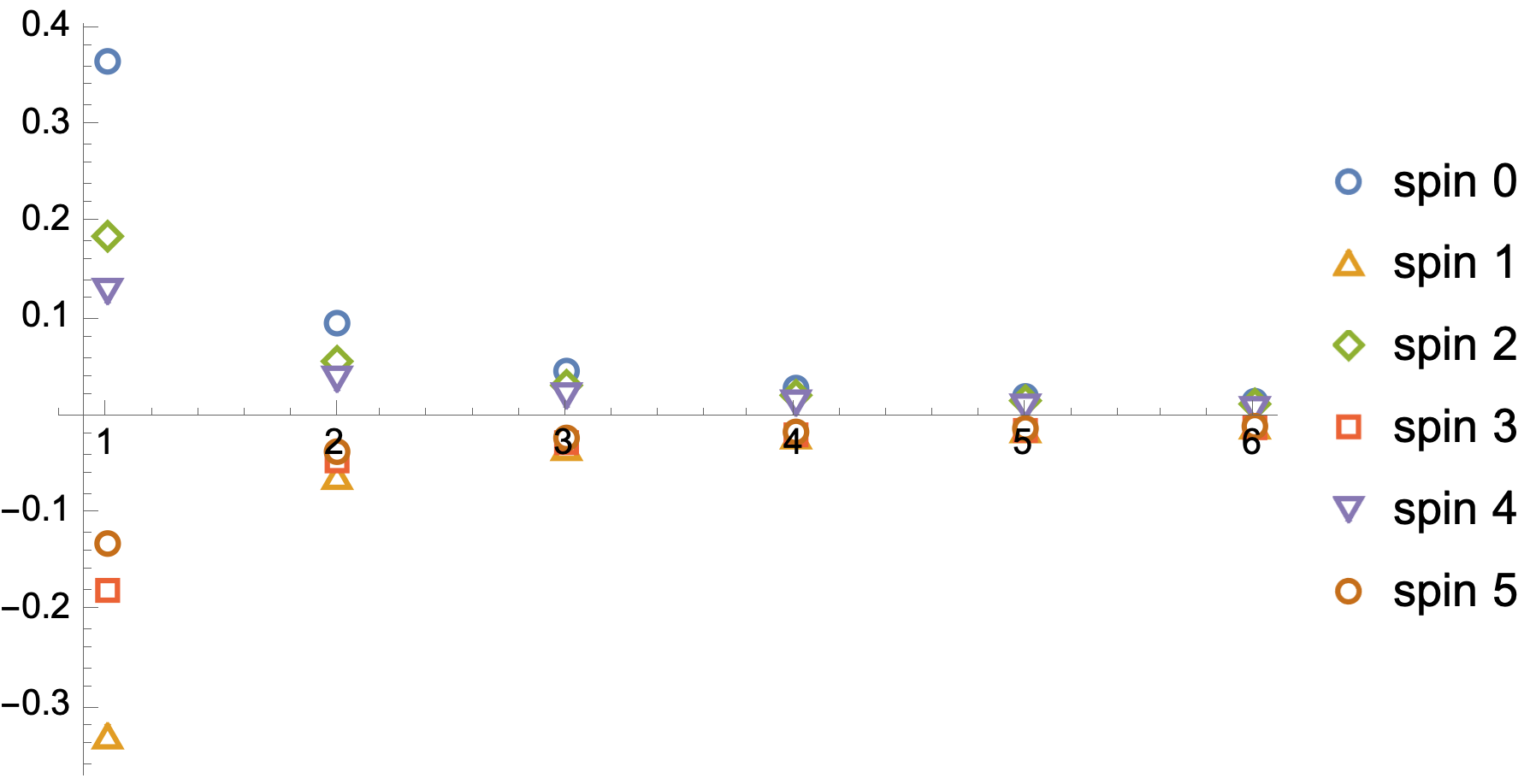}}
\setlength{\unitlength}{0.1\columnwidth}
\begin{picture}(0.3,0.4)(0,0)
\put(4,0.3){\makebox(0,0){$\ell$}}
\put(0,3){\makebox(0,0){$\Delta$}}
\put(5,3){\makebox(0,0){$\gamma$}}
\put(9,1.4){\makebox(0,0){$m$}}
\end{picture}
\caption{Plots of the spectrum of low lying operators with $\ell \leq 5$. On the left we plot the dimensions for various spins, while on the right we plot the anomalous dimension as a function of level number $m$, viz., $\gamma(m,\ell)  = \Delta - 2\Delta_\phi -2 m - \ell $ defined in \eqref{eq:anodim}.
}
\label{fig:lowlyingops}
\end{figure}
\begin{table}[htp]
\begin{center}
\begin{tabular}{|c|c|c|c|}
\hline
\shadeB{operator} & $\shadeB{\ell}$ & \shadeB{$\Delta$} & \shadeB{bootstrap bound}\\
\hline \hline
$(\Phib \Phi)$ & $0$& 1.6994 & $<1.9098$\\
\hline
$(\Phib \Phi)'$ & $0$& 3.4295 & $<5.3$\\
\hline
$J^\mu$ & $1$ & 2 & 2 \\
\hline
$(J')^\mu$ & $1$& 4.2676 & $<5.25$\\
\hline
\end{tabular}
\end{center}
\caption{Comparison between the actual dimension of the operators and their bounds from bootstrap.   $(\Phib \Phi)$ and $(\Phib \Phi)'$ are the lowest dimension and second-lowest dimension scalar superconformal primaries in the $\Phib\times \Phi$ OPE, respectively. $J^\mu$ is the  $R$-current, while $(J')^\mu$ is the second-lowest dimension spin-1 superconformal primary in the $\Phib\times \Phi$ OPE.}
\label{tab:cmpnum}
\end{table}%

The spectrum we have obtained is consistent with existing results from the numerical bootstrap of the $\mathcal{N}=2$ super-Ising model \cite{Bobev:2015vsa}. The latter is a  three dimensional (non-disordered) Wess-Zumino model and the numerical bounds obtained therein admit the spectrum we have obtained for the IR SCFT, see \cref{tab:cmpnum}. We also plot the spectrum of the low lying operators in \cref{fig:lowlyingops}.

Note that while we have exploited a $U(N)$ symmetry rotating the $\Phi_i$ into each other, there are no flavour current multiplets in the spectrum. They contain a dimension one scalar as their bottom component, and there are no such scalars in the singlet spectrum we have analyzed, and also the adjoint spectrum as we will see explicitly in \cref{sec:non-singletOPE}. We believe the symmetry is emergent in the IR, but has no corresponding Noether charges owing to the disorder average. We also highlight the absence of any marginal operators in the spectrum: in fact the only relevant operators are  in the scalar $(\Phib\Phi)$ multiplet and the $R$-current multiplet.

\subsubsection{Unitarity bounds}
\label{sec:unitarity}

The spectrum of superconformal primaries appearing in the $\Phib \times \Phi$ OPE is furthermore unitary, as would be desired for a sensible SCFT.  In the sector under consideration all operators must have $R$-charge $Q_{R}=0$. The unitarity bound for these superconformal primaries \cite{Minwalla:1997ka,Cordova:2016emh} is\footnote{ For $\ell=0$, there is also the special case $\Delta=0$ corresponding to the identity operator, which we have accounted for above.} 
\begin{equation}\label{eq:unitaritybd}
\begin{cases}\Delta >\ell+ 1 & \ell>1\,, \\ \Delta \geq 2 & \ell=1\,, \\ \Delta =0~{\rm or} ~\Delta >1 & \ell=0 \,.
\end{cases}
\end{equation}	
More explicitly, the superfield $\Phi$ is in the superconformal multiplet $L\overline B_1[0]^{(Q_{R})}_{\Delta}$ with $\Delta=Q_{R}=\frac{2}{3}$ in the notation of \cite{Cordova:2016emh}. We have the fusion rule
\begin{equation}
B_1\overline L[0]^{(-\frac{2}{3})}_{\frac{2}{3}}\times L\overline B_1[0]^{(\frac{2}{3})}_{\frac{2}{3}} = B_1\overline B_1[0]^{(0)}_0+ A_1\overline A_1[1]_{2}^{(0)} + \sum_{\ell\in{\mathbb Z}_{\ge0},\,\Delta> \ell+1 }L\overline L[2\ell]^{(0)}_{\Delta}\,,
\end{equation}
where $B_1\overline B_1[0]^{(0)}_0$ is the multiplet of the identity operator, $A_1\overline A_1[1]_{2}^{(0)}$ is the stress tensor multiplet, and $L\overline L[2\ell]^{(0)}_{\Delta}$ are long multiplets.

One can show that operators violating the unitary bound do not satisfy equation \eqref{eq:kernev1}. We will assume that all possible operators appearing in the spectrum have real conformal dimension $\Delta$ (which can independently be checked numerically). First consider $\ell=0$, when the kernel becomes
\begin{equation}\label{eq:kernl0}
k(\Delta,0) = 
	-\frac{2^{\frac{5}{3}}\pi}{3^{\frac{3}{2}}\Gamma(\frac{1}{3})^2} \; \frac{\Gamma(\frac{2}{3}-\frac{\Delta}{2}) \, \Gamma(\frac{\Delta}{2} +\frac{1}{6})}{\Gamma(\frac{4}{3}-\frac{\Delta}{2})\, \Gamma(\frac{\Delta}{2} + \frac{5}{6})}.
\end{equation}	
One observes that for $\Delta < 1$ all the Gamma functions have positive argument -- all such terms are positive, and hence $k(\Delta,0) < 0$. Thus, all operators violating the unitarity bound \eqref{eq:unitaritybd} for $\ell=0$ do not satisfy \eqref{eq:kernev1}.

Next, we consider the case $\ell > 0$. We can restate the unitarity bound \eqref{eq:unitaritybd} in terms of the twist $\tau$ as
\begin{equation}\label{eq:unitaritytwist}
\tau \geq 1\,, \qquad \tau = \Delta -\ell\,.
\end{equation}	
It is then convenient to rewrite the kernel eigenvalue in the terms of the twist as
\begin{equation}\label{eq:kerntwist}
k(\tau,\ell) = (-1)^{\ell+1} \, \mathcal{B} \, 
	\frac{\Gamma(\frac{2}{3}-\frac{\tau}{2})}{\Gamma(\frac{4}{3}-\frac{\tau}{2})} \ \frac{\Gamma(\frac{\tau}{2}+\ell+\frac{1}{6})}{\Gamma(\frac{\tau}{2} + \ell+\frac{5}{6})}\,,
\end{equation}	
where
\begin{equation}\label{eq:Bkdef}
\mathcal{B} = 
	-2^{2-2\Delta_\phi}\, (2\Delta_\phi-1)\, \frac{\Gamma(\Delta_\phi-1)\Gamma(2\Delta_\phi)}{\Gamma(\frac{\Delta_\phi}{2})^2} 
	= \frac{2^{\frac{5}{3}}\pi}{3^{\frac{3}{2}}\Gamma(\frac{1}{3})^2} > 0.
\end{equation}	
Due to the factor $(-1)^{\ell+1}$ appearing in \eqref{eq:kerntwist}, we break the argument into two cases: $\ell$ even and $\ell$ odd.

\begin{itemize}[wide,left=0pt]
\item \underline{$\ell$ even}: Here we employ the  same reasoning as the $\ell=0$ case. For $\tau < 1$, all Gamma functions in \eqref{eq:kerntwist} have positive argument so they are all positive, and hence $k(\tau,\ell) < 0$ leading to no states in the spectrum violating the unitarity bound.

\item 
\underline{$\ell$ odd}: This case is more involved,  and to proceed we will argue that  
\begin{equation}\label{eq:loddbound}
k(\tau,\ell) < 1 \qquad \mathrm{for\;all\;} \;  \tau < 1\,.
\end{equation}	
For $\ell=1$, this bound is saturated, namely $k(1,1) = 1$ owing to the presence of the R-current multiplet. For this reason, the bound is difficult to show analytically for $\ell=1$ and instead we check numerically that $k(\tau,1) < 1$ for all $\tau < 1$. For $\ell \geq 3$, we can demonstrate the bound analytically. We employ Wendel's inequality
\begin{equation}\label{eq:Wendel}
\frac{\Gamma(x)}{\Gamma(x+t)} \leq \frac{(x+t)^{1-t}}{x}, \qquad 0 < t < 1\,,\ 0  < x,
\end{equation}	
to bound each of the Gamma function ratios appearing in \eqref{eq:kerntwist} separately and obtain
\begin{equation}
k(\tau,\ell) \leq 
	\mathcal{B}\, \frac{(\frac{4}{3}-\frac{\tau}{2})^{\frac{1}{3}}}{(\frac{2}{3}-\frac{\tau}{2})}
	\frac{(\frac{\tau}{2} + \ell+\frac{5}{6})^{\frac{1}{3}}}{(\frac{\tau}{2} + \ell+\frac{1}{6})}.
\end{equation}	
For each $\ell$ the r.h.s.\ is a monotonically increasing function for $\tau < 1$, and hence obtains its maximum at $\tau =1$. The resulting function of $\ell$ is monotone decreasing and is bounded in turn by the value at $\ell =3$. 
Altogether, 
\begin{equation}
k(\tau,\ell) \leq k(1,\ell) \leq k(1,3) = \mathcal{B}\, \frac{18}{11}\bigg(\frac{65}{18}\bigg)^{\frac{1}{3}} \approx 0.67 < 1\,,
\end{equation}	
which indeed establishes the unitarity of the spectrum.
\end{itemize}
%

\subsubsection{Anomalous dimensions}
\label{sec:anomdim}

We now turn to the asymptotic part of the spectrum. The solutions to \eqref{eq:kernev1} are organized, for each value of spin, $\ell$, into the following sequence: 
\begin{equation}\label{eq:anodim}
\tau =2\,\Delta_\phi +2\, m + \gamma(m,\ell)\ .
\end{equation}	
Here $\gamma(m,\ell)$ parameterizes the anomalous dimensions and for each spin $\ell$ the solutions are labeled by an `oscillator level' $m \in \mathbb{Z}_+$.  This is in accord with the general expectations from the analytic bootstrap results of \cite{Fitzpatrick:2012yx,Komargodski:2012ek}, where it was argued that for a CFT whose spectrum contains a scalar operator of dimension $\Delta_\phi$, one must have a tower of operators for each value of spin $\ell$, with the twist $\tau$ accumulating towards $\tau \to 2\Delta_\phi + 2m$.  We give the anomalous dimensions for the operators of the first few spins and levels in Table~\ref{tab:anmdim}.

\begin{table}[htp]
\begin{center}
\begin{tabular}{|c||c|c|c|c|}
\hline
\shadeB{\diagbox{$\ell$}{$m$}} & $\shadeB{0}$ & \shadeB{1} & \shadeB{2} & \shadeB{3}\\
\hline \hline
$\shadeR{0}$ & 0.36611 & 0.09618 & 0.04680 & 0.02925\\
\hline
$\shadeR{1}$ & $-\frac{1}{3}$& $-0.06574$ & $-0.035945$ & $-0.02399$\\
\hline
$\shadeR{2}$ & 0.18578 & 0.05683 & 0.03207 & 0.02181 \\
\hline
\end{tabular}
\end{center}
\caption{The anomalous dimension $\gamma(m,\ell)$ for $\ell=0,\,1,\,2$ and $m=0,\,1,\,2,\,3$.}
\label{tab:anmdim}
\end{table}%

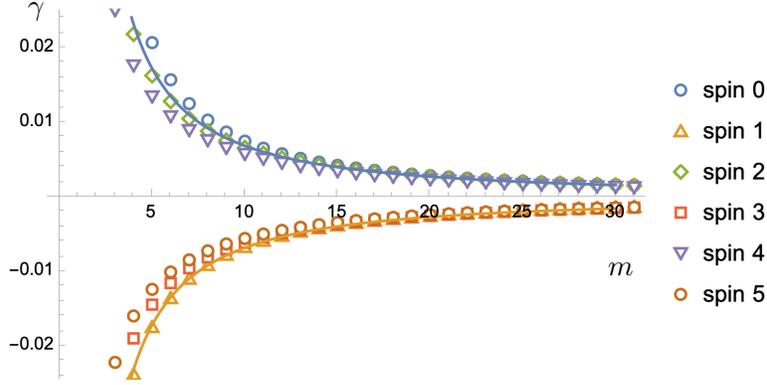
\begin{figure}[tp!]
\centerline{
\includegraphics[width=4in]{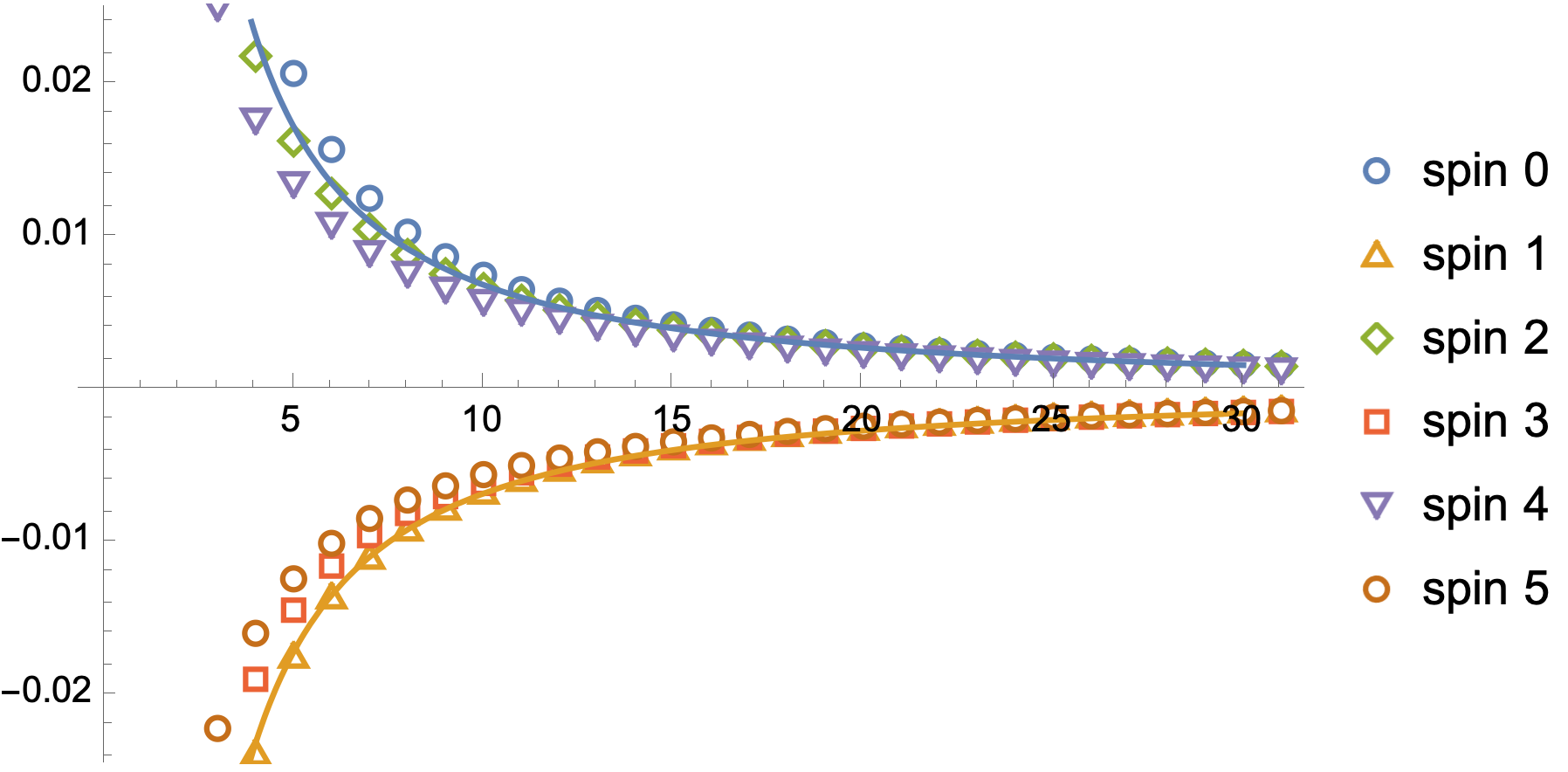}}
\setlength{\unitlength}{0.1\columnwidth}
\begin{picture}(0.3,0.4)(0,0)
\put(2,3.6){\makebox(0,0){$\gamma$}}
\put(7,1.4){\makebox(0,0){$m$}}
\end{picture}
\caption{A plot of the anomalous dimension $\gamma(m,\ell)$ as a function of level number $m$ displaying the convergence to the asymptotic behaviour $\pm 0.147\, m^{-\frac{4}{3}}$ predicted in \eqref{eq:gammam} which are indicated by the solid curves for even and odd spins, respectively.
}
\label{fig:gammasymptote}
\end{figure}

It is instructive to examine the behaviour of the spectrum at large $m$ for fixed spin, or for large spin. We find for large $m$ and fixed spin
\begin{equation}\label{eq:gammam}
\gamma(m,\ell) = (-1)^{\ell+1} \, \frac{\mathfrak{g}_3(\Delta_\phi)}{m^{2\Delta_\phi}} \,,\qquad m \gg \ell \sim 1 
\end{equation}	
while for large spin
\begin{equation}\label{eq:gammal}
\gamma(m,\ell) = (-1)^{\ell+1}\,  \frac{\mathfrak{g}_3(\Delta_\phi)}{\ell^{\Delta_\phi}} \ \frac{\Gamma(m-\Delta_\phi+1)}{\Gamma(m+1)} \,, \qquad \ell \gg 1
\end{equation}	
where we defined: 
\begin{equation}\label{eq:Fdef}
\begin{split}
\mathfrak{g}_3(\Delta_\phi) 
&= 
	 \frac{4^{2-\Delta _\phi} \left(2 \,\Delta_\phi -1\right) \sin \left(\pi  \Delta_\phi \right) \cos \left(\frac{\pi  \Delta _{\phi }}{2}\right) \Gamma \left(\Delta
   _{\phi }-1\right) \Gamma \left(2 \Delta _{\phi }\right) }{\pi 
   \Gamma \left(\tfrac{\Delta _{\phi }}{2}\right)^2 } \,, \\
\mathfrak{g}_3(\tfrac{2}{3}) 
&= 
	-\frac{3}{2^\frac{1}{3}\, \Gamma(-\frac{2}{3})^2} \simeq -0.147\,.   
\end{split}
\end{equation}	
One can check that the anomalous dimensions computed perfectly match the large $m$ asymptotics, as illustrated in \cref{fig:gammasymptote}.

The behaviour at large spin can be directly deduced from analytic bootstrap analysis 
\cite{Fitzpatrick:2012yx}. One expects on general grounds 
\begin{equation}\label{eq:gammagen}
\gamma(m,\ell) = (-1)^{\ell}\frac{\gamma_m}{\ell^{\tau_\text{min}}} \,,
\end{equation}	
where $\tau_\text{min}$ is the minimal twist among operators appearing in the $\Phi_j \times \Phi_k$ OPE. This can be deduced by comparing the $s$-channel expansion of the 4-point function with the corresponding $t$-channel expansion using crossing symmetry. 
By Nachtmann's theorem we are guaranteed that the minimal twist should be found for the operator with the smallest spin, which for us would correspond to a scalar operator. In fact, given the cubic Yukawa interaction in the superpotential, the $\psi_i\times \psi_j$ OPE contains the scalar $\phib_k$. Thus, we are led to conclude that 
\begin{equation}\label{eq:taumin}
\tau_\text{min} = \tau_\phi  = \Delta_\phi \,, \qquad \gamma_m =  0.147 \, \frac{\Gamma(m+\frac{1}{3})}{\Gamma(m+1)}\,.
\end{equation}	
consistent with the expression \eqref{eq:gammal}. In light of the arguments for the convexity of twists of double-trace operators at large spin \cite{Fitzpatrick:2012yx, Komargodski:2012ek}, it may seem puzzling that the anomalous dimensions we found are positive for even spin. However, the supersymmetric case is more subtle because there are cancellations between conformal blocks within a given superconformal block at leading order in the large $\ell$ expansion, as we will explain in detail in \S\ref{sec:neutraldtwist}.

\subsubsection{Regge intercept and hyperbolic chaos}
\label{sec:Regge}

We have thus far focused on the solution to the kernel eigenvalue equation \eqref{eq:kernev1} which have real integral spin $\ell$ and noted for each spin there are distinct solutions parameterized by the `oscillator level' $m$. There is however also another branch of solutions where the conformal dimension lies on the principal series representation of the conformal group, with 
\begin{equation}\label{eq:prinser}
\Delta  = \frac{d}{2} + i\, \nu \,.
\end{equation}	
For fixed $\nu$ one again has multiple solutions to the eigenvalue equation. Let us index the solutions by $n \in \mathbb{Z}_+$ with 
$\Re(\ell_0(\nu)) > \Re(\ell_1(\nu)) > \Re(\ell_2(\nu)) > \cdots $. For $\nu =0$ the solutions $\ell_n(0)$ are all real, and the leading Regge intercept is given by $\ell_0(0)$ \cite{Costa:2012cb}. 

Not only does the leading Regge intercept capture some of the spectral information, but it can also be related to the Lyapunov exponent obtained from the out-of-time-order four-point function in hyperbolic space \cite{Murugan:2017eto}. This follows from the fact that the vacuum correlation functions of a CFT in flat space, can be conformally mapped to thermal correlators on hyperbolic space (with curvature radius set by the inverse temperature). Geometrically one maps a Rindler wedge of flat spacetime, which is conformal to the domain of dependence of a spherically symmetric ball-shaped region of flat spacetime, to the hyperbolic cylinder, cf., \cite{Casini:2011kv} for an explicit map. Tracking the observables through the map, one finds that the Regge limit is equivalent to the chaos limit of the resulting thermal system in hyperbolic space with 
\begin{equation}\label{eq:lamhyp}
\lambda_L^\text{hyp} = \ell_0(0) -1 \,.
\end{equation}	

For the 3d SYK kernel \eqref{eq:kernelDl}, owing to the presence of the factor $(-1)^\ell$, even and odd spin superconformal primaries form different Regge trajectories. It is convenient to rewrite the kernel as
\begin{equation}
k(\Delta,\ell) = \frac{1+(-1)^\ell}{2} \, k_{\rm even}(\Delta,\ell)+  \frac{1 - (-1)^\ell}{2}k_{\rm odd}(\Delta,\ell) \,.
\end{equation}	
The dimensions of the even (odd) spin superconformal primary operators appearing in the $\Phi\times\Phib$ OPE are given by solutions to the equations
\begin{equation}
k_{\rm even}(\Delta,\ell)=1,\qquad k_{\rm odd}(\Delta,\ell)=1 \,.
\end{equation}	
There are also Regge trajectories formed by the conformal primary but superconformal descendant operators that appear in the $\overline\Phi\times\Phi$ OPE. In the long multiplet $L\overline L[2\ell]^{(0)}_{\Delta}$, those operators have dimensions and spins given by\footnote{This follows directly from the formula for the superconformal blocks in terms of the conformal blocks in (66) of \cite{Bobev:2015jxa} which we describe in \cref{sec:4pt}, see  Eqs.~\eqref{eq:sconfdecomp} and \eqref{eq:sconfdecompcoeffs}.}
\begin{equation}
(\Delta+1,\ell\pm 1)\,,\quad(\Delta+2,\ell)\,.
\end{equation}	
Hence, they are solutions to the equations
\begin{equation}
k_{\rm even/odd}(\Delta-1,\ell-1)=1,\qquad k_{\rm even/odd}(\Delta-1,\ell+1)=1\,,\quad k_{\rm even/odd}(\Delta-2,\ell)=1\,.
\end{equation}	

We find that the largest leading Regge intercept of the Regge trajectories occurs among the even spin operators (either superconformal primaries or the conformal primary superconformal descendants) to be
\begin{equation}
\ell_{{\rm even},0}(0)=-0.263329\,, \qquad{\rm for}\quad \Delta_\phi=\frac{2}{3}\,.
\end{equation}	
Likewise among the odd spin operators we find the largest leading Regge intercept 
\begin{equation}
\ell_{{\rm odd},0}(0)=1.15207 \,, \qquad{\rm for}\quad \Delta_\phi=\frac{2}{3}\,.
\end{equation}	

The hyperbolic chaos exponent is then determined from the odd spin sector and is thus
\begin{equation}\label{eq:hypLy}
\lambda_L^\text{hyp} = 0.15207\,, \quad{\rm for}\quad \Delta_\phi =\frac{2}{3}.
\end{equation}	
We note that this value is considerably smaller than the exponents obtained in two-dimensional SYK models. These can be deduced  from the computations of \cite{Murugan:2017eto}. For instance, in the model with $\mathcal{N} = (1,1)$ or $\mathcal{N}=(2,2)$ supersymmetry in $d=2$, one finds that $\lambda_L^\text{hyp} \sim 0.58$. In models with $\mathcal{N} = (0,2)$ supersymmetry in $d=2$, one finds a tunable exponent that has a maximum value $\lambda_L^\text{hyp} \sim 0.58$, cf., \cite{Peng:2018zap}.
In two dimensions however, there is no distinction between the Lyapunov exponent in hyperbolic space and the thermal chaos correlator in flat spacetime (a single spatial direction does not support curvature). This being no longer true in $d=3$ one is left with two possibilities:
\begin{itemize}
\item The fact that the $\lambda_L^\text{hyp} (d=3) < \lambda_L^\text{hyp} (d=2)$ could be taken to suggest that the model under consideration is very weakly chaotic. Random couplings constrained by spacetime locality are not effective scramblers. 
\item The exponential growth of spatial volume in hyperbolic spacetimes overwhelms in $d=3$ the effective scrambling of the infra-red fixed point theory. In other words, the drastic reduction in $\lambda_L^\text{hyp} $ is more due to the nature of the observable and the ambient geometry and not intrinsic to the fixed point. 
\end{itemize}
One would like to have a clear answer to which of these two possibilities characterizes the three-dimensional disordered fixed point. There is a straightforward way to proceed, one which involves studying the real-time thermal observables which can be done numerically. We hope to report on this elsewhere \cite{Chang:2021tch}.

\section{Euclidean four-point function}
\label{sec:4pt}

We now turn to an explicit calculation of the Euclidean four-point function $\widehat{\mathcal{W}}^{il}_{kj}$ defined in \eqref{eq:Wikjlnormal}.  For simplicity,  we will illustrate the general idea by working first with the singlet sector or equivalently with the connected part of the averaged correlator encoded in  $\widehat{\mathcal{F}}(u,v)$ introduced in  \eqref{eq:4ptfn}. The main technical tool involves expanding $\widehat{\mathcal{F}}$ in the basis of three-dimensional $\mathcal{N}=2$ superconformal partial waves $\Upsilon_{\Delta,\ell}$ as follows:
\begin{equation}\label{eq:4ptfn_scpw}
\widehat{\mathcal{F}}(u,v) = \frac{1}{1-K}\widehat{\mathcal{F}}_{0}(u,v) = \sum_{\ell = 0}^{\infty}\int_{0}^{\infty}ds\,\frac{\expval{\Upsilon_{\Delta,\ell},\widehat{\mathcal{F}}_{0}}}{1-k(\Delta,\ell)}\frac{\Upsilon_{\Delta,\ell}(u,v)}{\expval{ \Upsilon_{\Delta,\ell},\Upsilon_{\Delta,\ell}}},
\end{equation}
where $\widehat{\mathcal{F}}_{0}$ is the zero-rung ladder described earlier  (see \cref{fig:ladder}). The principal series for the $\mathcal{N}=2$ superconformal partial waves have conformal dimension $\Delta = \frac{1}{2}+is$.\footnote{Our discussion hitherto has been confined to representations of the Euclidean 3d conformal group $\mathrm{SO}(4,1)$ whose principal series has dimension $\Delta = \frac{3}{2} + i\, s$. We will henceforth travel back and forth between superconformal and conformal algebras and will indicate without introducing new notation the relevant representation labels (without hopefully causing any confusion). }

The goal is to find an inner product on the space of $\mathcal{N}=2$ superconformal partial waves and then use it to compute the desired inner products in \eqref{eq:4ptfn_scpw}. Armed with these quantities, we will obtain an analytic formula for the OPE coefficients in the $\Phib \times \Phi$ OPE. Finally, we compute the central charge of our model and interestingly demonstrate that it agrees exactly with that of the super-Ising (WZ) model  (modulo a trivial factor of $N$ from the species). 

\subsection{Superconformal partial waves}
\label{sec:scpw}

We will define the superconformal partial waves using the corresponding superconformal block and its supershadow block. This will allow us to rewrite them in terms of the bosonic conformal partial waves for which calculations are simpler. One can equivalently define them in the supershadow formalism as an integral over a product three-point functions (cf., \cref{sec:ssops}).

To define the superconformal partial waves, we use the Euclidean superconformal group $OSp(4|2,2)$ to fix coordinates. This group has $10$ bosonic generators coming from the conformal group, an $R$ symmetry generator, and $8$ fermionic generators. We use the bosonic generators in the usual way to fix $x_{1} = \mathbf{0}$, $x_{2} = (\frac{z+\zb}{2},\frac{z-\zb}{2i},0)$, $x_{3} = (1,0,0)$, $x_{4} = \infty$ and we use the fermionic generators to fix
\begin{equation}\label{eq:grassmanzero}
\thetab_{1} = \theta_{2} = \theta_{3} = \thetab_{4} = 0.
\end{equation}
By chirality, $\Upsilon_{\Delta,\ell}$ now has no dependence on the Grassmann variables and is simply a function of the complex variables $z,\zb$, or equivalently, the cross-ratios $u=z\zb$ and $v=(1-z)(1-\zb)$.\footnote{We adhere to the standard notation and hope that the use of $z_{ij}^\mu$ for the supertranslation invariant, cf.,  \eqref{eq:superz}, does not cause confusion. } The superconformal Casimir after fixing these coordinates acting on our four point function \eqref{eq:4ptfn} is given by \cite{Bobev:2015jxa}\footnote{Note that we are considering the four-point function $\expval{ \phib(0)\phi(z)\phi(1)\phib(\infty) }$ while \cite{Bobev:2015jxa} considered $\expval{ \phi(0)\phib(z)\phi(1)\phib(\infty)}$ so our superconformal Casimir and superconformal block differ from theirs by $z \to  \frac{z}{z-1}$.}
\begin{equation}\label{eq:sccasimirop}
\frac{1}{2}\, \mathcal{D}_{z,\zb} 
=  z^2(1-z)\,\partial^2 + \zb^2(1-\zb)\,\bar{\partial}^2 - (z^2\,\partial+\zb^2\,\bar{\partial})+ (z\partial+\zb\bar{\partial})+\frac{z\zb}{z-\zb}[(1-z)\partial-(1-\zb)\bar{\partial}].
\end{equation}
The superconformal partial waves are eigenfunctions of the superconformal Casimir \eqref{eq:sccasimir} with the same eigenvalue as the superconformal block $\mathcal{G}_{\Delta,\ell}$:
\begin{equation}\label{eq:sccasimir}
\mathcal{D}_{z,\zb}\,\Upsilon_{\Delta,\ell}(z,\zb) = \big[\Delta(\Delta-1)+\ell(\ell+1)\big]\Upsilon_{\Delta,\ell}(z,\zb).
\end{equation}
They can be defined as the unique solution to \eqref{eq:sccasimir} that is single-valued and symmetric under $z \leftrightarrow \zb$. This fixes $\Upsilon_{\Delta,\ell}$ to be the following linear combination of the superconformal block and its supershadow block (up to an overall normalization)
\begin{equation}\label{eq:scpw_scblock}
\Upsilon_{\Delta,\ell}(z,\zb) = A_{\widetilde{\Delta},\ell}\, \mathcal{G}_{\Delta,\ell}(z,\zb) + A_{\Delta,\ell} \, \mathcal{G}_{\widetilde{\Delta},\ell}(z,\zb),
\end{equation}
where $\widetilde{\Delta} \equiv 1-\Delta$ is the conformal dimension of the supershadow operator and $A_{\Delta,\ell}$ is the supershadow coefficient. This is the advantage of working with superconformal partial waves instead of superconformal blocks, viz., they are single-valued and thus allow for an inner product for which the superconformal Casimir is self-adjoint.

The superconformal partial wave $\Upsilon_{\Delta,\ell}$ can be expressed in terms of the bosonic conformal partial waves $\Psi_{\Delta,\ell}^{\Delta_{12},\Delta_{34}}$ which are more familiar (we review them in \cref{sec:cpw}), by observing a relation between the superconformal and conformal blocks as follows \cite{Bobev:2015jxa}:\footnote{ Our superconformal block is related to the superconformal block in \cite{Bobev:2015jxa} by
\begin{equation}
\mathcal{G}^{\mathrm{us}}_{\Delta,\ell}(z,\zb)=(-1)^{\ell}(\mathcal{G}^{0,0}_{\Delta,\ell})^{\mathrm{them}}\left(\frac{z}{z-1},\frac{\zb}{\zb-1}\right)\,.
\end{equation}
To get the formula \eqref{eq:superblockrelation}, we have used (64) in \cite{Bobev:2015jxa} and the relation
\begin{equation}
G^{-1,-1}_{\Delta,\ell}\left(\frac{z}{ z-1}, \frac{\zb}{ \zb-1}\right)=(-1)^{\ell} |1-z|^{-1}G_{\Delta,\ell}^{1,-1}(z,\zb)\,.
\end{equation}
}
\begin{equation}\label{eq:superblockrelation}
\mathcal{G}_{\Delta,\ell}(z,\zb) = \frac{1}{\abs{z}}\, G_{\Delta+1,\ell}^{1,-1}(z,\zb),
\end{equation}
where $G^{\Delta_{12},\Delta_{34}}_{\Delta,\ell}(z,\zb)$ is the conformal block of a four-point function of scalar primary operators of dimensions $\Delta_i$ for $i=1,\cdots,\,4$. This formula can be further unpacked as 
\begin{equation}\label{eq:sconfdecomp}
\mathcal{G}_{\Delta,\ell} = 
  G_{\Delta,\ell}+a_{1}(\Delta,\ell)\, G_{\Delta+1,\ell+1}+a_{2}(\Delta,\ell)\, G_{\Delta+1,\ell-1}+a_{3}(\Delta,\ell)\, G_{\Delta+2,\ell},
\end{equation}
where $G_{\Delta,\ell}\equiv G^{0,0}_{\Delta,\ell}$ and the coefficients $a_i(\Delta,\ell)$ are
\begin{equation}\label{eq:sconfdecompcoeffs}
\begin{split}
a_{1}(\Delta,\ell) 
&= 
  \frac{(\Delta+\ell)}{2(\Delta+\ell+1)}, \\
a_{2}(\Delta,\ell) 
&=
   \frac{\ell^2(\Delta-\ell-1)}{2(\Delta-\ell)(2\ell-1)(2\ell+1)}, \\
a_{3}(\Delta,\ell) 
&= 
  \frac{\Delta^2(\Delta+\ell)(\Delta-\ell-1)}{4(\Delta-\ell)(2\Delta-1)(2\Delta+1)(\Delta+\ell+1)}.
\end{split}
\end{equation}
We adopt the same normalization of the conformal block as in \cite{Bobev:2015jxa}, that in the $z\ll \zb\ll 1$ limit (or equivalently the $u\ll(1-v)\ll 1$ limit) the conformal block behaves as
\begin{equation}
G^{\Delta_{12},\Delta_{34}}_{\Delta,\ell}(z,\zb)\sim \frac{(-1)^\ell}{2^\ell}(z\zb)^\frac{\Delta}{2}\left(\frac{\zb}{z}\right)^\frac{\ell}{2}\,.
\end{equation}

The superconformal partial wave \eqref{eq:scpw_scblock} thus becomes the following linear combination of the conformal and shadow conformal blocks\footnote{ We use $\hat{\Delta} = 3-\Delta$ to denote the shadow conformal dimension in \cref{sec:cpw}, but write out the dimensions explicitly in this section.}  
\begin{equation}\label{eq:scpw_scblock2}
\Upsilon_{\Delta,\ell}(z,\zb) = \frac{1}{\abs{z}}\, \left(A_{\widetilde{\Delta},\ell}\, G_{\Delta+1,\ell}^{1,-1}(z,\zb) + A_{\Delta,\ell}\, G_{2-\Delta,\ell}^{1,-1}(z,\zb)\right).
\end{equation}
Observe that this takes the same form as $|z|^{-1}$ times the conformal partial wave $\Psi_{\Delta+1,\ell}^{1,-1}$ given in \eqref{eq:cpw}. Since the linear combination of blocks appearing in the superconformal partial wave is fixed by single-valuedness, we must have
\begin{equation}\label{eq:scpw_cpw}
\Upsilon_{\Delta,\ell}(z,\zb) =  \frac{ \mathfrak{f}(\Delta,\ell)}{\abs{z}}\, \, \Psi_{\Delta+1,\ell}^{1,-1}(z,\zb),
\end{equation}
where $\mathfrak{f}(\Delta,\ell)$ is a normalization that relates the supershadow coefficient to the shadow coefficient (see \eqref{eq:shadowcoeff}) by
\begin{equation}\label{eq:fdef}
A_{\widetilde{\Delta},\ell} = \mathfrak{f}(\Delta,\ell) \, S_{2-\Delta,\ell}^{\Delta_{34}=-1}.
\end{equation}
An explicit expression for the function $\mathfrak{f}(\Delta,\ell)$ will not be needed for our purposes because it cancels in the four-point function. Nevertheless,  since the $\mathcal{N}=2$ supershadow coefficients do not seem to have been reported in the literature, we provide a computation of $\mathfrak{f}(\Delta,\ell)$ in \cref{sec:ssops}.

Finally, it was shown in \cite{Simmons-Duffin:2017nub} that $\Psi_{\Delta,\ell}^{\Delta_{12},\Delta_{34}} \propto \Psi_{3-\Delta,\ell}^{\Delta_{12},\Delta_{34}}$, which implies that $\Upsilon_{\Delta,\ell} \propto \Upsilon_{\widetilde{\Delta},\ell}$. This allows us to restrict to $s>0$ since $s \to -s$ is equivalent to $\Delta \to \widetilde{\Delta}$. We shall see momentarily that the superconformal partial waves with $s>0$ form a complete basis with respect to the natural superconformally invariant inner product.  

\subsection{Superconformal inner product}
\label{sec:ip}

We want to find a superconformally invariant inner product on the space of $\mathcal{N}=2$ superconformal partial waves.   Let us briefly recall how this is done in the bosonic case. We start with the conformally invariant integral
\begin{equation}\label{eq:bosip_unfixed}
\expval{ F,G }_{0}^{\mathrm{unfixed}} = \int \frac{d^{3}x_{1}\,d^{3}x_{2}\,d^{3}x_{3}\,d^{3}x_{4}}{x_{12}^{6}x_{34}^{6}}\,\overline{F}(\{x_{i}\}) \, G(\{x_{i}\}),
\end{equation}
where $F$ and $G$ are conformally invariant functions. Due to conformal invariance, this inner product is proportional to the volume of the conformal group, and hence diverges. To remedy this, we gauge-fix the coordinates using the conformal group and include the associated determinant which is computed in \cref{sec:det}. If we fix our coordinates to the choice in the previous section, the inner product becomes\footnote{Note that our inner product differs from \cite{Simmons-Duffin:2017nub} by a factor of $4$: $\expval{,}_{0}^{\mathrm{us}} = 4\expval{,}_{0}^{\mathrm{them}}$.}
\begin{equation}\label{eq:bosip}
\expval{ F,G }_{0} = \int d^{2}z\,\frac{|z-\zb|}{|z|^{6}}\,\overline{F}(z,\zb)\,G(z,\zb).
\end{equation}
The functions $F$ and $G$ implicitly depend on the external dimensions $\Delta_{i}$ and the internal dimension and spin $(\Delta,\ell)$. This inner product actually only holds for external dimensions living in the bosonic principal series: $\Delta_{i} \in \frac{3}{2}+i\mathbb{R}$. After analytic continuation to real external dimensions, the inner product becomes\footnote{We explain this subtlety in \cref{sec:cpw}. }
\begin{equation}\label{eq:bosip_realext}
\expval{ F,G }_{0}^{\Delta_{12},\Delta_{34}} = \int d^{2}z\,\frac{|z-\zb|}{|z|^{6}} \, |1-z|^{-\Delta_{12}+\Delta_{34}}\,\overline{F}(z,\zb)G(z,\zb).
\end{equation}
Crucially, the bosonic conformal Casimir is self-adjoint with respect to this inner product. Due to the relation \eqref{eq:scpw_cpw}, we will only need the following bosonic inner product:
\begin{equation}\label{eq:bosip_1,-1}
\expval{ F,G }_{0}^{1,-1} = \int d^{2}z\,\frac{|z-\zb|}{|1-z|^{2} \, |z|^{6}}\,\overline{F}(z,\zb)G(z,\zb).
\end{equation}

The $\mathcal{N}=2$ inner product follows from a similar procedure. Consider two functions $f,g$ on superspace $\mathbb{R}^{3|4}$ that satisfy the chirality conditions
\begin{equation}\label{eq:chiralconds}
D_{1,4}f = D_{1,4}g = \overline{D}_{2,3}f = \overline{D}_{2,3}g = 0,
\end{equation}
where $D_{i,j},\overline{D}_{i,j}$ are the superderivatives acting on the $i^{\rm th}$ or $j^{\rm th}$ coordinate. We define the superconformally invariant integral
\begin{equation}\label{eq:ip_unfixed}
\expval{ f,g }^{\mathrm{unfixed}} = \int \frac{d^{3}x_{1}d^{3}x_{2}d^{2}\thetab_{1}d^{2}\theta_{2}}{|z_{12}|^{4}}\frac{d^{3}x_{3}d^{3}x_{4}d^{2}\theta_{3}d^{2}\thetab_{4}}{|z_{43}|^{4}}\; \overline{f}(\{x_{i}\},\{\theta_{i}\}) \, g(\{x_{i}\},\{\theta_{i}\}).
\end{equation}
Once again, we must gauge-fix the superspace coordinates and include the corresponding Berezinian in order to obtain a finite inner product. If we choose to gauge-fix the bosonic coordinates as we did above and gauge-fix the Grassmann coordinates to \eqref{eq:grassmanzero}, we obtain the inner product  (see \cref{sec:det}) 
\begin{equation}\label{eq:ip}
\expval{ f,g } = \int d^{2}z\,\frac{|z-\zb|}{|1-z|^{2} \, |z|^{4}} \,f(z,\zb)g(z,\zb).
\end{equation}
It is straightforward to check that the superconformal Casimir \eqref{eq:sccasimirop} is self-adjoint with respect to this inner-product. 

Observe that there is a simple relation between the $\mathcal{N}=2$ inner product \eqref{eq:ip} and the bosonic inner product \eqref{eq:bosip_1,-1}:
\begin{equation}\label{eq:ipbos_rel}
\expval{ f,g } = \expval{ \abs{z}\, f, \abs{z}\, g }_{0}^{1,-1}.
\end{equation}
We can now use this relation along with \eqref{eq:scpw_cpw} to compute the desired inner products in the four-point function \eqref{eq:4ptfn_scpw} using bosonic conformal partial waves and the bosonic inner product. In particular, the normalization of the superconformal partial waves now follows from the bosonic case:
\begin{equation}\label{eq:scpw_norm}
\begin{split}
\expval{ \Upsilon_{\Delta,\ell},\Upsilon_{\Delta',\ell'} } 
&= 
	\mathfrak{f}(\Delta,\ell) \, \mathfrak{f}(\Delta',\ell') \, 
	\expval{ \frac{1}{\abs{z}} \, \Psi_{\Delta+1,\ell}^{1,-1}\,,\frac{1}{\abs{z}} \, \Psi_{\Delta'+1,\ell'}^{1,-1} } \\
	&= 
		\mathfrak{f}(\Delta,\ell)\, \mathfrak{f}(\Delta',\ell')\,
		\expval{ \Psi_{\Delta+1,\ell}^{1,-1}\,,\Psi_{\Delta'+1,\ell'}^{1,-1} }_{0}^{1,-1} \\
	&= \mathfrak{f}(\Delta,\ell)^{2} \, n_{\Delta+1,\ell} \;2\pi\delta(s-s')\, \delta_{\ell\ell'},
\end{split}
\end{equation}
where the normalization constant $n_{\Delta,\ell}$ is given in \eqref{eq:normconst}. The superconformal partial waves thus obey the following completeness relation:
\begin{equation}\label{eq:scpw_comp}
\frac{1}{\mathfrak{f}(\Delta,\ell)^{2}\, n_{\Delta+1,\ell}}\; \sum_{\ell=0}^{\infty}\int_{0}^{\infty} ds\,\Upsilon_{\Delta,\ell}(z,\zb)\overline{\Upsilon}_{\Delta,\ell}(z',\zb') = \frac{|1-z|^{2}\, |z|^{4}}{|z-\zb|}\, \delta^{(2)}(z-z'),
\end{equation}
which confirms our previous statement that a complete basis of eigenfunctions of the superconformal Casimir is formed by restricting to the $s>0$ superconformal partial waves.

\subsection{Four-point functions}
\label{sec:4ptfn}

With the superconformal inner product at hand, we can finally turn to the evaluation of the inner product $\expval{\Upsilon_{\Delta,\ell},\widehat{\mathcal{F}}_{0}}$. The zero-rung ladder is
\begin{equation}\label{eq:zero-rung}
\widehat{\mathcal{F}}_{0} 
= \frac{\mathscr{G}(\Xb_{1},X_{3})\, \mathscr{G}(\Xb_{4},X_{2})}{\mathscr{G}(\Xb_{1},X_{2})\,   \mathscr{G}(\Xb_{4},X_{3})} \;\; \xrightarrow{\mathrm{fixing\;coords.}} \;\; |z|^{2\Delta_{\phi}}.
\end{equation}
Therefore,
\begin{equation}\label{eq:zero-rung_ip}
\expval{\Upsilon_{\Delta,\ell},\widehat{\mathcal{F}}_{0}} = \mathfrak{f}(\Delta,\ell)\expval{ \Psi_{\Delta+1,\ell}^{1,-1},|z|^{2\Delta_{\phi}+1} }_{0}^{1,-1}.
\end{equation}
We can compute this by writing $\Psi_{\Delta+1,\ell}^{1,-1}$ in the shadow formalism \eqref{eq:cpw_explicit}. With the standard gauge fixing the $x_{5}$ integral is difficult to compute. This can be avoided by undoing the gauge-fixing of the bosonic coordinates and choosing the gauge $x_{1} = \mathbf{0}$, $x_{2} = \mathbf{1}$, $x_{5} = \infty$ instead. With this new gauge-fixing, we arrive at
\begin{equation}\label{eq:zero-rung_ip_final}
\begin{split}
\expval{\Upsilon_{\Delta,\ell},\widehat{\mathcal{F}}_{0}} &= \mathfrak{f}(\Delta,\ell)\int d^{3}x_{3}\,d^{3}x_{4}\,\frac{|x_{34}|^{2\Delta_{\phi}-\Delta-3}}{|x_{4}|^{2}|x_{3}|^{2\Delta_{\phi}}|1-x_{4}|^{2\Delta_{\phi}}}(-1)^{\ell}\widehat{C}_{\ell}\left(\frac{\mathbf{1} \cdot x_{34}}{|x_{34}|}\right) \\
  &= \mathfrak{f}(\Delta,\ell)\frac{3\sqrt{3}\pi^{\frac{5}{2}}(-1)^{\ell}\Gamma(\ell+1)}{2^{\ell-2}\Gamma(\ell+\frac{1}{2})(\Delta+\ell)(\Delta-\ell-1)}k(\Delta,\ell),
\end{split}
\end{equation}
where the integral can be evaluated using equation (B.8) in \cite{Murugan:2017eto} twice.

We now have all the pieces we need to compute the four-point function \eqref{eq:4ptfn_scpw}. Putting them all together,
\begin{equation}\label{eq:4ptfn_final}
\begin{split}
\widehat{\mathcal{F}}(u,v) 
&= 
	\sum_{\ell = 0}^{\infty}\int_{0}^{\infty}ds \; 
	\frac{\expval{\Upsilon_{\Delta,\ell},\widehat{\mathcal{F}}_{0}}}{1-k(\Delta,\ell)}\frac{\Upsilon_{\Delta,\ell}(u,v)}{\expval{ \Upsilon_{\Delta,\ell},\Upsilon_{\Delta,\ell}}} \\
&=
	 \sum_{\ell = 0}^{\infty}\int_{-\infty}^{\infty}\frac{ds}{2\pi}\,
	 \frac{k(\Delta,\ell)}{1-k(\Delta,\ell)} \;
	 \frac{3\sqrt{3}\, \pi^{\frac{5}{2}}\,(-1)^{\ell}\,\Gamma(\ell+1)}{2^{\ell-2}\, n_{\Delta+1,\ell}(\Delta+\ell)(\Delta-\ell-1)\,\Gamma(\ell+\frac{1}{2}) } \;
	 S_{2-\Delta,\ell}^{\Delta_{34}=-1}\,\mathcal{G}_{\Delta,\ell}(u,v) \\
&= 
	\sum_{\ell = 0}^{\infty}\oint_{\Delta=\frac{1}{2}+is}\frac{d\Delta}{2\pi i} \; \rho(\Delta,\ell)\, \mathcal{G}_{\Delta,\ell}(u,v),
\end{split}
\end{equation}
where we have defined the spectral coefficient function
\begin{equation}\label{eq:rho}
\begin{split}
\rho(\Delta,\ell) &\equiv 
	\frac{\rho_{_\mft}(\Delta,\ell)}{1-k(\Delta,\ell)} \\
\rho_{_\mft}(\Delta,\ell) &= 
	k(\Delta,\ell)\frac{2^{\ell}\, 3\sqrt{3}\, (-1)^{\ell}\,\Gamma\left(\ell+\frac{3}{2}\right)}{(\Delta-\ell-1) \, \Gamma(\ell+1)} \; 
	\frac{\Gamma(\Delta)\,\Gamma(1-\Delta+\ell)\,\Gamma\left(\frac{\Delta+\ell}{2}\right)^{2}}{\Gamma\left(\Delta-\frac{1}{2}\right) \Gamma(\Delta+\ell) \, \Gamma\left(\frac{1-\Delta+\ell}{2}\right)^{2}}
\end{split}	
\end{equation}
with $\rho_{_\mft}(\Delta,\ell)$ the coefficient function for mean field theory of a single chiral superfield. In the second line of \eqref{eq:4ptfn_final}, we used the fact that $\rho(\Delta,\ell) = \rho(\widetilde{\Delta},\ell)$ to write the contribution to $\Upsilon_{\Delta,\ell}$ coming from the superconformal shadow block $\mathcal{G}_{\widetilde{\Delta},\ell}$ in terms of the superconformal block $\mathcal{G}_{\Delta,\ell}$ integrated from $-\infty$ to $0$.

While we have illustrated the computation using the singlet sector, we in fact now have all the data at hand to discuss the general structure of the correlation functions and the OPE coefficients, for both the singlet sector \eqref{eq:Wsingletav} and the non-singlet sectors \eqref{eq:Wadj}.

\subsection{The non-chiral OPE coefficients}
\label{sec:OPEcoeffs}

To obtain the superconformal block expansion \eqref{eq:4ptfn_sconfblocks} in our theory, we deform the contour in \eqref{eq:4ptfn_final} to the right of the principal series line $\Delta = \frac{1}{2}+is$. This will allow us to write the OPE decomposition of the 4-point function, with the residues at the poles of $\rho(\Delta, \ell)$ giving us the product of the OPE coefficients.

In general, we expect the normalized four-point function $\widehat{\mathcal{W}}^{il}_{kj}$ to be a sum over the superconformal blocks of all the superconformal primaries appearing in the $s$-channel  $\Phib^i \times \Phi_k$ OPE:\footnote{We use $\cope{ABC}$ to denote the OPE coefficient between three superconformal primary operators reserving $C_{abc}$ for the OPE coefficient of conformal primary operators.}
\begin{equation}\label{eq:4ptfn_sconfblocksij}
\widehat{\mathcal{W}}^{il}_{kj}(z,\zb) = \sum_{\mathcal{V}_{\Delta,\ell,({\bf r},a)} \in \Phib^i \times \Phi_k}\;
  \cope{\Phib^i\Phi_k\mathcal{V}_{\Delta,\ell,({\bf r},a)} } \, \cope{\Phib^l\Phi_j\mathcal{V}_{\Delta,\ell,({\bf r},a)} }
  \,\mathcal{G}_{\Delta,\ell}(z,\zb)\,.
\end{equation}
The OPE coefficient is proportional to the Clebsch-Gordan coefficient, and we write
\begin{equation}
\cope{\Phib^i\Phi_k\mathcal{V}_{\Delta,\ell,({\bf r},a)}} = \left(c_{\bf r}\right)^i_{k,a}\;\cope{\Phib\Phi\mathcal{V}_{\Delta,\ell,{\bf r}}} \,.
\end{equation}
Equivalently, the functions $\mathcal{W}_\mathbf{1}$ and $\mathcal{W}_\mathbf{adj}$ in \eqref{eq:4ptikjl}, after normalizing as in \eqref{eq:Wikjlnormal}, admit the expansion
\begin{equation}\label{eqn:SBE_1_adj}
\widehat{\mathcal W}_{\bf r}(z, \zb)=\sum_{\mathcal{V}_{\Delta,\ell,{\bf r}}}\left|\cope{\Phib\Phi\mathcal{V}_{\Delta,\ell,{\bf r}}}\right|^2 \mathcal{G}_{\Delta,\ell}(z,\zb)\quad{\rm for}\quad{\bf r}={\bf 1},{\bf adj}\,.
\end{equation}

Since the superconformal blocks are independent of the dimensions of the external operators, the singlet contribution isolated by averaging over the external operators $\widehat{\mathcal{W}}$ can similarly be expanded in terms of the superconformal blocks as
\begin{equation}\label{eq:4ptfn_sconfblocks}
\widehat{\mathcal{W}}(z,\zb) =  \frac{1}{N}\, \widehat{\mathcal{W}}_\mathbf{1}(z,\zb)= \sum_{\mathcal{V}_{\Delta,\ell} \in \Phib \times \Phi}\;
   \abs{\cope{\Phib\Phi\mathcal{V}_{\Delta,\ell} } }^{2}\,\mathcal{G}_{\Delta,\ell}(z,\zb).
\end{equation}
We let $\Phib \times \Phi$ denote the set of superconformal primaries that appear in any of the $\Phib^i \times \Phi_i$ OPE, and  $\cope{\Phib\Phi\mathcal{V}_{\Delta,\ell} }$ is the operator averaged OPE coefficient, viz., 
\begin{equation}\label{eq:opeav}
\cope{\Phib\Phi\mathcal{V}_{\Delta,\ell}}  \equiv \sum_{{\bf r}={\bf 1},{\bf adj}} 
\left(\frac{1}{N} \, \sum_{i=1}^N\right)\left(\frac{1}{{\rm dim}({\bf r})} \, \sum_{a=1}^{{\rm dim}({\bf r})}\right)\, \cope{\Phib^i\Phi_i\mathcal{V}_{\Delta,\ell,({\bf r},a)}}= \frac{1}{\sqrt{N}} \, \cope{\Phib \Phi \,\mathcal{V}_{\Delta,\ell, \mathbf{1}} }\,.
\end{equation}  
This is naturally mapped to OPE coefficients for operators appearing in the singlet channel, $\mathcal{V}_{\Delta,\ell, \mathbf{1}}$ (using the $SU(N)$ representation label for characterization), due to the relations of the Clebsch-Gordan coefficients
\begin{equation}
\frac{1}{N}\sum_{i=1}^N \left(c_{\bf r}\right)^i_{i,a} = \frac{1}{\sqrt{N}} \, \delta_{{\bf r},{\bf 1}}\delta_{a,1}\,.
\end{equation}  
%

\subsubsection{The singlet sector OPE decomposition} 

The contour deformation in \eqref{eq:4ptfn_final} to the right picks up the poles from the spectrum at $k(\Delta,\ell) =1$. The residue of a pole gives the OPE coefficient of the corresponding superconformal primary $\mathcal{V}_{\Delta,\ell}$ in the $\Phib \times \Phi$ OPE:
\begin{equation}\label{eqn:OPEcoeffs}
\abs{\cope{\Phib\Phi\mathcal{V}_{\Delta,\ell}}}^{2} 
  = 
    -\frac{1}{N}\; \underset{\Delta=\Delta_{\mathcal{V}}}{\Res}\,\rho(\Delta,\ell) = \frac{\rho_{ _\mft} (\Delta_{\mathcal{V}},\ell)} {N \,k'(\Delta_{\mathcal{V}},\ell)} \,,
\end{equation}
where $k'(\Delta,\ell)= \pdv{k(\Delta,\ell)}{\Delta}$.
The negative sign arises because of the contour running clockwise while the factor of $\frac{1}{N}$ originates from the relation between $\widehat{\mathcal{F}}$ and the full four-point function $\widehat{\mathcal{W}}$, \eqref{eq:4ptfn}.

When we deform the contour to the right, we also pick up any poles in $\rho(\Delta,\ell)$ or $\mathcal{G}_{\Delta,\ell}$ with $\Re(\Delta) > \frac{1}{2}$. In order for the spectrum to localize on the physical states, $k(\Delta, \ell)=1$, we must also check that the contribution from the  poles of $\rho(\Delta,\ell)$ disappear.  A general theory independent argument  (for non-supersymmetric theories) showing the cancellation of the contributions from the spurious poles in the coefficient function and the poles coming from the conformal block was given in \cite{Simmons-Duffin:2017nub}. Their argument can be applied to our four-point function $\mathcal{W}$ after rewriting it in terms of bosonic conformal blocks using \eqref{eq:superblockrelation}.  Hence the  expression for the four-point function \eqref{eq:4ptfn_final} is a sum over only the physical operators in our theory. Furthermore, note that $k(\Delta, \ell)$ itself has poles when $\Delta = 2\Delta_\phi+\ell +2n$ with $n\in \mathbb{Z}_{\geq 0}$, but these are not poles of the integrand in \eqref{eq:4ptfn_final}.

Let us analyze our formula \eqref{eqn:OPEcoeffs} for the OPE coefficients in the $\Phib \times \Phi$ OPE. Firstly, one can extract the OPE coefficients of the low dimension operators discussed in \cref{tab:cmpnum}: 
\begin{equation}\label{eq:lowdimOPEcoeffs}
\abs{\cope{\Phib\Phi(\Phib \Phi)} }^{2} \approx \frac{1}{N} 0.6601, \qquad \abs{\cope{\Phib\Phi(\Phib \Phi)'} }^{2} \approx \frac{1}{N} 4.216 \times 10^{-3} , \qquad \abs{\cope{\Phib\Phi J'} }^{2} \approx \frac{1}{N} 1.329 \times 10^{-3}.
\end{equation}
We have not found any bootstrap bounds in the literature for these OPE coefficients; it would be interesting to investigate if our OPE coefficients satisfy them.

One can also compute the OPE coefficient of the supercurrent multiplet \eqref{eq:supercurrent}:
\begin{equation}\label{eq:supercurrent_OPE}
\abs{\cope{\Phib\Phi \mathcal{R}} }^2 = \frac{1}{N} \, \frac{9\sqrt{3}\,\pi}{32\left(\frac{2\pi}{\sqrt{3}}-\frac{9}{8}\right)}.
\end{equation}
We will use this OPE coefficient to compute the central charges of the theory below.

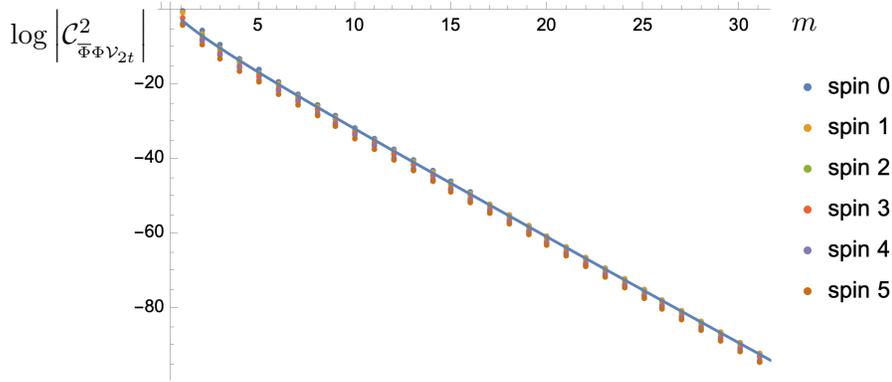
\begin{figure}[tp!]
\centerline{
\includegraphics[width=4in]{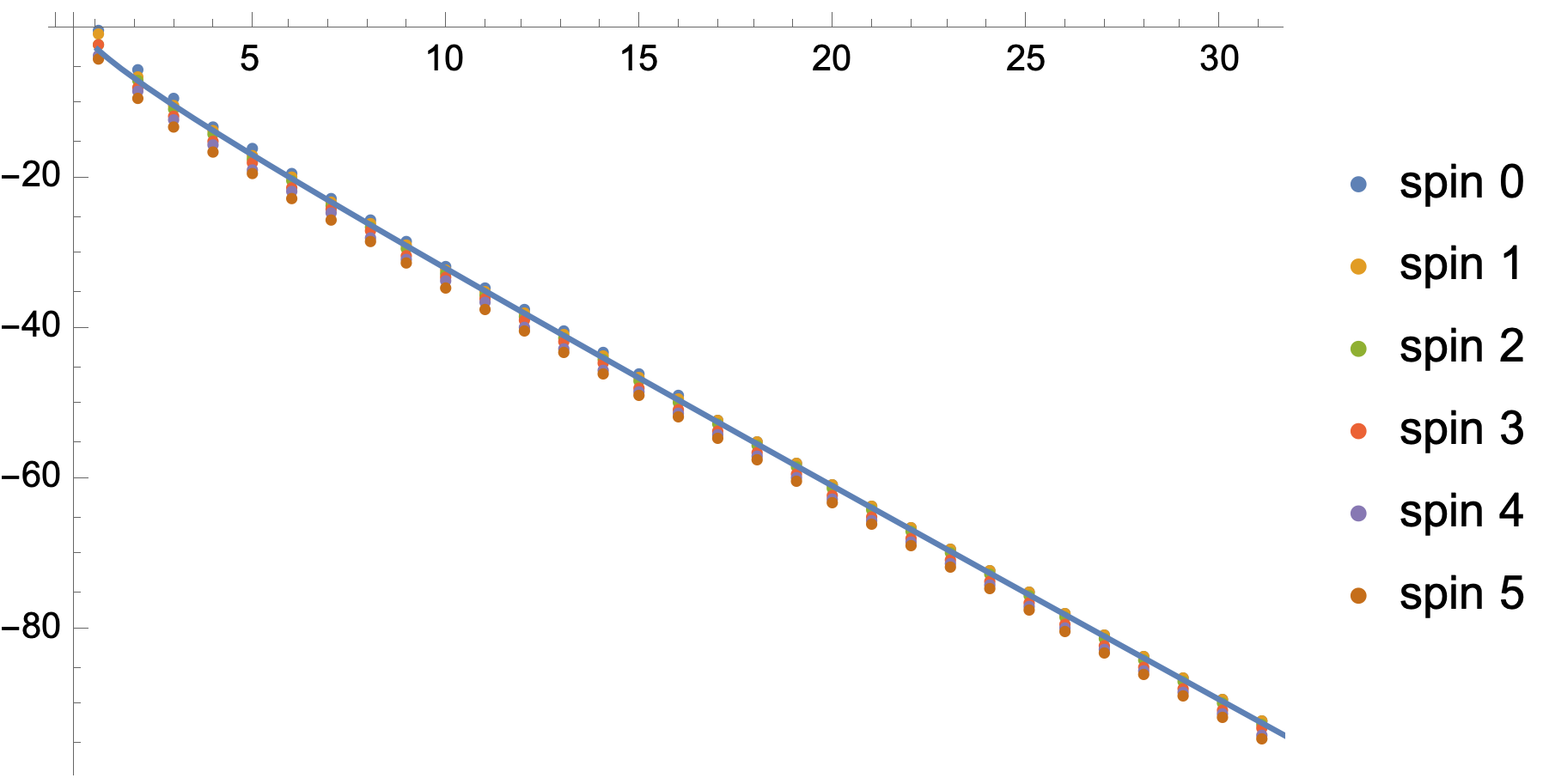}}
\setlength{\unitlength}{0.1\columnwidth}
\begin{picture}(0.3,0.4)(0,0)
\put(1.3,3.4){\makebox(0,0){$\log\abs{\cope{\Phib\Phi\mathcal{V}_{2t}}^2}$}}
\put(7.5,3.5){\makebox(0,0){$m$}}
\end{picture}
\caption{A plot of the product of OPE coefficients $\abs{\cope{\Phib\Phi\mathcal{V}_{\mathrm{2t}}}}^{2}$ against the level number $m$ displaying the convergence to the asymptotic behaviour $-4 \log 2\, m - \frac{11}{6} \, \log m$ predicted in \eqref{eq:largemOPE} (indicated by the solid line).
}
\label{fig:opeasymptote}
\end{figure}

It is interesting to examine the OPE coefficients of the double-twist (2t) operators $\mathcal{V}_{\mathrm{2t}}$ with twist $\tau = 2\Delta_{\phi}+2m+\gamma(m,\ell)$ whose anomalous dimensions we computed in \cref{sec:anomdim}. At large spin $\ell$ and fixed twist $\tau$, one finds
\begin{equation}\label{eq:largelOPE}
\begin{split}
\abs{\cope{\Phib\Phi\mathcal{V}_{\mathrm{2t}}}}^2 
&= 
  \frac{\ell^{2\Delta_{\phi}-\frac{3}{2}}}{N} \, \frac{2\, \sqrt{\pi}}{2^{2\Delta_{\phi}+2m+\ell}} \, 
  \frac{\Gamma\left(\Delta_{\phi}-\frac{1}{2}+m\right)^{2}\,  \Gamma(2\Delta_{\phi}+m-2)}{\Gamma(m+1)\, \Gamma(\Delta_{\phi})^{2}\, \Gamma\left(\Delta_{\phi}-\frac{1}{2}\right)^{2} \,\Gamma(2\Delta_{\phi}+2m-2)} \\
  &\times
   \bigg[1+(-1)^{\ell}\frac{\gamma_m}{\ell^{\Delta_{\phi}}}
   \bigg(
    \psi(1+m-\Delta_{\phi})+\psi(2-2m-2\Delta_{\phi})+\psi\left(\frac{1}{2}-m-\Delta_{\phi}\right) \\
  & \qquad 
  -2\psi(1-2m-2\Delta_{\phi})-\psi(m+1)-\pi\cot(\pi\Delta_{\phi})-\log(2)\bigg)  \\
  &\qquad 
    + \order{\ell^{-2\Delta_{\phi}} } \bigg], \qquad \ell \gg 1,
\end{split}
\end{equation}
where $\psi(x)$ is the digamma function. Notice that the leading contribution is equal to the large $\ell$ limit of the mean field theory (MFT) OPE coefficients, which can already be seen from the way we wrote $\rho(\Delta,\ell)$ in \eqref{eq:rho}, while the subleading correction behaves as $\ell^{-\Delta_{\phi}}$. We shall see in the next section that these results all agree with the predictions of the analytic bootstrap. For large twist $\tau$ and fixed spin $\ell$ the double-twist OPE coefficients become (for  $m \gg \ell \sim 1$)
\begin{equation}\label{eq:largemOPE}
\begin{split}
\abs{\cope{\Phib\Phi\mathcal{V}_{\mathrm{2t}}}}^{2} 
&= 
  -\frac{3\sqrt{2}\, \Gamma\left(\ell+\frac{3}{2}\right) \, \mathfrak{g}_{3}(\Delta_{\phi})}{2^{4\Delta_{\phi}+\ell}\, 
  \Gamma(\ell+1)} \; m^{-2\Delta_{\phi}-\frac{1}{2}}  \, e^{-4\log(2)m} 
  \left(1+\order{m^{-2\Delta_{\phi}} }\right),
\end{split}
\end{equation}
These have the expected exponential decay of heavy operators in any unitary CFT \cite{Pappadopulo:2012jk}.\footnote{Note that it is not actually required that the OPE coefficients of a given tower of operators decay exponentially at large $\Delta$. It is only necessary that this be true on average in the CFT, i.e., that the integral over all OPE coefficients in the CFT lying within some finite $\Delta$-window must decay exponentially.}  The approach to the asymptotic value is clearly demonstrated in \cref{fig:opeasymptote}.

\subsubsection{The non-singlet sector OPE}
\label{sec:non-singletOPE}

To understand the non-singlet sector spectrum we go back to \eqref{eq:Wadj} from which we conclude that 
\begin{equation}\label{eq:Wadjnormal}
\widehat{\mathcal{W}}_\textbf{adj}(u,v) = \widehat{\mathcal{F}}_0(u,v) + \frac{1}{N}\, \frac{\mathscr{G}(\Xb_{1},X_{3})\, \mathscr{G}(\Xb_{4},X_{2})}{\mathscr{G}(\Xb_{1},X_{2})\,   \mathscr{G}(\Xb_{4},X_{3})} \left(\frac{K}{1-K} \widehat{\mathcal{F}}_0\right)\left(\frac{1}{u},\frac{v}{u}\right)  + \order{N^{-2}} \,. 
\end{equation}  
Computing the inner product of the terms on the r.h.s with the superconformal partial waves we end up with
\begin{equation}\label{eq:Wadjdecompose}
\widehat{\mathcal{W}}_\textbf{adj}(u,v) =  \widehat{\mathcal{F}}_0(u,v)+ \frac{1}{ N}\,
  u^{\Delta_\phi}\sum_{\ell = 0}^{\infty}\oint_{\Delta=\frac{1}{2}+is}\frac{d\Delta}{2\pi i} \; \rho_\textbf{adj}(\Delta,\ell)\, \mathcal{G}_{\Delta,\ell}\left(\frac{1}{u}, \frac{v}{u}\right) + \order{N^{-2}} \,,
\end{equation}  
where 
\begin{equation}\label{eq:rhoadj}
 \rho_\textbf{adj}(\Delta,\ell)  = 
 \frac{k(\Delta,\ell)}{1-k(\Delta,\ell)}  \rho_{_\mft}(\Delta,\ell)\,.
\end{equation}  

The leading term can be expanded in terms fo the $s$-channel superconformal blocks by
\begin{equation}
\begin{split}
 \widehat{\mathcal{F}}_0(u,v)&=\sum_{\ell = 0}^{\infty}\int_{0}^{\infty}ds \; 
	\expval{\Upsilon_{\Delta,\ell},\widehat{\mathcal{F}}_{0}}\frac{\Upsilon_{\Delta,\ell}(u,v)}{\expval{ \Upsilon_{\Delta,\ell},\Upsilon_{\Delta,\ell}}} 
\\
&= \sum_{\ell = 0}^{\infty}\oint_{\Delta=\frac{1}{2}+is}\frac{d\Delta}{2\pi i} \; \rho_{_\mft}(\Delta,\ell)\, \mathcal{G}_{\Delta,\ell}(u,v)\,.
\end{split}
\end{equation}
Deforming the contour to the right, we encounter the poles of $\rho_{_\mft}$ located at 
\begin{equation}\label{eq:adjmftpoles}
\Delta = \Delta_{_\mft} \equiv 2\Delta_\phi + \ell + 2 n \,, \qquad n \in \mathbb{Z}_{\geq 0} \,.
\end{equation}  
We will refer to these as the mean field theory double-twist operators $\mathcal{V}^\mft_{\Delta,\ell,{\bf adj}}$. 
For these MFT poles \eqref{eq:adjmftpoles} we find the OPE coefficients to be
\begin{equation}\label{eqn:OPEcoeffsMFT}
\begin{split}
\abs{\cope{\Phib\Phi\mathcal{V}^{\text{\tiny MFT}}_{\Delta,\ell,{\bf adj}}}}^2
  &= 
    -\underset{\Delta=\Delta_{_\mft}}{\Res}\,\rho_{ _\mft }(\Delta,\ell) \,.
\end{split}
\end{equation}
The residues at these mean field theory poles  can be evaluated explicitly to be 
\begin{equation}\label{eq:MFTresidue}
\begin{split}
&\underset{\Delta=2\Delta_\phi+\ell+2n}{\Res}\,\rho_{ _\mft }(\Delta,\ell) 
 = 
 (-1)^{n+1}\, \frac{3\sqrt{3}\; 2^{3+\ell-2\Delta_\phi}\, \Gamma\left(\ell+\frac{3}{2}\right)}{\Gamma(n+1)\, \Gamma(\ell+1)} \, \frac{2\Delta_\phi-1}{2\Delta_\phi-1+2n}\,  \frac{\Gamma(\Delta_\phi-1)\, \Gamma(2\Delta_\phi)}{\Gamma\left(\frac{\Delta_\phi}{2}\right)^2}  \\
&\qquad
\times \frac{\Gamma(1-2n-2\Delta_\phi)}{\Gamma(\Delta_\phi-n)\, \Gamma\left(\frac{1}{2}-n-\Delta_\phi\right)^2} \, \frac{\Gamma\left(\frac{1+\Delta_\phi}{2} + \ell +n\right) \, \Gamma(\ell+n+\Delta_\phi)^2 \, \Gamma(\ell+2n+2\Delta_\phi)}{\Gamma(2\ell +2n +2 \Delta_\phi)\, \Gamma\left(\frac{1+3\Delta_\phi}{2}+\ell+n\right)\, \Gamma\left(-\frac{1}{2}+ \ell+2n + 2 \Delta_\phi\right)} \,.
\end{split}
\end{equation}
This expression can be shown to be equivalent to the MFT OPE coefficients which were computed in \cite{Bobev:2015jxa}, who obtained 
\begin{equation}\label{eq:MFTOPEcoeffs}
\begin{split}
\abs{\cope{\Phib\Phi\mathcal{V}^{\text{\tiny MFT}}_{\Delta,\ell,{\bf adj}}}}^2
&= 
  \frac{2^{\ell}}{\Gamma(m+1)\Gamma(\ell+1) \left( \ell+\frac{3}{2}\right)_{m} } \\
& \quad \times 
  \frac{\left(\Delta_{\phi}\right)_{m+\ell}^{2} \left(\Delta_{\phi}-\frac{1}{2}\right)_{m}^{2}}{  \left(2\Delta_{\phi}+2m+\ell\right)_{\ell} \left(2\Delta_{\phi}+m-1\right)_{m} \left(2\Delta_{\phi}+m+\ell-\frac{1}{2}\right)_{m}} \,.
\end{split}
\end{equation}

The $\order{N^{-1}}$ term in \eqref{eq:Wadjdecompose} can be expanded in terms of the $u$-channel superconformal blocks by deforming the contour to the right and picking up the residue of the poles of the integrand. However, it is not straightforward to re-expand it in terms of the $s$-channel superconformal blocks. We leave this for future work.

\subsection{Central charges}
\label{sec:central}

Given the canonical normalization of the $R$-current $J^\mu$ and the stress-tensor $T_{\mu\nu}$, the coefficients in their two-point functions are physical quantities known as the central charges $C_{J}$ and $C_{T}$, respectively. We can compute them using the $R$-current OPE coefficient $\abs{\cope{\Phib\Phi \mathcal{R}}}^{2}$ given in \eqref{eq:supercurrent} as follows. The Ward identity for an infinitesimal  conformal transformation (a diffeomorphism along a conformal Killing vector)  fixes the OPE coefficient
$C_{\phib\phi T}$, while the Ward identity for the $R$-current fixes $C_{\phib\phi J}$. One has 
\begin{equation}
C_{\phib\phi T} = -\frac{1}{V^2_{_{\mathbf{S}^2}}}  \,, \qquad C_{\phib\phi J} = -\frac{2}{3} \, \frac{1}{V^2_{_{\mathbf{S}^2}}} 
\end{equation}  
where $V_{_{\mathbf{S}^2}}$is the volume of a unit $\mathbf{S}^2$.  Extracting the contribution of the $R$-current $J^\mu$ to the $\phib \times \phi$ OPE in the limit $u \ll (1-v) \ll 1$ gives the following contribution to the four-point function
\begin{equation}\label{eq:Rcurr_4ptfn}
\frac{\expval{\phib(x_{1})\phi(x_{2})\phi(x_{3})\phib(x_{4})}}{\expval{\phib(x_{1})\phi(x_{2})}\expval{\phi(x_{3})\phib(x_{4})}} \supset 
	-\frac{|C_{\phib\phi J}|^{2}}{2\, C_J}\, V_{_{\mathbf{S}^{2}}}^2\, u^{\frac{1}{2}}\, (1-v).
\end{equation}
Comparing to \eqref{eq:4ptfn_sconfblocks} and using the behavior of the superconformal block in the same limit
\begin{equation}\label{eq:sconfblockOPElimit}
\mathcal{G}_{2,1}(u,v) \sim -\frac{1}{2} \, u^{\frac{1}{2}}\, (1-v)
\end{equation}
gives a simple relation between $C_{J}$ and $\abs{\cope{\Phib\Phi \mathcal{R}} }^2$:\footnote{
With our choice of normalization, the free central charge is $C_T^{\mathrm{free}} = 2\, C_T^{(b)}+C_T^{(f)} = 6$, where $C_{T}^{(b)} = \frac{3}{2}$ and $C_{T}^{(f)} = 3$ are the central charges for the free boson and free Dirac fermion, respectively. Similarly, $C_{J}^{\mathrm{free}} = 2\, C_{J}^{(b)}+C_{J}^{(f)} = 1$, where $C_{J}^{(b)} = Q(b)^2 = \frac{1}{4}$ for the free boson and $C_{J}^{(f)} = 2Q(f)^2 = \frac{1}{2}$ for the free Dirac fermion.}
\begin{equation}\label{eq:CJOPErelation}
C_{J} = \frac{4}{9|c_{\overline{\Phi}\Phi\mathcal{R}}|^2} = N\frac{2^{7}}{3^{4}\sqrt{3}\pi}\left(\frac{2\pi}{\sqrt{3}}-\frac{9}{8}\right).
\end{equation}
The relationship between $C_{T}$ and $\abs{\cope{\Phib\Phi \mathcal{R}} }^{2}$ follows from supersymmetry since $J_\mu$ and $T_{\mu\nu}$ live in the same multiplet, with the result
\begin{equation}\label{eq:CT}
C_{T} = 6\,C_{J}.
\end{equation}
This procedure was worked out in \cite{Chang:2017xmr} and we refer the reader there for the details.\footnote{Note that our conformal blocks have a different normalization from \cite{Chang:2017xmr}: $(G_{\Delta,\ell}^{\Delta_{i}})^{\mathrm{us}} = \frac{(-1)^{\ell}}{2^{\ell}}(G_{\Delta,\ell}^{\Delta_{i}})^{\mathrm{them}}$.}

Alternatively, the central charges can be computed from the squashed sphere partition function $Z_{\mathbf{S}_{b}^{3}}$ with squashing parameter $b$. They are proportional to the second derivative of the free energy $F = -\log Z_{\mathbf{S}_{b}^{3}}$ with respect to the squashing parameter $b$:
\begin{equation}\label{eqn:tauRR}
C_{T} = \frac{48}{\pi^2}\, \Re\frac{\partial^2 F}{\partial b^2}\bigg|_{b=1}.
\end{equation}	
The squashed sphere partition function for the 3d $\mathcal{N}=2$ Wess-Zumino model has been computed via localization \cite{Nishioka:2013gza,Gang:2019jut} and one finds exact agreement with our value for the central charges (up to the ubiquitous factor of $N$).\footnote{One can extend this comparison to arbitrary $q$ with $\Delta_{\phi} = \frac{2}{q}$ whence one finds the simple relation $C_{T}^{\mathrm{us}} = \frac{q^{2}}{9}C_{T}^{\mathrm{localization}}$. It is unclear how to interpret this since $q > 3$ models do not admit a low energy fixed point, while the $q=2$ theory is gapped in the IR.}

\section{Chiral sector and the analytic bootstrap}
\label{sec:chiral}

The diagrammatic analysis that we have used thus far to compute the two- and four-point functions only gave us access to the non-chiral sector of our theory, that is, to operators appearing in the $\Phib^i \times \Phi_k$ OPE. However, crossing symmetry relates this OPE to the $\Phi_j \times \Phi_k$ OPE which allows us to extract information about the chiral sector of the theory using what we have computed for the non-chiral sector. We will use the analytic bootstrap to determine the leading order anomalous dimensions and OPE coefficients of the charged double-twist operators at large spin and to deduce the existence of a special operator in the chiral sector along with its OPE coefficient. Furthermore, we will find that our result for the OPE coefficients of the neutral double-twist operators \eqref{eq:largelOPE} agrees with the prediction of the analytic bootstrap.

Let us first summarize the allowed set of operators in the chiral sector of any 3d $\mathcal{N}=2$ SCFT \cite{Bobev:2015jxa}. Due to the chirality condition $\overline{Q}_{\alpha}\Phi_j = 0$, any conformal primary operator $\mathcal{O}_{\Delta,\ell}$ appearing in the $\Phi_j \times \Phi_k$ OPE must also satisfy this condition. This condition turns out to be highly constraining, in particular it implies that only one conformal primary in each superconformal multiplet can appear in the OPE. The possible dimension and spin of the conformal primary operators in the chiral sector are required to be the following:
\begin{itemize}
\item
$\Delta_{\mathcal{O}} = 2\, \Delta_\phi+\ell, \qquad  \ell \in \mathbb{Z}_{\geq 0}$
\item
$\Delta_{\mathcal{O}} = 3-2\, \Delta_\phi, \qquad  \ell = 0,\;\Delta_\phi \leq \frac{3}{4}$
\item
$\Delta_{\mathcal{O}} \geq 2\, \abs{\Delta_\phi-1} + \ell + 2, \qquad  \ell \in \mathbb{Z}_{\geq 0}$.
\end{itemize}
Observe that the operators in the first category, which correspond to the charged double-twist operators at level $m=0$, have vanishing anomalous dimension. The operators in the second category are in general given by  $\xi = \epsilon^{\alpha\beta}\bar{Q}_{\alpha}\bar{Q}_{\beta}\overline{\Xi}$ where $\overline{\Xi}$ is an anti-chiral field with dimension $\Delta_{\overline{\Xi}} = 2(1-\Delta_{\phi})$. Numerical bootstrap results suggest the existence of such an operator in the Wess-Zumino model \cite{Bobev:2015vsa}. We will see that such an operator must exist in our theory due to the anomalous dimension of neutral double-twist operators found in \eqref{eq:gammal}.  In fact, from the dimensions it is clear that  $\overline{\Xi} = \Phib$ and thus $\xi = \Fb$.  The third category contains the charged double-twist operators at level $m>0$.

\subsection{Charged double-twist operators}
\label{sec:chargeddtwist}

The analytic bootstrap \cite{Fitzpatrick:2012yx, Komargodski:2012ek, Li:2015rfa} uses crossing symmetry to relate the anomalous dimensions and OPE coefficients of double-twist operators at large spin $\ell$ in one channel to the twists and OPE coefficients of the minimal twist operators in the cros$s$-channel. Here we will be interested in the charged double-twist operators appearing in the $\Phi_j \times \Phi_k$ OPE which take the schematic form $[\phi\phi]_{m,\ell} \sim \phi_j\,\partial^{2m}\partial^{\mu_{1}}\ldots\partial^{\mu_{\ell}}\phi_k$ and have dimension $\Delta_{[\phi\phi]_{m,\ell}} = 2\Delta_{\phi}+\ell+2m+\gamma_{c}(m,\ell)$.\footnote{It is interesting to ask whether the $\ell=m=0$ operator, corresponding to $\Phi\Phi$, exists in our theory. For the WZ model, the chiral ring relations set $\Phi\Phi = 0$, but for our theory the chiral ring relations for each choice of couplings give a set of linear relations between the $\Phi_{i}\Phi_{j}$ and it is unclear what happens after the disorder average. Our bootstrap analysis only pertains to large $\ell$ so we will be unable to determine the existence of this operator.}

In the following analysis, we will focus on the averaged four-point function \eqref{eq:Wsingletav}, which have been expanded in terms of the $s$-channel superconformal blocks in \eqref{eq:4ptfn_sconfblocks}, with the OPE coefficients computed in \cref{sec:OPEcoeffs}. We reproduce the expansion formula  here,
\begin{equation}\label{eq:4ptfn_sconfblocksB}
\widehat{\mathcal{W}}^{(s)}(z,\zb)=\sum_{\mathcal{V}_{\tau,\ell}}\left|\cope{\Phib\Phi\mathcal{V}_{\tau,\ell}}\right|^2 \mathcal{G}_{\tau,\ell}(z,\zb)\,,
\end{equation}
where we added the superscript $(s)$ to emphasize that it is in $s$-channel, and relabeled the blocks in terms of twist $\tau=\Delta-\ell$.

To study the OPE in the $t$-channel, let us return to the general four-point functions $\widehat{\mathcal{W}}^{il}_{kj}(z,\zb)$ introduced in \eqref{eq:4ptikjl}. Using the $U(N)$ symmetry, the four-point function can be decomposed as
\begin{equation}\label{eqn:general4ptT}
\widehat{\mathcal{W}}^{il}_{kj}(z,\zb)  = \tensor{\left(P_{\bf S}\right)}{_{jk}^{il}}\,  \widehat{\mathcal{W}}_{\bf S}(1-z,1-\zb)+ \tensor{\left(P_{\bf A}\right)}{_{jk}^{il}} \widehat{\mathcal{W}}_{\bf A}(1-z,1-\zb)\,,
\end{equation}
where the projection matrices $\tensor{\left(P_{\bf S}\right)}{_{jk}^{il}}$ and $\tensor{\left(P_{\bf A}\right)}{_{jk}^{il}}$ are
\begin{equation}
\tensor{\left(P_{\bf S}\right)}{_{jk}^{il}}=\delta_{(j}^i\delta_{k)}^l\,, \quad  \tensor{\left(P_{\bf A}\right)}{_{jk}^{il}} =\delta_{[j}^i\delta_{k]}^l\,.
\end{equation}
The functions $\mathcal{W}_{\bf S}(1-z,1-\zb)$ and $\mathcal{W}_{\bf A}(1-z,1-\zb)$ receive contribution from the operators in the $\Phi_j \times \Phi_k$ OPE that transform in the symmetric or anti-symmetric representations and admit the expansion
\begin{equation}
\widehat{\mathcal{W}}_{\bf r}(1-z,1- \zb)=  \frac{(z\zb)^{\Delta_{\phi}}}{(1-z)^{\Delta_\phi}\, (1-\zb)^{\Delta_\phi}}\sum_{\mathcal{O}_{\tau,\ell,{\bf r}}}\left|\cope{\Phi\Phi\mathcal{O}_{\tau,\ell,{\bf r}}}\right|^2 G_{\tau,\ell}(1-z,1-\zb)\,,
\end{equation}
for ${\bf r}={\bf S},{\bf A}$ and $G_{\tau,\ell}(z,\zb)$ is the conformal block for the four-point function of identical scalars. Note that by the Bose symmetry the conformal primaries $\mathcal{O}_{\tau,\ell,{\bf S}}$ are spin-even and the conformal primaries $\mathcal{O}_{\tau,\ell,{\bf A}}$ are spin-odd. Comparing with the $s$-channel expansion \eqref{eqn:Wikjldecompose}, we find the crossing equations
\begin{equation}
\begin{split}
\frac{1}{2}\left[\widehat{\mathcal W}_{\bf S}(1-z,1-\zb)-\widehat{\mathcal W}_{\bf A}(1-z,1-\zb)\right]&= \frac{1}{N}  \left[\widehat{\mathcal W}_{\bf 1}(z,\zb)-\widehat{\mathcal W}_{\bf adj}(z,\zb)\right]\,,
\\
\frac{1}{2}\left[\widehat{\mathcal W}_{\bf S}(1-z,1-\zb)+\widehat{\mathcal W}_{\bf A}(1-z,1-\zb)\right]&=\widehat{\mathcal W}_{\bf adj}(z,\zb)\,.
\end{split}
\end{equation}

Let us focus on the operator averaged four-point function, which admits the $t$-channel expansion as
\begin{equation}\label{eq:t-channel_avg}
\begin{split}
\widehat{\mathcal{W}}^{(t)}(z,\zb) 
& =
  \frac{N+1}{2N}\, \widehat{\mathcal{W}}_{\bf S}(1-z,1-\zb)- \frac{N-1}{2N} \, 
  \widehat{\mathcal{W}}_{\bf A}(1-z,1-\zb)
\\
&=\frac{(z\zb)^{\Delta_{\phi}}}{(1-z)^{\Delta_\phi}\, (1-\zb)^{\Delta_\phi}}\sum_{\mathcal{O}_{\tau,\ell}}(-1)^\ell \left|\cope{\Phi\Phi\mathcal{O}_{\tau,\ell}}\right|^2 G_{\tau,\ell}(1-z,1-\zb)\,.
\end{split}
\end{equation}
where the OPE coefficient $\cope{\Phi\Phi\mathcal{O}_{\tau,\ell}}$ is
\begin{equation}
\cope{\Phi\Phi\mathcal{O}_{\tau,\ell}}=
\begin{cases}
  \sqrt{\frac{N+1}{2N}}\, \cope{\Phi\Phi\mathcal{O}_{\tau,\ell,{\bf S}}} 
    \quad{\rm for}\quad \ell\in2{\mathbb Z}\,,
\\
\sqrt{\frac{N-1}{2N}}\, \cope{\Phi\Phi\mathcal{O}_{\tau,\ell,{\bf A}}}
  \quad{\rm for}\quad \ell\in2{\mathbb Z}+1\,.
\end{cases}
\end{equation}

The crossing equation for the averaged four-point function reads
\begin{equation}\label{eqn:s-t_crossing}
\sum_{\mathcal{V}_{\tau,\ell}}\left|\cope{\Phib\Phi\mathcal{V}_{\tau,\ell}}\right|^2 \mathcal{G}_{\tau,\ell}(z,\zb)=\frac{(z\zb)^{\Delta_{\phi}}}{(1-z)^{\Delta_\phi}\, (1-\zb)^{\Delta_\phi}}\sum_{\mathcal{O}_{\tau,\ell}}(-1)^\ell \left|\cope{\Phi\Phi\mathcal{O}_{\tau,\ell}}\right|^2 G_{\tau,\ell}(1-z,1-\zb)\,.
\end{equation}
The first step of the analytic bootstrap is to consider the kinematic regime $z \ll 1-\zb \ll 1$ where the conformal blocks simplify. To understand the behavior of the superconformal block in this regime, it is useful to decompose the superconformal block in terms of the conformal blocks of the primary operators as in \eqref{eq:sconfdecomp}.  The conformal blocks in the $s$-channel in the desired limit become
\begin{equation}\label{eq:blocklimit}
\begin{split}
G_{\tau,\ell}(z,\zb) 
&\approx 
 (-1)^\ell \, z^{\frac{\tau}{2}}\left(\frac{\zb}{2}\right)^{\ell}{}_{2}{F}_{1}\left(\frac{\tau}{2}+\ell,\frac{\tau}{2}+\ell,\tau+2\ell;\zb\right) \\
	&= (-1)^\ell\, z^{\frac{\tau}{2}}\left(\frac{\zb}{2}\right)^{\ell} \, 
    \frac{\Gamma(\tau+2\ell)}{\Gamma\left(\frac{\tau}{2}+\ell\right)^{2}}\, 
  \sum_{m=0}^{\infty} \, 
      \left[\frac{\left(\frac{\tau}{2}+\ell \right)_m^2}{\Gamma(m+1)^2}  \,   (1-\zb)^{m}\right. 
      \\ 
&\left. \qquad\qquad 
  \times      \, 
   \bigg(2\,\psi(m+1)-2\psi\bigg(\frac{\tau}{2}+\ell\bigg)-\log(1-\zb)\bigg)\,  \right] .
\end{split}
\end{equation}
Therefore, the leading contribution after the identity in the $s$-channel comes from the minimal twist operator in the $\Phib \times \Phi$ OPE, which in our theory is the $R$-current  multiplet $\mathcal{R}_{\mu}$. Observe that the first two conformal primaries in the supermultiplet \eqref{eq:sconfdecomp} have the same twist as the superconformal primary, but the last two have larger twist so these can be ignored in our limit. We conclude that the four-point function in the $s$-channel can be approximated by
\begin{equation}\label{eq:schannelapprox}
\widehat{\mathcal{W}}^{(s)}(z,\zb) \approx 1 + \abs{\cope{\Phib\Phi\mathcal{R}}}^2\bigg(G_{\tau=1,\ell=1}(z,\zb) + \frac{3}{8}G_{\tau=1,\ell=2}(z,\zb)\bigg).
\end{equation}

The crux of the analytic bootstrap is the observation that the sum over conformal blocks in the $t$-channel expansion \eqref{eqn:general4ptT} (and likewise for $\widehat{\mathcal{W}}^{(t)}$) must have the divergent behavior $z^{-\Delta_{\phi}}$ and $z^{\frac{1}{2}\, \tau_{\scriptscriptstyle{R}}-\Delta_{\phi}}$ in order to match the $z^{0}$ and $z^{\frac{1}{2}\, \tau_{\scriptscriptstyle{R}}}$ behavior of the two contributions, respectively, in the $s$-channel. However, the conformal blocks have power law and logarithmic behavior in $z$ so naively such divergent behavior can never be produced by the $t$-channel blocks. The resolution is that the conformal blocks have divergent behavior at large $\ell$ with $z\,\ell^2$ held fixed and the infinite sum at large $\ell$ can indeed reproduce the desired behavior in $z$. 

The identity contribution in the $s$-channel can only be reproduced by the $t$-channel if the OPE coefficients of the charged double-twist operators at leading order in large $\ell$ are equal to the OPE coefficients of double-twist operators in the mean field theory of two non-identical scalar fields (nb: the $\Phi_i$ all have the same  dimension) 
\begin{equation}\label{eq:MFTOPEcoeffs_scalar}
\begin{split}
\abs{\cope{\Phi\Phi[\phi\phi]_{m,\ell}}}^{2} 
&= 
  \frac{2^{\ell} }{\Gamma(m+1)\,\Gamma(\ell+1)   \left(\ell+\frac{3}{2}\right)_{m} } \\
 & \qquad
  \times 
    \frac{ \left(\Delta_{\phi}\right)_{m+\ell}^{2} \left(\Delta_{\phi}-\frac{1}{2}\right)_{m}^{2}}{  \left(2\Delta_{\phi}+2m+\ell-1\right)_{\ell}   \left(2\Delta_{\phi}+m-2\right)_{m} \left(2\Delta_{\phi}+m+\ell-\frac{3}{2}\right)_{m}} \\   
&
  \xrightarrow{\ell \gg 1} \; \frac{4\sqrt{\pi}}{2^{2\Delta_{\phi}+2m+\ell}\,\Gamma(\Delta_{\phi})^{2}} \; 
    \frac{\left(\Delta_{\phi}-\frac{1}{2}\right)_{m}^{2}}{\Gamma(m+1) \left(2\Delta_{\phi}+m-2\right)_{m}} \; \ell^{2\Delta_{\phi}-\frac{3}{2}}.
\end{split}
\end{equation}
Note that this is the chiral superfield MFT OPE coefficient in contrast to \eqref{eq:MFTOPEcoeffs} where we reported the scalar field MFT  coefficients. 

First, to verify that we do reproduce the identity contribution in the $s$-channel, we expand the conformal blocks in the $t$-channel in $1-\zb$ and large $\ell$
\begin{equation}\label{eq:tchannelapprox}
\begin{split}
G_{\tau,\ell}(1-z,1-\zb) 
&= 
  G_{\tau,\ell}(1-\zb,1-z) \\
&\approx 
    \frac{(-1)^{\ell}}{2^{\ell}}(1-\zb)^{\frac{\tau}{2}} \; k_{\tau+2\ell}(1-z) \\
&\approx 
    (-1)^{\ell}\frac{2^{\tau+\ell}}{\sqrt{\pi}} \, \ell^{\frac{1}{2}} \, (1-\zb)^{\frac{\tau}{2}} \, K_{0}(2\ell\sqrt{z}),
\end{split}
\end{equation}
where $k_{2a}(z) = z^{\frac{\tau}{2}} \, {}_{2}{F}_{1}(a,a,2a;z)$ is the conformal block in one dimension, i.e., the $\mathrm{SL}(2,\mathbb{R})$ block. If we focus on the level $m=0$ double-twist operators, the $t$-channel expansion now becomes
\begin{equation}\label{eq:idfromsum}
\begin{split}
\widehat{\mathcal{W}}^{(t)} 
&\supset 
  \left(\frac{z\zb}{(1-z) \, (1-\zb)}\right)^{\Delta_{\phi}} \, 
  \sum_{\ell \gg 1}
    \abs{\cope{\Phi\Phi[\phi\phi]_{0,\ell}}^{\phi\mft}}^{2}\,G_{2\Delta_{\phi},\ell}(1-\zb,1-z) \\
&\approx 
  \frac{4}{\Gamma(\Delta_{\phi})^{2}}\, z^{\Delta_{\phi}}\,
  \sum_{\ell \gg 1} \, K_{0}(2\ell\sqrt{z})\,
    \ell^{2\Delta_{\phi}-1} \approx 1,
\end{split}
\end{equation}
where we approximated the sum over $\ell$ as an integral. One can show that the higher order terms in $(1-\zb)$ get cancelled by the level $m>0$ contributions.

Next, we need to reproduce the $R$-current contribution in the $s$-channel from the $t$-channel decomposition. Let us  expand the contribution to the $t$-channel at large $\ell$ of the charged double-twist operators around their MFT conformal dimensions and OPE coefficients. Denoting the shift away from the MFT value of the OPE coefficients by $\delta$ and letting $\gamma_c(m,\ell)$ be the anomalous dimensions, we find 
\begin{equation}\label{eq:MFTcorrection}
\begin{split}
\widehat{\mathcal{W}}^{(t)} 
&\supset 
  \frac{4}{\Gamma(\Delta_{\phi})^{2}}\,
  \frac{\left(\Delta_{\phi}-\frac{1}{2}\right)_{m}^{2}}{\Gamma(m+1) \left(2\Delta_{\phi}+m-2\right)_{m}}\, z^{\Delta_{\phi}} \, (1-\zb)^{m}\\	
  &\qquad 
    \times
      \sum_{\ell \gg 1}\, 
       K_{0}(2\ell\sqrt{z}) \, \ell^{2\Delta_{\phi}-1}
        \left(\delta\abs{\cope{\Phi\Phi[\phi\phi]_{m,\ell}}^{\phi\mft}}^{2}+\log(2)\gamma_{c}(m,\ell)+\frac{\gamma_{c}(m,\ell)}{2}\log(1-\zb)\right)\,,
\end{split}
\end{equation}
where $\delta|\cope{\Phi\Phi[\phi\phi]_{m,\ell}}^{\phi\mft}|^{2}$ denotes the deviation of the OPE coefficients away from the MFT OPE coefficients.
To reproduce the $z^{\frac{1}{2} \tau_R}$ behavior in the $s$-channel, the anomalous dimensions and shift in the OPE coefficients of the charged double-twist operators at leading order in large $\ell$ must take the following form
\begin{equation}\label{eq:anomOPEcharged}
\gamma_{c}(m,\ell) = \frac{\gamma_c(m)}{\ell^{\tau_R}}, \qquad 
\delta\abs{\cope{\Phi\Phi[\phi\phi]_{m,\ell}}^{\phi\mft}}^{2} = \frac{\mathfrak{C}_c(m)}{\ell^{\tau_R}}.
\end{equation}
Here the subscript $c$ indicates the contribution from charged double-twist operators.

Focusing on the level $m=0$ case, the anomalous dimension $\gamma_{c}(0,\ell)$ is obtained by matching the $(1-\zb)^{0}\log(1-\zb)$ on the two sides of the crossing equation. One finds that in the $s$-channel the $(1-\zb)^{0}\log(1-\zb)$ terms cancel between the two conformal blocks appearing in \eqref{eq:schannelapprox}, and hence 
\begin{equation}
\gamma_{c}(0,\ell) = 0 \,.
\end{equation}  
This is exactly what is required by supersymmetry for these protected operators.\footnote{This bootstrap argument only shows that the leading contribution to $\gamma_{c}(0,\ell)$ vanishes, but one can show that the cancellation of the $(1-\zb)^{0}\log(1-\zb)$ terms between the two conformal blocks is actually true for any superconformal primary appearing in the $s$-channel, and hence $\gamma_{c}(0,\ell)=0$ exactly.}  The OPE coefficient for level $m=0$ can be 
obtained by matching the $(1-\zb)^{0}$ term on each side of the crossing equation with the result
\begin{equation}\label{eq:chargedOPEcoeffs_m=0}
\mathfrak{C}_{c}(0) = {-}\frac{16}{3\pi} \, \abs{\cope{\Phib\Phi\mathcal{R}}}^2.
\end{equation}

The case $m>0$ is much more involved, but it has been worked out for the bosonic case in \cite{Kaviraj:2015xsa}, which we can adapt to our supersymmetric theory. Denoting the anomalous dimension due to an $s$-channel conformal primary of twist $\tau$ and spin $\ell$ by $\gamma_{m}^{(\tau,\ell)}$ and taking into account our normalization of the conformal blocks, we find
\begin{equation}
\begin{split}
\gamma_c(m) 
&= 
  -\frac{1}{2}\gamma_{m}^{(1,1)}+\frac{3}{32}\gamma_{m}^{(1,2)} \\
&= 
  (-1)^{m}\, \frac{32}{\pi^{2}}\, 
  \abs{\cope{\Phib\Phi\mathcal{R}}}^2 \, 
    \frac{\Gamma(m+1)\Gamma(\Delta_{\phi})^{2}}{\Gamma\left(\Delta_{\phi}-\frac{1}{2}\right)^{2}}\\
&  \quad \times   
   \sum_{k=0}^{m}\, \frac{(2\Delta_{\phi}+m-2)_{k}}{\Gamma(m-k+1)} \, 
    \frac{\Gamma\left(k+\frac{3}{2}\right)^{2}}{\Gamma(k+1)^{2}} 
  \Bigg[{}_{3}{F}_{2}\left(
    \begin{array}{c}
    {-}k,-k,\,\Delta_{\phi}-2 \\
    -\left(k+\frac{1}{2}\right) ,-\left(k+\frac{1}{2}\right) 
    \end{array}; 1\right)
\\	&\qquad\qquad \qquad
  -\left(\frac{2k}{3}+1\right)^2
  {}_{3}{F}_{2}\left(
    \begin{array}{c}
    -k,-k\,,\Delta_{\phi}-3 \\
    -\left(k+\frac{3}{2}\right) ,- \left(k+\frac{3}{2}\right) 
    \end{array} ; 1\right)\Bigg].
\end{split}
\end{equation}
We have not analyzed $\mathfrak{C}_c(m)$ for $m\neq 0$ which while  in principle feasible is technically more challenging.

\subsection{Neutral double-twist operators}
\label{sec:neutraldtwist}

We would now like to repeat these analytic bootstrap arguments for the neutral double-twist operators whose anomalous dimensions and OPE coefficients at large $\ell$ were computed for our theory in \eqref{eq:gammal}, \eqref{eq:Fdef} and \eqref{eq:largelOPE}, respectively. Recall that we argued below \eqref{eq:gammagen} that the behavior $\gamma(0,\ell) \sim \ell^{-\Delta_{\phi}}$ of the anomalous dimension comes from the existence of $\phi_{k}$ in the $\psi_{i} \times \psi_{j}$ OPE in the $t$-channel which is a consequence of the cubic superpotential. However, one does not have explicit access to the $\psi_i \times \psi_j$ OPE in the superconformal four-point function so, at the level of the superconformal bootstrap, there must exist some conformal primary in the $\Phi_j \times \Phi_k$ OPE that implies this behavior for the anomalous dimension. It will turn out, rather non-trivially, that the operator $\xi$ (introduced in the beginning of this section) must exist in our theory in order to produce this behavior; we will identity this operator to be $\Fb$. 

First, we must determine the leading large $\ell$ behavior of the OPE coefficients of the neutral double-twist operators, which come from reproducing the identity contribution in the cross-channel. However, one cannot use the previous crossing equation \eqref{eqn:s-t_crossing} because the identity operator does not appear in the $\Phi_j \times \Phi_k$ OPE due to $R$-charge conservation. Therefore, we need a crossing equation that relates the $\Phib^i \times \Phi_k$ OPE to the $\Phib^l \times \Phi_k$ OPE.  We will begin with this and then revert back to the previous crossing equation to compute the corrections to the OPE coefficients. 

\subsubsection{The leading neutral OPE data}
\label{sec:nOPE1}

The superconformal block expansion of the four-point function \eqref{eq:4ptikjl} in the $u$-channel reads
\begin{equation}\label{eqn:Wikjldecompose_u-channel}
\widehat{\mathcal{W}}^{il}_{kj}(z,\zb)
  = |z|^{2\Delta_\phi}\left[ \left(P_\mathbf{1}\right)^{il}_{jk} \;  \widehat{\mathcal{W}}_\mathbf{1}\left( \frac{1}{z},\frac{1}{\zb}\right) + \left(P_{\bf adj}\right)^{il}_{jk} \; \widehat{\mathcal{W}}_\mathbf{adj} \left( \frac{1}{z},\frac{1}{\zb}\right)\right]
\end{equation}
with the expansions of $\widehat{\mathcal{W}}_\mathbf{1}$ and $\widehat{\mathcal{W}}_\mathbf{adj}$ given in \eqref{eqn:SBE_1_adj}. Comparing with the $s$-channel expansion, we find the crossing equation
\begin{equation}
\frac{1}{N}\left[\widehat{\mathcal{W}}_\mathbf{1}(z)-\widehat{\mathcal{W}}_\mathbf{adj}(z)\right] = |z|^{2\Delta_{\phi}}\widehat{\mathcal{W}}_\mathbf{adj}\left( \frac{1}{z},\frac{1}{\zb}\right).
\end{equation}
Now let us focus on the averaged four-point function, which admits the $u$-channel expansion 
\begin{equation}
\widehat{\mathcal{W}}^{(u)}(z,\zb) =|z|^{2\Delta_\phi}\left[ \frac{1}{N^{2}} \;  \widehat{\mathcal{W}}_\mathbf{1}\left( \frac{1}{z},\frac{1}{\zb}\right) +  \frac{N^2-1}{N^2}  \; \widehat{\mathcal{W}}_\mathbf{adj} \left( \frac{1}{z},\frac{1}{\zb}\right)\right] .
\end{equation}
The crossing equation for the averaged four-point function reads
\begin{equation}
\frac{1}{ N}\widehat{\mathcal{W}}_\mathbf{1}(z,\zb) =|z|^{2\Delta_\phi}\left[ \frac{1}{N^{2}} \;  \widehat{\mathcal{W}}_\mathbf{1}\left( \frac{1}{z},\frac{1}{\zb}\right) +  \frac{N^2-1}{N^2}  \; \widehat{\mathcal{W}}_\mathbf{adj} \left( \frac{1}{z},\frac{1}{\zb}\right)\right] .
\end{equation}
It is convenient to change the variable $z\to \frac{z}{z-1}$ and write the above as
\begin{equation}
\widehat{\mathcal{W}}_\mathbf{1}\left(\frac{z}{z-1},\frac{\zb}{\zb-1}\right) =\left|\frac{z}{z-1}\right|^{2\Delta_\phi}\left[ \frac{1}{N} \;  \widehat{\mathcal{W}}_\mathbf{1}\left(1- \frac{1}{z},1-\frac{1}{\zb}\right) +  \frac{N^2-1}{N}  \; \widehat{\mathcal{W}}_\mathbf{adj} \left(1- \frac{1}{z},1-\frac{1}{\zb}\right)\right]  .
\end{equation}

Now, in the limit $1-z \ll \zb \ll 1$, the crossing equation at leading order becomes
\begin{equation}\label{eq:idneutral}
\frac{1}{N}\frac{\zb^{\Delta_{\phi}}}{(1-z)^{\Delta_{\phi}} } \approx  \sum_{\ell \gg 1} 
   \abs{\cope{\Phi\Phib\mathcal{V}_{\mathrm{2t}}}}^{2}\mathcal{G}_{2\Delta_{\phi},\ell}\left(\frac{z}{z-1},\frac{\zb}{\zb-1}\right)=
  \sum_{\ell \gg 1} 
   (-1)^{\ell}\abs{\cope{\Phi\Phib\mathcal{V}_{\mathrm{2t}}}}^{2}\mathcal{G}_{2\Delta_{\phi},\ell}^{(-)}(z,\zb)\,,
\end{equation}
where the l.h.s is the identity contribution to the $u$-channel and r.h.s is the sum over neutral double-twist operators in the $s$-channel. In the second equality, we have used the relation of the conformal block
\begin{equation}
G_{\Delta,\ell} (z,\zb)
  = (-1)^\ell \,G_{\tau,\ell}\left(\frac{z}{z-1},\frac{\zb}{\zb-1}\right)\,.
\end{equation}
The superconformal block $\mathcal{G}_{\Delta,\ell}^{(-)}$ is expanded in terms of conformal blocks as
\begin{equation}\label{eq:sconfdecomp-}
\mathcal{G}_{\Delta,\ell}^{(-)} 
  = G_{\tau,\ell}-a_{1}(\tau,\ell)\, G_{\tau,\ell+1}-
  a_{2}(\tau,\ell)\, G_{\tau+2,\ell-1}+a_{3}(\tau,\ell)\, G_{\tau+2,\ell}\,.
\end{equation}
In the desired kinematic regime and large $\ell$ limit, the $(-)$ superconformal blocks at leading order in $\zb$ become
\begin{equation}\label{eqn:sconfblockdiff_limit}
\mathcal{G}_{\tau,\ell}^{(-)}(\zb,z) 
\approx G_{\tau,\ell}-a_{1}(\Delta,\ell) \, G_{\tau,\ell+1}
   \xrightarrow{\ell \gg 1}  (-1)^{\ell}\, \frac{2^{\tau+\ell+1}}{\sqrt{\pi}}\, \ell^{\frac{1}{2}} \, \zb^{\frac{\tau}{2}} \,
   K_{0}\left(2\ell\sqrt{1-z}\right).
\end{equation}
The crossing equation \eqref{eq:idneutral} now implies that the OPE coefficients of the neutral double-twist operators at leading order in large $\ell$ are equal to those of mean field theory for a chiral superfield. 
These are the same as those given in \eqref{eq:MFTOPEcoeffs} (up to a factor of $1/N$)  
\begin{equation}\label{eq:MFTOPEcoeffs_chiral}
\begin{split}
\abs{\cope{\Phib\Phi\mathcal{V}_{\mathrm{2t}}}^{\Phi\mft}}^{2} 
&= 
\frac{1}{N}\frac{2^{\ell}\,\left(\Delta_{\phi}\right)_{m+\ell}^{2} \left(\Delta_{\phi}-\frac{1}{2}\right)_{m}^{2}}{\Gamma(m+1)\Gamma(\ell+1) \left( \ell+\frac{3}{2}\right)_{m}   \left(2\Delta_{\phi}+2m+\ell\right)_{\ell} \left(2\Delta_{\phi}+m-1\right)_{m} \left(2\Delta_{\phi}+m+\ell-\frac{1}{2}\right)_{m}} \\
&\xrightarrow{\ell \gg 1} \;
\frac{1}{N}\frac{2\sqrt{\pi}}{2^{2\Delta_{\phi}+2m+\ell}\Gamma(\Delta_{\phi})^{2}} \; 
  \frac{\left(\Delta_{\phi}-\frac{1}{2}\right)_{m}^{2}}{\Gamma(m+1) \left(2\Delta_{\phi}+m-1\right)_{m}} \; 
  \ell^{2\Delta_{\phi}-\frac{3}{2}}.
\end{split}
\end{equation}
A very similar calculation to \eqref{eq:idfromsum} verifies that the sum over $m=0$ neutral double-twist operators at large $\ell$ in \eqref{eq:idneutral} indeed produces the l.h.s of \eqref{eq:idneutral} and the terms with $m>0$ cancel the higher order terms $(1-\zb)$.

\subsubsection{The subleading neutral OPE data}
\label{sec:nOPE2}

Armed with the leading large $\ell$ OPE coefficients, we would now like to compute the subleading corrections to these OPE coefficients as well as the anomalous dimensions. If one applies the analytic bootstrap arguments to the crossing equation between the $s$- and $u$-channel, one will find $\gamma(m,\ell) \sim \delta |\cope{\Phib\Phi\mathcal{V}_{\mathrm{2t}}}^{\Phi\mft}|^{2} \sim \ell^{-\tau_R}$ since the $R$-current $\mathcal{R}_{\mu}$ is the minimal twist operator in the $s$-channel. 

However, in our theory, such behavior is subdominant -- the leading order behavior is actually $\ell^{-\Delta_{\phi}}$. Physically it means that the $\ell^{-\Delta_{\phi}}$ behavior must come from the minimal twist operator in the $\Phi_j \times \Phi_k$ OPE. We must therefore return to the crossing equation \eqref{eqn:s-t_crossing}.

Since the neutral double-twist operators now appear in the $s$-channel instead of the $t$-channel, we consider the kinematic regime $1-z \ll \zb \ll 1$. The $t$-channel is then dominated by the minimal twist operators appearing in the $\Phi_j \times \Phi_k$ OPE which are restricted by supersymmetry to be $[\phi\phi]_{0,\ell}$ and $\xi$. The r.h.s of the crossing equation is thus
\begin{equation}\label{eq:tchannel_chiral}
\begin{split}
\mathcal{W}^{(t)}(1-z,1-\zb) 
&\approx 
  \frac{(z\zb)^{\Delta_{\phi}}}{(1-z)^{\Delta_\phi}\, (1-\zb)^{\Delta_{\phi}}}
    \sum_{\ell=0}^{\infty} \left[ \abs{\cope{\Phi\Phi[\phi\phi]_{0,\ell}}}^{2} \, G_{2\Delta_{\phi},\ell}(1-z,1-\zb)  \right. \\
& \left. \hspace{5cm}  
    +\;  \abs{\cope{\Phi\Phi\xi}}^{2} \, G_{3-2\Delta_{\phi},0}(1-z,1-\zb)\right]. 
\end{split}
\end{equation}
We want to relate this to the sum over large $\ell$ neutral double-twist operators in the $s$-channel, which contribute (with $\gamma'(\ell)= \gamma(0,\ell)$)
\begin{equation}\label{eq:schannel_dtwist}
\begin{split}
\mathcal{W}^{(s)}(z,\zb) \supset \sum_{\ell \gg 1} \abs{\cope{\Phib\Phi\mathcal{V}_{\mathrm{2t}}}}^{2}\mathcal{G}_{2\Delta_{\phi}+\gamma'(\ell),\ell}(z,\zb).
\end{split}
\end{equation}

We previously stated that the analytic bootstrap matches divergent sums at large $\ell$ of double-twist operators in one channel to minimal twist operators in the cross-channel that give a divergent contribution in the chosen kinematic limit. However, it was shown in \cite{Alday:2015eya} that the more precise statement is that each low twist operator whose contribution can be made divergent by repeated application of the one-dimensional conformal Casimir $\mathcal{D}_{d=1} \equiv z^{2}(1-z)\partial_{z}^{2}-z^{2}\partial_{z}$ can be matched by a sum over double-twist operators at large $\ell$. A contribution of this type is called Casimir-singular. The contribution of each such low twist operator in one channel maps to a given term in the asymptotic expansion at large $\ell$ of the anomalous dimension of double-twist operators in the cross-channel. 

Let us now illustrate this for our crossing equation. We want to find the lowest twist Casimir-singular contribution to the $t$-channel. While the charged double-twist operators at level $m=0$ are the minimal twist operators appearing in the $t$-channel, they have power law and logarithmic behavior in $(1-z)$ so they are not Casimir-singular. On the other hand, the contribution of $\xi$ behaves as $(1-z)^{\frac{\Delta_{\xi}}{2}-\Delta_{\phi}}$ which becomes singular after applying the Casimir operator.

Next, we need to find the leading Casimir-singular contribution to the $s$-channel. At leading order in large $\ell$, the sum over neutral double-twists gives
\begin{equation}\label{eq:neutraldtwist_leadingorder}
\sum_{\ell \gg 1} \abs{\cope{\Phib\Phi\mathcal{V}_{\mathrm{2t}}}^{\Phi\mft}}^{2}\mathcal{G}_{2\Delta_{\phi},\ell}(z,\zb) \approx \sum_{\ell \gg 1} \frac{(-1)^\ell}{N}\frac{2\sqrt{\pi} \;  \ell^{2\Delta_{\phi}-\frac{3}{2}} \ \zb^{\Delta_{\phi}} }{2^{2\Delta_{\phi}+2\ell}\Gamma(\Delta_{\phi})^{2}}
  \left[k_{2\Delta_{\phi}+2\ell}(z)-\frac{1}{4}k_{2\Delta_{\phi}+2(\ell+1)}(z)\right].
\end{equation}
One can see that this cannot be Casimir-singular by the following argument. Consider the MFT relation
\begin{equation}\label{eq:MFTrelation}
\sum_{\ell = 0}^{\infty} \abs{\cope{\Phib\Phi\mathcal{V}_{\mathrm{2t}}}^{\Phi\mft}}^{2}\mathcal{G}_{2\Delta_{\phi},\ell}(z,\zb) = (z\zb)^{\Delta_{\phi}},
\end{equation}
which demonstrates that the lefthand side is Casimir-regular. Furthermore, observe that the blocks $k_{2a}(z)$ are eigenfunctions of the $\mathrm{SL}(2,\mathbb{R})$ Casimir operator:
\begin{equation}\label{eq:1dCasimir_eigenfn}
\mathcal{D}_{d=1}k_{2a}(z) = a(a-1)\, k_{2a}(z),
\end{equation}
and hence the $\mathrm{SL}(2,\mathbb{R})$ blocks are Casimir-regular. Now the large $\ell$ sum in \eqref{eq:neutraldtwist_leadingorder} differs from the Casimir-regular sum in \eqref{eq:MFTrelation} by a finite sum of $\mathrm{SL}(2,\mathbb{R})$ blocks so the large $\ell$ sum must also be Casimir-regular. 

The fact that the sum over neutral double-twists at leading order in large $\ell$ is not Casimir-singular can be understood as due to the $(-1)^{\ell}$ appearing in the sum coming from the normalization of the superconformal blocks. However, the sub-leading contribution in $\ell$ involves the anomalous dimension and the correction of the OPE coefficient to MFT, each of which comes with a factor of $(-1)^{\ell}$ cancelling the $(-1)^{\ell}$ coming from the block. As a result, this sub-leading contribution is in fact Casimir-singular.

Therefore, matching the leading Casimir-singular terms between the $s$- and $t$-channels gives
\begin{equation}\label{eq:neutralcrossing}
\begin{split}
&\delta \, \sum_{\ell \gg 1} \abs{\cope{\Phib\Phi\mathcal{V}_{\mathrm{2t}}}}^{2}
\left\{ \mathcal{D}_{d=1}\mathcal{G}_{2\Delta_{\phi}+\gamma'(\ell),\ell}(\zb,z)  \right\}\\
& \qquad \quad 
\approx 
\abs{\cope{\Phi\Phi \xi}}^{2} \left\{\mathcal{D}_{d=1}\frac{(z\zb)^{\Delta_{\phi}}}{(1-z)^{\Delta_\phi} \, (1-\zb)^{\Delta_{\phi}}} \; G_{\Delta_{\xi},0}(1-z,1-\zb) \right\},
\end{split}
\end{equation}
where the $\delta$ on the l.h.s indicates that we consider the correction from MFT at large $\ell$ coming from the anomalous dimension and the correction to the OPE coefficients, analogous to \eqref{eq:MFTcorrection} and we let $\gamma'(\ell) = \gamma(0,\ell)$. This can be written  out more explicitly
\begin{equation}\label{eq:neutralcrossing_explicit}
\begin{split}
&\delta\sum_{\ell \gg 1} \frac{(-1)^\ell}{2^\ell} \,
  \abs{\cope{\Phib\Phi\mathcal{V}_{\mathrm{2t}}}}^{2}
  \zb^{\Delta_{\phi}+\frac{\gamma'(\ell)}{2}} \\
&  \qquad \quad 
\times \left\{ \mathcal{D}_{d=1} \, 
  \left[k_{2\Delta_{\phi}+2\ell+\gamma'(\ell)}(z)-\frac{1}{2}\, a_{1}(2\Delta_{\phi}+\ell+\gamma'(\ell),\ell) \, 
  k_{2\Delta_{\phi}+2(\ell+1)+\gamma'(\ell)}(z)\right] \right\} \\
&\approx 
  -\abs{\cope{\Phi\Phi \xi}}^{2} \, \zb^{\Delta_{\phi}} \, \left(\frac{\Delta_{\xi}}{2}-\Delta_{\phi}\right)^{2}(1-z)^{\frac{\Delta_{\xi}}{2}-\Delta_{\phi}-1}
  \ \frac{\Gamma(\Delta_{\xi})}{\Gamma(\frac{\Delta_{\xi}}{2})^{2}} 
  \left[2\gamma_{_\text{E}}+2\psi\bigg(\frac{\Delta_{\xi}}{2}\bigg)+\log\zb\right].
\end{split}
\end{equation}
Let us examine the l.h.s of this equation more closely. At leading order in large $\ell$, the $\mathrm{SL}(2,\mathbb{R})$ block $k_{2a}(z)$ can be approximated by a Bessel function, as given in \eqref{eq:tchannelapprox}. However, this leading contribution cancels between the two blocks on the l.h.s of \eqref{eq:neutralcrossing_explicit}, and hence we need the subleading correction to the block at large $\ell$. This can be extracted from the integral representation of the hypergeometric function, as explained in \cite{Alday:2015eya}, with the result 
\begin{equation}\label{eq:1dblocksubleading}
\begin{split}
k_{2\Delta_{\phi}+2\ell+\gamma'(\ell)}(z)  
&= 
  \frac{2^{2\Delta_{\phi}+2\ell+\gamma'(\ell)}}{\sqrt{\pi}} \, \ell^{\frac{1}{2}} \left[ A(\eta) + \frac{B(\eta)}{\ell} + \order{\ell^{-2}}  \right] \\
A(\eta) 
&= 
   K_{0}(2\eta) \\ 
B(\eta)
&=
  \eta^{2}\, K_2(2\eta)-\eta\big(2\Delta_{\phi}+\gamma'(\ell)\big) \, K_{1}(2\eta)+
   \frac{1}{4} \, (2\Delta_{\phi}+\gamma'(\ell)-\frac{1}{2}) \, K_{0}(2\eta) 
\end{split}
\end{equation}
where $\eta \equiv \ell\sqrt{1-z}$ is held fixed in the small $1-z$ and large $\ell$ limit.

We now have all the pieces that we need to match the two sides of the crossing equation to determine the anomalous dimensions and the correction to MFT of the OPE coefficients of $\mathcal{V}_{\mathrm{2t}}$. The behavior $(1-z)^{\frac{\Delta_{\xi}}{2}-\Delta_{\phi}-1}$ in the $t$-channel can be produced by the sum over neutral double-twist operators at large $\ell$ in the $s$-channel if their anomalous dimensions and OPE coefficients at large $\ell$ are given by (subscript $n$ denoting contribution from neutral double-twist operators)
\begin{equation}\label{eqn:anomOPEneutral}
\gamma(m,\ell) = (-1)^{\ell}\, \frac{\gamma_n(m)}{\ell^{\Delta_{\xi}-1}}, \qquad 
\delta \abs{\cope{\Phib\Phi\mathcal{V}_{\mathrm{2t}}}}^{2} = (-1)^{\ell} \, \frac{\mathfrak{C}_n(m)}{\ell^{\Delta_{\xi}-1}} \,.
\end{equation}
Matching the $\zb^{\Delta_{\phi}}\log \zb$ term on each side of the crossing equation gives
\begin{equation}\label{eq:xiOPEcoeff}
\abs{\cope{\Phi\Phi \xi}}^{2} 
= -\frac{\gamma_n(0)}{N} \,    (4\Delta_{\phi}-2\Delta_{\xi}-1)\, 
  \frac{\Gamma\left(\frac{\Delta_{\xi}}{2}\right)^2}{8\,\Gamma(\Delta_{\xi})}
\, 
  \frac{ \Gamma\left(\Delta_{\phi}-\frac{\Delta_{\xi}}{2}\right)^{2}}{\Gamma(\Delta_{\phi})^2 } \,, 
\end{equation}
and matching the $\zb^{\Delta_{\phi}}$ term gives
\begin{equation}\label{eqn:neutraldoubletwOPE}
\mathfrak{C}_n(0) = \gamma_n(0) 
  \left(\gamma_{_\text{E}}+\psi\bigg(\frac{\Delta_{\xi}}{2}\bigg)-\log(2)\right) .
\end{equation}

Let us see the consequences of this rather general analysis for our theory. We found that the existence of the operator $\xi$ is equivalent to the behavior of the anomalous dimension $\gamma(0,\ell) \sim \ell^{1-\Delta_{\xi}}$ of the neutral double-twist operators. Since in our theory  we know that $\gamma(0,\ell) \sim \ell^{-\Delta_{\phi}}$ we must have  $\Delta_{\phi} = \Delta_{\xi}-1$. Moreover, its OPE coefficient can be computed from \eqref{eq:xiOPEcoeff} since we computed $\gamma(m,\ell)$ explicitly in \eqref{eq:gammal}, \eqref{eq:Fdef}.  One finds
\begin{equation}\label{eq:xiOPEcoeffus}
\abs{\cope{\Phi\Phi \xi}^{\mathrm{us}}}^{2} 
= 
  \frac{\mathfrak{g}_{3}(\Delta_{\phi})}{N}\, (4\Delta_{\phi}-2\Delta_{\xi}-1) \, 
  \frac{\Gamma\left(\frac{\Delta_\xi}{2}\right)^2\,  \Gamma\left(\Delta_\phi-\frac{\Delta_\xi}{2}\right)^2\, 
  \Gamma(1-\Delta_\phi)}{8\, \Gamma(\Delta_\phi)^2\,\Gamma(\Delta_\xi) }
 \approx \frac{2.91}{N}.
\end{equation}
The leading order large $\ell$ OPE coefficients of $\mathcal{V}_{\mathrm{2t}}$ and their subleading correction \eqref{eqn:neutraldoubletwOPE} agree exactly with what we found for our model \eqref{eq:largelOPE}. Therefore $\xi$ indeed must exist in our theory -- in fact from the derived data it is clear that $\xi = \Fb$. 

\section{Discussion}
\label{sec:discuss}

We have constructed a  large $N$ superconformal fixed point in three dimensions exploiting the advantages accorded by the random couplings. The physical data of the low-energy theory, specifically the spectrum and OPE coefficients, are  obtained directly by solving the Schwinger-Dyson equations and decomposing the result for 4-point functions into superconformal blocks (along with suitable use of crossing symmetry). The analytic control  arising from the disorder averaging enabled us to scrutinize the model in some detail. 
We were able to obtain not just the spectrum of the non-chiral operators, but also the explicit OPE data which were confirmed to obey the known conformal bootstrap bounds. We also have some additional information (eg., non-chiral OPE coefficients) which have not  been constrained  thus far in the bootstrap literature. Our analysis not only captured the singlet sector of the OPE (i.e., the data from the operator averaged correlators) but also aspects of the non-singlet states.

There are several interesting directions that are worth exploring.  A simple variant of the model we studied wherein we single out one of the chiral superfields  should allow one to make contact with  vector-like large $N$ theories. This class of  models have been analyzed in the $\epsilon$ expansion in \cite{Ferreira:1996az,Ferreira:1997hx} and were critically investigated recently in \cite{Chester:2015qca,Chester:2015lej}. We will only describe some key features here leaving a more detailed analysis for the future  \cite{Chang:2021vm}.

To be precise, consider taking $N$ chiral superfields $\Phi_i$ and and additional $(N+1)^{\rm st}$ field, $\AS$, distinguished from the rest.  One takes the $\Phi_i$ and $\AS$ to have canonical K\"ahler terms. There are various possibilities for the superpotential (see eg., \cite{Chester:2015qca}) but the simplest choice of interest is one where we have a cubic monomial $h^{ij} \Phi_i \,\Phi_j \, \AS$ with $h^{ij}$ now drawn from a Gaussian random ensemble. This model has a $\mathbb{Z}_2 \times U(1)$ global symmetry where the  $\mathbb{Z}_2$ acts by reflection $\Phi_i \to -\Phi_i$ and $q(X = -2)$ and $q(\Phi) =1$ under the global $U(1)$. To see these are the disordered analog of the vector-like models analyzed in \cite{Chester:2015qca} we simply note that $h_{ij} = h\, \delta_{ij}$ gives us a global $O(N)$ symmetric models with $\Phi_i$ transforming in the vector representation.

Since there are $N$ $\Phi_i$ and only one $\AS$ it is reasonable to expect that the $\Phi$ affect $\AS$ more than the other way around. Indeed a quick diagrammatic check indicates that the diagrams where $\AS$ runs in the loops are sub-dominant. The theory at large $N$ flows to a fixed point where $\Phi$ stays free but $\AS$ picks up an anomalous dimension; $\Delta(\phi) = \frac{1}{2}$ and $\Delta(\Ax) = 1$ where $\Ax$ is the bottom component of $\AS$.  For the undisordered $O(N) $ symmetric model this was confirmed numerically in  \cite{Chester:2015lej}. We expect this theory to have higher spin conserved currents at large $N$, which would be interesting to investigate.
 
One motivation for us to  analyze the 3d model was to understand the finite temperature dynamics in the system. For one, it is interesting to understand the analytic structure of real-time thermal Green's functions. We recall that strongly coupled systems such as those with holographic duals exhibit quasinormal poles (resulting from the thermal density matrix being dual to a black hole geometry), in contrast to weakly coupled examples which exhibit branch cuts \cite{Hartnoll:2005ju}. Relatedly, it would be interesting to analyze transport properties  (see eg., \cite{Romatschke:2019ybu,Romatschke:2019gck} for recent studies of the $O(N)$ vector model) and the nature of chaotic scrambling dynamics of the fixed point. In quantum mechanics, one doesn't have spatial profile and hence no notion of a butterfly velocity for scrambling. In two dimensional critical systems, the underlying conformal invariance fixes the butterfly velocity to be the speed of light (though there can non-trivial momentum dependence).   We refer the reader to \cite{Mezei:2019dfv} for a detailed analysis from a wide class of models studied in the literature.  In higher dimensions, one expects a butterfly cone and perhaps even  a non-trivial velocity dependence for the Lyapunov exponent. Thus, a-priori, it is possible that the higher dimensional examples exhibit features that are quite distinct.  

The non-trivial part in computing the thermal properties of the IR SCFT is that one has to numerically solve the Schwinger-Dyson equations to obtain the thermal real-time propagators (as explained in \cite{Maldacena:2016hyu}). Unlike in lower dimensional examples thermal 2-point functions are no longer determined by a conformal map from the vacuum 2-point function. A related consequence is a fact that we already discussed in \cref{sec:Regge}: the chaos exponent is no longer simply related to the Regge intercept of the vacuum 4-point function. Our preliminary numerical investigations \cite{Chang:2021tch} suggests that the model will not provide an example of a system exhibiting maximal chaos, and is thus unlikely to have a sparse enough spectrum to admit a classical gravitational dual. We anticipate that the theory has a finite string tension and the semiclassical worldsheet dynamics would be more strongly coupled in the three dimensional case than in the two dimensional example analyzed in \cite{Murugan:2017eto}. Nevertheless much more needs to be done to confirm these preliminary findings -- we hope to report on these issues in the not too distant future.

\acknowledgments
It is a pleasure to thank R. Loganayagam, Mark Mezei,  David Simmons-Duffin, and  Douglas Stanford for discussions.
CC is partly supported by National Key R\&D Program of China (NO. 2020YFA0713000).
SCE and CP were  supported by U.S.\ Department of Energy grant {DE-SC0019480} under the HEP-QIS QuantISED program and funds from the University of California. The work of CP after he returned to China is supported by the Fundamental Research Funds for the Central Universities, by funds from the University of Chinese Academy of Science (UCAS), and funds from the Kavli Institute for Theoretical Science (KITS). MR  was supported by  U.S. Department of Energy grant DE-SC0009999 and by funds from the University of California.

\appendix
\section{Conventions} 
\label{sec:conventions}

We will find it convenient to work in $\mathcal{N}=2$ superspace, which we parameterize by a pair of complex spinor coordinates $\theta_\alpha$, $\thetab_\alpha$.  The two supercharges  
\begin{equation}
\Q_\alpha  =-\pdv{}{\theta^\alpha}+ i\,C^{\beta\gamma} \, \thetab_\gamma\, \left(\sigma^\mu\right)_{\alpha \beta}\, \pdv{}{x^\mu} \,,
\qquad
\Qb_\alpha  =-\pdv{}{\thetab^\alpha}- i\,C^{\beta\gamma} \, \theta_\gamma\, \left(\sigma^\mu\right)_{ \beta \alpha}\, \pdv{}{x^\mu} \,,
\end{equation}	
satisfy the supersymmetry algebra
\begin{equation}
\{\Q_\alpha, \Q_\beta\}=  \{\Qb_\alpha,\Qb_\beta\}=0\,,\qquad
\{\Q_\alpha,\Qb_\beta\}=2i\,(\sigma^\mu)_{\alpha\beta} \pdv{}{x^\mu} \,.
\end{equation}	
Here $\sigma$ are the Pauli matrices and we take $C^{\alpha\beta}$, a Hermitian matrix, which defines the inner product between fermions via
\begin{equation}
\chi \psi = \chi^\alpha \psi_\alpha = C^{\alpha\beta}\,  \chi_\beta \psi_\alpha = \psi \chi \,,
\end{equation}
to be
\begin{equation}
C^{\alpha\beta} = 
\begin{pmatrix}
0 & -i\\
i & 0 
\end{pmatrix} \ .
\end{equation}	
This matrix satisfies some identities which are useful to record:
\begin{equation}
\begin{split}
C^{\alpha\beta} & =-C_{\alpha\beta}\,, 
\qquad \qquad \qquad \;
C_{\alpha\beta}=-C_{\beta\alpha}\,. \\
C_{\alpha\beta}C^{\sigma\gamma}&=\delta_\alpha^\sigma\delta_\beta^\gamma - \delta_\alpha^\gamma\delta_\beta^\sigma\,,
\qquad C_{\alpha\beta}C^{\gamma\beta}=\delta_\alpha^\gamma\,,
\qquad  C_{\alpha\beta}C^{\alpha\beta}=2\ .
\end{split}
\end{equation}	
Finally, for functions on superspace the following identities are useful in expansion:
\begin{equation}\label{eq:susyexp}
\mathfrak{f}(y^\mu) 
= 
	e^{-i\, \theta \sigma^\mu \thetab \, \partial_\mu} \mathfrak{f}(x^\mu)\,, \quad
\overline{\mathfrak{f}}(y^{\dagger\mu}) 
= 
	e^{i\, \theta \sigma^\mu \thetab \, \partial_\mu} \overline{\mathfrak{f}}(x^\mu) \,,
\end{equation}	
where
\begin{equation}
y^\mu = x^\mu - i\, \theta\, \sigma^\mu \thetab\,,\quad  y^{\dagger \mu}  = x^\mu + i\, \theta\, \sigma^\mu \thetab\,.
\end{equation}
%

\section{Generalization to \texorpdfstring{$q$}{q}-body superpotential in \texorpdfstring{$d$}{d} dimensions}
\label{sec:genqd}

Let us consider supersymmetric SYK model with 4 supercharges in $d$ (Euclidean) spacetime dimensions. We will consider the cases $d=1,2,3$ where the low energy theory exists; cf., \cite{Anninos:2016szt,Gates:2021jdm} for the quantum mechanical model and \cite{Murugan:2017eto} for the two dimensional system.\footnote{Notice that in this appendix we only consider models with the non-chiral supersymmetry. For the $d=2$ results in the theory with chiral (${\cal N}=(0,2)$) supersymmetry, see \cite{Peng:2018zap}. For $d=1$ SYK models with lower ($\mathcal{N}=1$ or 2) supersymmetry, see \cite{Fu:2016vas}.}
However,  it is helpful to write the formal expressions for arbitrary dimensions. The supersymmetric Lagrangian density is given by 
\begin{equation}\label{eq:susyLdq}
\begin{split}
{\cal L}
&=
	\int d^2\theta d^2\thetab \,\Phib^i\Phi_i - \int d^2\theta \, 
	\frac{1}{q} \, g^{i_1\cdots i_q }\Phi_{i_1}\cdots\Phi_{i_q} -\int d^2\thetab \ \frac{1}{q} \,\overline{g}_{i_1\cdots i_q} \Phib^{i_1}\cdots \Phib^{i_q}. \\
& =
	-i \psib^{i \alpha}(\sigma^\mu)_\alpha{}^\beta \partial_\mu\psi_{i \beta} 
	+\partial_\mu \phib^i\partial^\mu\phi_i -\Fb^i F_i
\\
&\qquad 
-g^{i_1\cdots i_q}\left(F_{i_1}\phi_{i_2}\cdots \phi_{i_q}- \frac{q-1}{2}\, \psi_{i_1}^\alpha \psi_{i_2 \alpha} \phi_{i_3}\cdots \phi_{i_q}\right) + \text{c.c}
\end{split}
\end{equation}	
The coupling constants are  Gaussian distributed with zero mean and variance given by 
\begin{equation}
\expval{ g^{i_1\cdots i_q}\overline g_{j_1\cdots j_q}}
=  \frac{qJ^{2(d-1)-(d-2)q}}{ N^{q-1}}  \,\delta^{i_1}_{(j_1}\cdots\delta^{i_q}_{j_q)}
\equiv   \frac{q\hat{J}_{d,q}}{N^{q-1}}\,\delta^{i_1}_{(j_1}\cdots\delta^{i_q}_{j_q)}\,,
\end{equation}	
the choice of normalization being made to ensure  the existence of a large $N$ limit. $\hat{J}_{d,q}$ sets the mass scale of the interaction.

The Schwinger-Dyson equations for the propagators are as in \eqref{eq:3dSD} with the only change being in the equations for the self-energies which are now given by 
\begin{equation}\label{eq:Sigmadq}
\begin{split}
\Sigma_\phi(x)
& 
	= \hat{J}_{d,q}\, (q-1)
	 \left[G_F(x)\, G_\phi(x) 
	 - \frac{1}{2} \, (q-2)\,G\indices{_\alpha^\beta}(x) G\indices{^\alpha_\beta}(x) \right] G_\phi(x)^{q-3}\,,\\
\bm{\Sigma}(x)
& =
	\hat{J}_{d,q} \, (q-1) \mathbf{G}(x)\, G_\phi(x)^{q-2}, \\
\Sigma_F(x) 
& = 
	\hat{J}_{d,q} \, G_\phi(x)^{q-1}.
\end{split}
\end{equation}

We can solve the Schwinger-Dyson equations in the strong coupling limit $|p|^{-1} J\to\infty$. As before we drop the UV contribution (the free propagator), and consider the conformal ansatz:
\begin{equation}
G^*_\phi(x)
	= \frac{b_\phi }{ |x|^{2\Delta_\phi}},
\qquad  
\mathbf{G}^*(x)
	= b_\psi \, \frac{x^\mu\sigma_\mu}{ |x|^{2\Delta_\psi+1}},
\qquad 
G^*_F(x)= \frac{b_F}{ |x|^{2\Delta_F}}.
\end{equation}	
The self-energies evaluate to
\begin{equation}
\begin{split}
\Sigma^*_\phi(x)
&=
	 \hat{J}_{d,q} \, (q-1) \left[\frac{b_F\, b_\phi^{q-2}}{ |x|^{2\Delta_F+2(q-2)\Delta_\phi}} -
	(q-2) \, \frac{b_\psi^2\, b_\phi^{q-3}}{ |x|^{4\Delta_\psi+2(q-3)\Delta_\phi}}\right], \\
\bm{\Sigma}^*(x)
&=
	\hat{J}_{d,q}\, (q-1) \, b_\psi\, b_\phi^{q-2} \, \frac{x^\mu\sigma_\mu}{ |x|^{2\Delta_\psi+1+2(q-2)\Delta_\phi}},\\
\Sigma^*_F(x)
&=
	\hat{J}_{d,q}\,  b_\phi^{q-1} \, \frac{1}{ |x|^{2(q-1)\Delta_\phi}}.
\end{split}
\end{equation}
In deriving these expressions (and the analogous ones in the main text) we have made use of the following Fourier transformations:
\begin{equation}\label{eq:propFT}
\begin{split}
\int\, d^dx\, \frac{e^{ikx}}{|x|^{2\Delta}} 
&= 
	\frac{\pi^\frac{d}{2}\, \Gamma(\frac{d}{2}-\Delta)}{2^{2\Delta-d}\, \Gamma(\Delta)} \, \frac{1}{|k|^{d-2\Delta}} \,, \\
\int\, d^dx\, \frac{x^\mu \, e^{ikx}}{|x|^{2\Delta+1}} 
&= 
	i\, \frac{\pi^\frac{d}{2}\, \Gamma(\frac{d}{2}-\Delta + \frac{1}{2})}{2^{2\Delta-d}\, \Gamma(\Delta+\frac{1}{2})} \, \frac{k^\mu}{|k|^{d-2\Delta+1}} \,.	
\end{split}
\end{equation}

On the supersymmetric vacuum, the equations \eqref{eq:susyWard} give the relations:
\begin{equation}
\begin{split}
\Delta_\psi 
&=
	\Delta_\phi+ \frac{1}{2} \,, \qquad  
	\Delta_F=\Delta_\phi+1\,,  \\
b_\psi 
&= 
	2i\,\Delta_\phi\, b_\phi\,, 
	\qquad  \quad \!  b_F=2\Delta_\phi \, (2\Delta_\phi+2-d)\,b_\phi\,.
\end{split}
\end{equation}	
Consequently, the solution to the Schwinger-Dyson equations can be readily obtained and reads:
\begin{equation}
\begin{split}
\Delta_\phi 
& =
	 \frac{d-1}{q} \,,\\
b_\phi^q \, \hat{J}_{d,q} 
&= 
	\frac{1}{4\pi^{d+1}}  \, \cos\left(\frac{(d-1)(q-2) \pi}{2q}\right)\, \Gamma\left(\frac{d-1}{q}\right) \, \Gamma\left(\frac{(d-1)(q-1)}{q}\right)
\end{split}
\end{equation}	
which reduces to \eqref{eq:Dbphi} for $d=3$ and $q=3$ which is the situation of primary focus.

Given the general solution above, we can compare this against the results obtained in $d=2$ by \cite{Murugan:2017eto} and in $d=1$ by \cite{Anninos:2016szt}.\footnote{For a complete analysis  of the ${\cal N}=2$ model and more detailed results, see \cite{Peng:2017spg,Peng:2020euz}.} For example the asymptotics at large twist and spin for the two dimensional model can be deduced from the above (see also the results in \cite{Bulycheva:2018qcp}). Specifying to $q=3$ we find the conformal dimensions are organized into the sequence 
\begin{equation}
h = \ell+ \frac{t}{2}\,, \qquad  \bar{h} = \frac{t}{2}\,,  \qquad t= 2\Delta_\phi+ 2m + \gamma(m,\ell)
\end{equation}	
the anomalous dimensions $\gamma(m,\ell)$ are captured by 
\begin{equation}
\begin{split}
\gamma(m,\ell) &= \frac{\mathcal{F}_2(\Delta_\phi)}{m^{2-4\Delta_\phi}} \,, \qquad \qquad \qquad \quad m \gg \ell \sim 1 \,,  \\
\gamma(m,\ell) &= \frac{\mathcal{F}_2(\Delta_\phi)}{s^{1-2\Delta_\phi}} \, \frac{\Gamma(m+2\Delta_\phi)}{\Gamma(m+1)} \,, \qquad \ell \gg 1 \,,
\end{split}
\end{equation}	
with 
\begin{equation}
\begin{split}
\mathcal{F}_2(\Delta_\phi) &= \frac{2(\Delta_\phi -1)\, \Delta_\phi \, \sin(2\pi\Delta_\phi) \, \Gamma(-\Delta_\phi)^2}{\pi\, \Gamma(\Delta_\phi)^2}  \\
\mathcal{F}_2(\Delta_\phi = \frac{1}{3}) &= -\frac{2\, \Gamma(-\frac{1}{3})^2}{3\sqrt{3}\, \Gamma(\frac{1}{3})^2} = -0.281 \,.
\end{split}
\end{equation}	
On the other hand in $d=1$ we simply have 
\begin{equation}
\begin{split}
\Delta &= 2\, \Delta_\phi + 2m + \gamma(m) \\ 
\gamma(m) &= \frac{16^{\Delta_\phi}\, \sin^2(2\pi\Delta_\phi) \, \Gamma(2-2\Delta_\phi)\, \Gamma(-2\Delta_\phi)}{\pi^2\, m^{1-4\Delta_\phi}} \,.
\end{split}
\end{equation}	
%
\section{Conformal partial waves: review}
\label{sec:cpw}

We review  bosonic conformal partial waves briefly in this appendix, for many of these results enter our calculation of the four-point function in $\S$\ref{sec:4pt}. A detailed discussion can be found in the early work \cite{Dolan:2011dv} and in the recent analysis of Lorentzian OPEs in \cite{Caron-Huot:2017vep}, \cite{Simmons-Duffin:2017nub}.

The conformal partial waves $\Psi_{\Delta,\ell}^{\Delta_{12},\Delta_{34}}$ are labelled by four external dimensions $\Delta_{i}$ and an internal dimension and spin $(\Delta,\ell)$. Here $\Delta_{ij} = \Delta_i -\Delta_j$. 
They can be written as the following linear combination of a conformal block and its shadow block\footnote{Note that our conformal blocks have a different normalization from \cite{Simmons-Duffin:2017nub}: $(G_{\Delta,\ell}^{\Delta_{i}})^{\mathrm{us}} = \frac{(-1)^{\ell}}{2^{\ell}}(G_{\Delta,\ell}^{\Delta_{i}})^{\mathrm{them}}$.}
\begin{equation}\label{eq:cpw}
\Psi_{\Delta,\ell}^{\Delta_{12},\Delta_{34}} = 
	S_{\widehat{\Delta},\ell}^{\Delta_{34}}
	\, G_{\Delta,\ell}^{\Delta_{12},\Delta_{34}} +
	S_{\Delta,\ell}^{\Delta_{12}}\, 
	G_{\widehat{\Delta},\ell}^{\Delta_{12},\Delta_{34}}
\end{equation}
where $\Delta = \frac{3}{2}+is$ is the principal series for the conformal group $SO(4,1)$ and $\widehat{\Delta}=3-\Delta$ is the conformal dimension of the shadow operator. A complete set of conformal partial waves can be constructed by restricting to $s>0$ and $\ell \geq 0$. The shadow coefficients $S_{\Delta,\ell}^{\Delta_{12}}$ are given by
\begin{equation}\label{eq:shadowcoeff}
S_{\Delta,\ell}^{\Delta_{12}} = 
\pi^{\frac{3}{2}}\frac{\Gamma\left(\Delta-\frac{3}{2}\right)\Gamma\left(\Delta+\ell-1\right)\Gamma\left(\frac{\widehat{\Delta}+\Delta_{12}+\ell}{2}\right)\Gamma\left(\frac{\widehat{\Delta}-\Delta_{12}+\ell}{2}\right)}{\Gamma\left(\Delta-1\right)\Gamma(\widehat{\Delta}+\ell)\, \Gamma\left(\frac{\Delta+\Delta_{12}+\ell}{2}\right)\Gamma\left(\frac{\Delta-\Delta_{12}+\ell}{2}\right)}.
\end{equation}
The normalization of the conformal partial waves can be computed directly from the expansion of the conformal blocks around the origin with the result
\begin{equation}\label{eq:cpw_norm}
\expval{\Psi_{\Delta,\ell}^{\Delta_{12},\Delta_{34}},\Psi_{\Delta',\ell'}^{\Delta_{12},\Delta_{34}}}_{0}^{\Delta_{12},\Delta_{34}} = n_{\Delta,\ell} \;  2\pi\delta(s-s')\delta_{\ell\ell'},
\end{equation}
where we have used the inner product defined in \eqref{eq:bosip_realext} and the normalization constant is
\begin{equation}\label{eq:normconst}
n_{\Delta,\ell} = \frac{\pi^{4}\,(2\ell+1)}{2^{2\ell-1}\, (\Delta+\ell-1)\,(2-\Delta+\ell)} \ 
	\frac{\Gamma\left(\Delta-\frac{3}{2}\right)\Gamma\left(\frac{3}{2}-\Delta\right)\Gamma\left(\ell+1\right)^{2}}{\Gamma\left(\Delta-1\right)\Gamma\left(2-\Delta\right)\Gamma\left(\ell+\frac{3}{2}\right)^{2}}.
\end{equation}

Let us now explain the subtlety regarding the bosonic inner product for real external dimensions that gives \eqref{eq:bosip_realext}. When the external dimensions live in the principal series ($\Delta_{i} \in \frac{3}{2}+is$), the inner product \eqref{eq:bosip} gives ($\widehat{\Delta}_{ij} \equiv \widehat{\Delta}_{i} - \widehat{\Delta}_{j}$)
\begin{equation}\label{eq:bosip_cpw}
\begin{split}
\expval{ \Psi_{\Delta,\ell}^{\Delta_{12},\Delta_{34}},\Psi_{\Delta',\ell'}^{\Delta_{12},\Delta_{34}} }_{0} 
	&= \int d^{2}z\,\frac{|z-\zb|}{|z|^{6}}\,
		\Psi_{\Delta',\ell'}^{\Delta_{12},\Delta_{34}}(z,\zb) \; 
		\overline{\Psi_{\Delta',\ell'}^{\Delta_{12},\Delta_{34}}}(z,\zb) \\
	&= 
		\int d^{2}z\,\frac{|z-\zb|}{|z|^{6}}\,
			\Psi_{\Delta',\ell'}^{\Delta_{i}}(z,\zb) \; 
			\overline{\Psi}_{\widehat{\Delta}',\ell'}^{\widehat{\Delta}_{12},\widehat{\Delta}_{34}}(z,\zb).
\end{split}
\end{equation}
After analytic continuation to $\Delta_{i} \in \mathbb{R}$, one now has 
$\overline{\Delta}_i \neq \widehat{\Delta}_{i}$. Using the identity
\begin{equation}\label{eq:confblock_id}
G_{\Delta,\ell}^{-\Delta_{12},-\Delta_{34}} = 
 |1-z|^{-\Delta_{12}+\Delta_{34}} \; G_{\Delta,\ell}^{\Delta_{12},\Delta_{34}},
\end{equation}
one can see that the correct analytic continuation of the inner product is
\begin{equation}\label{eq:bosip_correct}
\expval{ \Psi_{\Delta,\ell}^{\Delta_{12},\Delta_{34}},\,
	\Psi_{\Delta',\ell'}^{\Delta_{12},\Delta_{34}} }_{0}^{\Delta_{12},\Delta_{34}} = 
\int d^{2}z\,\frac{|z-\zb|}{|z|^{6}}|1-z|^{-\Delta_{12}+\Delta_{34}}\,
	\Psi_{\Delta',\ell'}^{\Delta_{12},\Delta_{34}}(z,\zb)\;
	\overline{\Psi}_{\widehat{\Delta}',\ell'}^{\Delta_{12},\Delta_{34}}(z,\zb).
\end{equation}

Finally, we review an alternate definition of the conformal partial wave using the shadow formalism. In this formalism, the conformal partial wave $\Psi_{\Delta,\ell}^{\Delta_{i}}$ is constructed from the three-point function $\expval{ \mathcal{O}_{1}\mathcal{O}_{2}\mathcal{O}_{\Delta,\ell} }$ and the corresponding three-point function for the shadow operator
\begin{equation}\label{eq:shadowop}
\widehat{\mathcal{O}}_{\widehat{\Delta},\ell}(x) = \int d^{3}y\, \frac{1}{|x-y|^{2\widehat{\Delta}}}\mathcal{O}_{\Delta,\ell}(y).
\end{equation}
To wit,
\begin{equation}\label{eq:cpw_int}
\Psi_{\Delta,\ell}^{\Delta_{12},\Delta_{34}}(\{x_{i}\}) = \int d^{3}x_{5}\frac{\expval{\mathcal{O}_{1}(x_{1})\mathcal{O}_{2}(x_{2})\mathcal{O}_{\Delta,\ell}^{\mu_{1} \ldots \mu_{\ell}}(x_{5})}\expval{\widehat{\mathcal{O}}_{\widehat{\Delta},\ell;\mu_{1} \ldots \mu_{\ell}}(x_{5})\mathcal{O}_{3}(x_{3})\mathcal{O}_{4}(x_{4})}}{\expval{\mathcal{O}_{1}(x_{1})\mathcal{O}_{2}(x_{2})} \expval{\mathcal{O}_{3}(x_{3})\mathcal{O}_{4}(x_{4})}}.
\end{equation}
The three-point function appearing in \eqref{eq:cpw_int} is
\begin{equation}\label{eq:bos3ptfn}
\expval{\mathcal{O}_{1}(x_{1})\mathcal{O}_{2}(x_{2})\mathcal{O}_{\Delta,\ell}^{\mu_{1}\ldots\mu_{\ell}}(x_{5})} = \frac{Z_{12}^{\mu_{1}}\ldots Z_{12}^{\mu_{\ell}} - \mathrm{traces}}{|x_{12}|^{\Delta_{1}+\Delta_{2}-\Delta}|x_{25}|^{\Delta_{2}+\Delta-\Delta_{1}}|x_{15}|^{\Delta_{1}+\Delta-\Delta_{2}}},
\end{equation}
where
\begin{equation}\label{eq:spinvar}
Z_{12}^{\mu} = \frac{|x_{15}||x_{25}|}{|x_{12}|}\left(\frac{x_{15}^{\mu}}{x_{15}^2}-\frac{x_{25}^{\mu}}{x_{25}^2}\right), \qquad Z_{12}^2 = 1.
\end{equation}
Explicitly, the conformal partial wave is 
\begin{equation}\label{eq:cpw_explicit}
\Psi_{\Delta,\ell}^{\Delta_{12},\Delta_{34}} = \int d^{3}x_{5}\,\frac{|x_{12}|^{\Delta}}{|x_{25}|^{\Delta_{2}+\Delta-\Delta_{1}}|x_{15}|^{\Delta_{1}+\Delta-\Delta_{2}}}\frac{|x_{34}|^{\widehat{\Delta}}}{|x_{35}|^{\Delta_{3}+\widehat{\Delta}-\Delta_{4}}|x_{45}|^{\Delta_{4}+\widehat{\Delta}-\Delta_{3}}}\widehat{C}_{\ell}(\eta),
\end{equation}
where
\begin{equation}\label{eq:shadowinv}
\eta = \frac{|x_{15}||x_{25}|}{|x_{12}|}\frac{|x_{45}||x_{35}|}{|x_{34}|}\bigg(\frac{x_{15}^{\mu}}{x_{15}^2}-\frac{x_{25}^{\mu}}{x_{25}^2}\bigg)\bigg(\frac{x_{35,\mu}}{x_{35}^2}-\frac{x_{45,\mu}}{x_{45}^2}\bigg)
\end{equation}
and $\widehat{C}_{\ell}$ is given by
\begin{equation}\label{eq:Chat}
\widehat{C}_{\ell}(x) \equiv \frac{\sqrt{\pi} \,\Gamma(\ell+1)}{2^{\ell} \,\Gamma\left(\ell+\frac{1}{2}\right)} \;
		{}_{2}{F}_{1}\left(-\ell,\ell+1,1;\frac{1-x}{2}\right).
\end{equation}
%

\section{Superconformal three-point function}
\label{sec:s3pt}

The form of the three-point function of general primary superfields in 3d CFTs with ${\cal N}$-extended supersymmetry is derived in \cite{Buchbinder:2015qsa}. In this appendix, we first review the result in \cite{Buchbinder:2015qsa}, and then specialize it to the three-point function of our interest -- that of a chiral superfield, an anti-chiral superfield and a general spin-$\ell$ superfield in ${\cal N}=2$ SCFT. 

Let us consider the superspace ${\mathbb R}^{3|2{\cal N}}$ with bosonic coordinates $x^\mu$ for $\mu=1,\cdots,\,3$, and fermionic coordinates $\theta^{i\alpha}$ for $i=1,\cdots,\,{\cal N}$ and $\alpha = 1,\,2$.  We define the two-point structures 
\begin{equation}
{\bf x}_{12}^{\alpha\beta}=(x_1-x_2)^\mu (\sigma^\mu)^{\alpha\beta}+2i\theta_1^{i(\underline{\alpha}}\theta_2^{i\underline{\beta})}-i \theta_{12}^{i\alpha}\theta_{12}^{i\beta}\,,\quad  u_{12}^{ij}=\delta^{ij}+2i\theta^{i\alpha}_{12}({\bf x}_{12}^{-1})_\alpha{}^\beta\theta^{j}_{12,\beta}
\end{equation}
and
\begin{equation}
|{\bf x}_{12}|^2=-\frac{1}{2}{\bf x}_{12,\alpha}{}^\beta {\bf x}_{21,\beta}{}^\alpha\,,\quad
\underline{{\bf x}}_{12,\alpha}{}^\beta=\frac{{\bf x}_{12,\alpha}{}^\beta}{ |{\bf x}_{12}|}\,.
\end{equation}
The two-point structures $\underline{{\bf x}}_{12,\alpha}{}^\beta$ and $u_{12}^{ij}$ transform only under the local Lorentz and  the local ${\rm SO}({\cal N})$ R-symmetry transformations of the superconformal algebra in the way indicated by their Lorentz and R-symmetry indices. The structure $|{\bf x}_{12}|^2$  transforms only under the local scaling transformation with scaling dimension $-1$.\footnote{The local Lorentz, scaling and ${\rm SO}({\cal N})$ R-symmetry transformations are the corresponding transformations with the local parameters defined in equations (4.7a), (4.7b) and (4.7c) of \cite{Buchbinder:2015qsa}.}
The two-point structures satisfy
\begin{equation}
{\bf x}_{12}{}^\beta{}_\alpha=-{\bf x}_{21,\alpha}{}^\beta\,,\quad u_{21}^{ij}=u_{12}^{ji}\,,
\quad\underline{\bf x}_{12,\alpha}{}^\gamma\underline{\bf x}_{21,\gamma}{}^\beta=\delta^\beta_\alpha\,,\quad u_{12}^{ik}u_{12}^{jk}=\delta^{ij}\,.
\end{equation}
Next, we define the three-point structures
\begin{equation}
\begin{split}
&{\bf X}_{3,\alpha}{}^\beta=({\bf x}_{13}^{-1})_\alpha{}^\gamma {\bf x}_{12,\gamma}{}^\delta({\bf x}_{32}^{-1})_\delta{}^\beta\,,\quad \Theta^i_{3,\alpha} = ({\bf x}_{13}^{-1})_\alpha{}^\beta \theta^i_{31,\beta} -({\bf x}_{23}^{-1})_\alpha{}^\beta \theta^i_{32,\beta}\,,
\\
&U_3^{ij}=u_{31}^{ik}u_{12}^{kl}u_{23}^{lj}=\delta^{ij} + 2 i \Theta_3^{i\alpha} ({\bf X}_3^{-1})_\alpha{}^\beta \Theta_{3,\beta}^{j}\,,
\end{split}
\end{equation}
and
\begin{equation}
\begin{aligned}
  |{\bf X}_3|^2
  &=\frac{1}{2}{\bf X}_{3,\alpha}{}^\beta{\bf X}_{3,\beta}{}^\alpha\,, & \qquad 
  \underline{\bf X}_{3,\alpha}{}^\beta
  &=\frac{{\bf X}_{3,\alpha}{}^\beta}{ |{\bf X}_3|}\,,
\\
  |\Theta_3|^2&=
  \Theta_3^{i\alpha}\Theta_{3,\alpha}^i\,,&\qquad 
  \underline\Theta_{3,\alpha}^i
  &=
  \frac{\Theta_{3,\alpha}^i}{ |{\bf X}_3|^\frac{1}{2}}\,,
\end{aligned}
\end{equation}
and also the cyclic permutations of the $1,\,2,\,3$ of the above structures. The structures $\underline{\bf X}_{3,\alpha}{}^\beta$, $\underline\Theta_{3,\alpha}^i$ and $U_3^{ij}$ transform only under the local Lorentz and the local ${\rm SO}({\cal N})$ $R$-symmetry transformations in the way indicated by their Lorentz and R-symmetry indices. The structures $|{\bf X}_3|$ and $|\Theta_3|$ transform only under the local scaling transformation with scaling dimension $1$ and $\frac{1}{2}$, respectively. Hence, the combination
\begin{equation}
\frac{|\Theta_3|^2}{ |{\bf X}_3|}
\end{equation}
is a three-point invariant. There are relations between the three-point structures
\begin{equation}
{\bf x}_{13,\alpha}{}^\gamma {\bf X}_{3,\gamma}{}^\delta {\bf x}_{31,\delta}{}^\beta=
\frac{{\bf X}_{1,\alpha}{}^\beta}{ |{\bf X}_1|^2}\,,\quad \Theta_1^{i\gamma}{\bf x}_{13,\gamma}{}^\delta {\bf X}_{3,\delta}{}^\beta=u_{13}^{ij}\Theta^{j\beta}_3\,,\quad U^{ij}_3=u^{ik}_{31}U_1^{kl}u_{13}^{lj}\,,
\end{equation}
and the cyclic permutations of the $1,\,2,\,3$ of them. We can choose to work with the set of three-point structures
\begin{equation}
{\bf x}_{13,\alpha}{}^\beta\,,\quad {\bf x}_{23,\alpha}{}^\beta\,,\quad u_{13}^{ij}\,,\quad u_{23}\,,\quad {\bf X}_{3,\alpha}{}^\beta\,,\quad \Theta_{3,\alpha}^i\,,\quad U_3^{ij}\,.
\end{equation}
The rest of the structures can be generated by the structures in this set.

Consider a primary superfield $\Phi^{\cal I}_{\cal A}(X)$ that transforms in the Lorentz representation $T$ with indices denoted by ${\cal A},\,{\cal B},\,\cdots$ and the ${\rm SO}({\cal N})$ $R$-symmetry representation $D$ with indices ${\cal I},\,{\cal J},\,\cdots$. The transformation law of primary superfields under the superconformal algebra is given in (5.1) of \cite{Buchbinder:2015qsa}, which constrains the two- and three-point functions of them to only depend on the two- and three-point structures introduced above. For example, the two-point function of a primary superfield $\Phi^{\cal I}_{\cal A}(X)$ and its conjugate $\overline\Phi_{\cal I}^{\cal A}(X)$ takes the form
\begin{equation}\label{eq:generalSUSY2pt}
\langle \Phi^{\cal I}_{\cal A}(X_1) \overline\Phi_{\cal J}^{\cal B}(X_2)\rangle = 
\, \frac{ {T_{\cal A}{}^{\cal B}(\underline{\bf x}_{12})}{D^{\cal I}{}_{\cal J}(u_{12})}}{ |{\bf x}_{12}|^{2\Delta}}\,,
\end{equation}
where $X=(x,\theta)$ is the superspace coordinate and $\Delta$ is the conformal dimension of $\Phi^{\cal I}_{\cal A}(X)$. The three-point function of three different primary superfields takes the form
\begin{equation}\label{eq:generalSUSY3pt}
\begin{split}
\langle \Phi^{{\cal I}_1}_{1,{\cal A}_1}(X_1)  \Phi^{{\cal I}_2}_{2,{\cal A}_2}(X_2)  \Phi^{{\cal I}_3}_{3,{\cal A}_3}(X_3)\rangle& = 
\frac{T^{(1)}{}_{{\cal A}_1}{}^{{\cal B}_1}(\underline{\bf x}_{13})T^{(2)}{}_{{\cal A}_2}{}^{{\cal B}_2}(\underline{\bf x}_{23})D^{(1)}{}^{{\cal I}_1}{}_{{\cal J}_1}(u_{13})D^{(2)}{}^{{\cal I}_2}{}_{{\cal J}_2}(u_{23})}{ |{\bf x}_{13}|^{2\Delta_1}  |{\bf x}_{23}|^{2\Delta_2} }
\\
&\quad\times H^{{\cal J}_1{\cal J}_2 {\cal I}_3}_{{\cal B}_1{\cal B}_2{\cal A}_3}({\bf X}_3,\Theta_3,U_3)\,,
\end{split}
\end{equation}
where the function $H^{{\cal J}_1{\cal J}_2 {\cal I}_3}_{{\cal B}_1{\cal B}_2{\cal A}_3}({\bf X},\Theta,U)$ satisfies the scaling condition
\begin{equation}\label{eq:scalingH}
H^{{\cal J}_1{\cal J}_2 {\cal I}_3}_{{\cal B}_1{\cal B}_2{\cal A}_3}(\lambda^2{\bf X},\lambda\Theta,U)=\lambda^{2\Delta_3-2\Delta_2-2\Delta_1}H^{{\cal J}_1{\cal J}_2 {\cal I}_3}_{{\cal B}_1{\cal B}_2{\cal A}_3}({\bf X},\Theta,U)\,,\quad{\rm for}\quad\lambda>0.
\end{equation}

The general three-point function \eqref{eq:generalSUSY3pt} can be specialized to the three-point function of superfields in the short multiplets. The shortening conditions usually involve the superderivatives
\begin{equation}
D^{i}_\alpha = \pdv{\theta^{i\alpha}}+i\theta^{i\beta} \sigma^\mu_{\beta\alpha} \, \pdv{x^\mu}
\end{equation}
acting on the superfields, and turn to differential equations on the function $H^{{\cal J}_1{\cal J}_2 {\cal I}_3}_{{\cal B}_1{\cal B}_2{\cal A}_3}({\bf X},\Theta,U)$. When deriving the differential equations, it would be useful to note the identities
\begin{equation}\label{eq:chiralconF}
\begin{split}
&D_{1,\gamma}^i f^{\mu_1\cdots\mu_\ell}({\bf X}_3,\Theta_3)=-({\bf x}_{31}^{-1})_\gamma{}^\alpha u_{13}^{ij}{\cal D}^j_{3,\alpha} f^{\mu_1\cdots\mu_\ell}({\bf X}_3,\Theta_3),
\\
&D_{2,\gamma}^i f^{\mu_1\cdots\mu_\ell}({\bf X}_3,\Theta_3)=-i({\bf x}_{32}^{-1})_\gamma{}^\alpha u_{23}^{ij}{\cal Q}^j_{3,\alpha} f^{\mu_1\cdots\mu_\ell}({\bf X}_3,\Theta_3),
\end{split}
\end{equation}
where the differential operators ${\cal D}^{i}_\alpha$ and ${\cal Q}^{i}_\alpha$ are
\begin{equation}
{\cal D}^{i}_\alpha = \pdv{\Theta^{i\alpha}}+i\Theta^{i\beta} \sigma^\mu_{\beta\alpha}\pdv{{\bf X}^\mu}\,,\quad {\cal Q}^{i}_\alpha =i \pdv{\Theta^{i\alpha}}+\Theta^{i\beta} \sigma^\mu_{\beta\alpha}\pdv{{\bf X}^\mu}\,,
\end{equation}
where ${\bf X}^\mu=\frac{1}{2}{\bf X}_\alpha{}^\beta (\sigma^\mu)_\beta{}^\alpha$.

Now, we specialize to ${\cal N}=2$, and introduce the complex fermionic coordinates and the complex superderivatives
\begin{align}
&\theta^\alpha=\frac{1}{\sqrt{2}} \, (\theta^{1\alpha}+i\theta^{2\alpha})\,,&& \bar \theta^\alpha=\frac{1}{\sqrt{2}} \, (\theta^{1\alpha}-i\theta^{2\alpha})\,.
\\
&D_\alpha = \frac{1}{\sqrt{2}} \, (D^1_\alpha-iD^2_\alpha)\,,&& \overline D_\alpha = -\frac{1}{\sqrt{2}} \, (D^1_\alpha+iD^2_\alpha)\,,
\end{align}
Consider a chiral superfield $\Phi(X)$ that satisfies the chiral condition
\begin{equation}
\overline D_\alpha \Phi(X)=0,
\end{equation}
which in particular constrains the conformal dimension $\Delta$ to be equal to the $R$-charge $q$.
To write down the correlation functions involving chiral superfields, it is convenient to introduce the chiral two-point structure
\begin{equation}
{\bf z}_{12}^{\alpha\beta}={\bf x}_{12}^{\alpha\beta}-2i\overline{\theta}^\alpha_{12}{\theta}_{12}^\beta={\bf x}^{\alpha\beta}_{12}-i(\theta_{12}^{i\alpha}\theta_{12}^{i\beta}+i\epsilon_{ij}\theta_{12}^{i\alpha}\theta_{12}^{j\beta}),
\end{equation}
which is related to the superspace translation invariant combination $z_{ 12}$ defined in \eqref{eq:superz} by 
\begin{equation}
{\bf z}_{1 2}^{\alpha\beta} = - z_{ \overline 2 1}^\mu(\sigma^\mu)^{\alpha\beta}.
\end{equation}
We also have the identity
\begin{equation}
|z_{21}|^2 = |{\bf x}_{12}|^2(u_{12}^{11}+iu_{12}^{12})\,.
\end{equation}
Specializing the general two-point function \eqref{eq:generalSUSY2pt} to the two-point function of the chiral superfield $\Phi(X)$ with its conjugate $\overline\Phi(X)$, we find
\begin{equation}
\langle \Phi(X_1) \overline\Phi(X_2)\rangle = \frac{ (u_{12}^{11}-i u_{12}^{12})^{\Delta} }{ |{\bf x}_{12}|^{2\Delta}}= \frac{1}{ |z_{2 1}|^{2\Delta}}\,.
\end{equation}
Specializing the general three-point function \eqref{eq:generalSUSY3pt} to the three-point function of the chiral superfield $\Phi(X)$, its conjugate $\overline\Phi(X)$ and a general spin-$\ell$ superfield ${\cal V}_{\Delta,\ell}^{\mu_1\cdots \mu_\ell}$, we find
\begin{equation}\label{eq:chiralSUSY3ptA}
\langle \Phi(X_1) \overline \Phi(X_2)  {\cal V}_{\Delta,\ell}^{\mu_1\cdots \mu_\ell}(X_3)\rangle = 
\frac{(u_{13}^{11}-i u_{13}^{12})^{\Delta_\Phi} (u_{23}^{11}+i u_{23}^{12})^{\Delta_\Phi} }{ |{\bf x}_{13}|^{2\Delta_\Phi}  |{\bf x}_{23}|^{2\Delta_\Phi} }H^{\mu_1\cdots\mu_\ell}({\bf X}_3,\Theta_3)\,,
\end{equation}
where the function $H^{\mu_1\cdots\mu_\ell}({\bf X}_3,\Theta_3)$ satisfies the chiral conditions
\begin{equation}\label{eq:chiralcondH}
\overline D_{1,\alpha}H^{\mu_1\cdots\mu_\ell}({\bf X}_3,\Theta_3)=0= D_{2,\alpha}H^{\mu_1\cdots\mu_\ell}({\bf X}_3,\Theta_3)\,.
\end{equation}
The useful identities \eqref{eq:chiralconF} give
\begin{equation}
\begin{split}
&\overline D_{1,\gamma} f^{\mu_1\cdots\mu_\ell}({\bf X}_3,\Theta_3)=-({\bf x}_{31}^{-1})_\gamma{}^\alpha (u_{13}^{11}-iu_{13}^{12})\overline {\cal D}_{3,\alpha} f^{\mu_1\cdots\mu_\ell}({\bf X}_3,\Theta_3),
\\
&D_{2,\gamma} f^{\mu_1\cdots\mu_\ell}({\bf X}_3,\Theta_3)=({\bf x}_{32}^{-1})_\gamma{}^\alpha (u_{23}^{11}+iu_{23}^{12}){\cal Q}_{3,\alpha} f^{\mu_1\cdots\mu_\ell}({\bf X}_3,\Theta_3),
\end{split}
\end{equation}
where the differential operators $\overline{\cal D}_\alpha$ and ${\cal Q}_\alpha$ are
\begin{equation}
\overline {\cal D}_\alpha = - \frac{1}{\sqrt{2}}\, ({\cal D}^1_\alpha+i{\cal D}^2_\alpha)\,,\quad{\cal Q}_\alpha=- \frac{1}{\sqrt{2}}\, (i {\cal Q}^1_\alpha+{\cal Q}^2_\alpha)\,.
\end{equation}
The chiral conditions \eqref{eq:chiralcondH} become differential equations
\begin{equation}\label{eq:diffH}
\overline {\cal D}_{3,\alpha}H^{\mu_1\cdots\mu_\ell}({\bf X}_3,\Theta_3)=0= {\cal Q}_{3,\alpha}H^{\mu_1\cdots\mu_\ell}({\bf X}_3,\Theta_3)\,.
\end{equation}
The solution to the scaling condition \eqref{eq:scalingH} and the differential equations \eqref{eq:diffH} is
\begin{equation}
H^{\mu_1\cdots\mu_\ell}({\bf X}_3,\Theta_3)=C_{\Phi\overline\Phi {\cal V}} |{\bf Y}_3|^{\Delta-\ell-2\Delta_\Phi}({\bf Y}_3^{\mu_1}\cdots{\bf Y}_3^{\mu_\ell}-{\rm traces})\,,
\end{equation}
where $C_{\Phi\overline\Phi {\cal V}}$ is the OPE coefficient and the vector ${\bf Y}^\mu_3$ is
\begin{equation}
{\bf Y}^\mu_3={\bf X}^\mu_3-i\Theta_3^\alpha (\sigma^\mu)_\alpha{}^\beta\overline\Theta_{3\beta}\,.
\end{equation}
We have the identity\footnote{This can be argued by the fact that both sides satisfy the chiral conditions and transform in the same way under the superconformal algebra.}
\begin{equation}
|{\bf Y}_3|^2 =\frac {|z_{2 1}|^2}{ |z_{ 3 1}|^2 |z_{ 2 3}|^2}\,.
\end{equation}
In summary, the three-point function is
\begin{equation}\label{eq:chiralSUSY3pt}
\langle \Phi(X_1) \overline \Phi(X_2)  {\cal V}_{\Delta,\ell}^{\mu_1\cdots \mu_\ell}(X_3)\rangle = C_{\Phi\overline\Phi {\cal V}} \, \frac{|z_{ 2 1}|^{\Delta-\ell-2\Delta_\Phi} }{ |z_{ 3 1}|^{\Delta-\ell}  |z_{ 23}|^{\Delta-\ell} }({\bf Y}_3^{\mu_1}\cdots{\bf Y}_3^{\mu_\ell}-{\rm traces})\,.
\end{equation}
%

\section{Supershadow coefficients}
\label{sec:ssops}

We discuss the computation of the supershadow coefficients $A_{\Delta,\ell}$ used to define the superconformal partial waves in terms of superconformal blocks \eqref{eq:scpw_scblock} in this appendix. This can be done using an alternate definition of the superconformal partial waves $\Upsilon_{\Delta,\ell}$ given by the supershadow formalism for 3d $\mathcal{N}=2$ SCFTs, which to our knowledge has received little attention in the literature (for analysis of the 4d $\mathcal{N}=1$ case, see \cite{Fitzpatrick:2014oza, Khandker:2014mpa}).

In the supershadow formalism, the superconformal partial wave $\Upsilon_{\Delta,\ell}$ corresponding to a superconformal primary $\mathcal{V}_{\Delta,\ell}$ is constructed from the three-point function of $\mathcal{V}_{\Delta,\ell}$ and the three-point function of its supershadow $\widetilde{\mathcal{V}}_{\widetilde{\Delta},\ell}$ in the following way:
\begin{equation}\label{eq:scpw_ssform}
\Upsilon_{\Delta,\ell} = \int d^{3}x_{5}\,d^{2}\theta_{5}\,d^{2}\thetab_{5}\,\frac{\expval{ \overline{\Phi}(\overline{X}_{1})\Phi(X_{2})\mathcal{V}_{\Delta,\ell}^{\mu_{1}\ldots\mu_{\ell}}(x_{5},\theta_{5},\thetab_{5})}\expval{ \widetilde{\mathcal{V}}_{\widetilde{\Delta},\ell;\mu_{1}\ldots\mu_{\ell}}(x_{5},\theta_{5},\thetab_{5})\Phi(X_{3})\overline{\Phi}(\overline{X}_{4})}}{\expval{\overline{\Phi}(\overline{X}_{1})\Phi(X_{2})}\expval{\Phi(X_{3})\overline{\Phi}(\overline{X}_{4})}}.
\end{equation}
We have restricted our attention here to the set of superconformal partial waves for the particular four-point function of interest, but one can easily generalize the above to arbitrary external operators. The three-point function is given in \eqref{eq:chiralSUSY3pt}.

In practice, this definition of the superconformal partial waves is less useful than the one in terms of superconformal blocks given in \eqref{eq:scpw_scblock} because it requires integrating over $\theta_{5},\thetab_{5}$ which is fairly involved and leads to quite complicated expressions. This is the reason we chose to work with the formulation in terms of superconformal blocks in the main text. However, the latter definition is incomplete because we never computed the supershadow coefficients $A_{\Delta,\ell}$. One could in principle compute these by directly performing the integrals \eqref{eq:scpw_ssform} and rewriting the result in terms of superconformal blocks. Instead, we will employ the simpler method of computing $\expval{ \Upsilon_{\Delta,\ell}, \mathcal{F}_{0}}$ using \eqref{eq:scpw_ssform} and comparing the result with the calculation of this inner product in \eqref{eq:zero-rung_ip_final}. This will give us an expression for $\mathfrak{f}(\Delta,\ell)$ and hence $A_{\Delta,\ell}$ via \eqref{eq:fdef}.

We start by gauge-fixing the bosonic coordinates to $x_{1} = \mathbf{0}$, $x_{2} = \mathbf{1}$, $x_{5} = \infty$ and the Grassmann coordinates to $\thetab_{1} = \theta_{2} = \theta_{5} = \thetab_{5}=0$. Let $V_{\Delta,\ell}$ be the conformal primary given by the bottom component of $\mathcal{V}_{\Delta,\ell}$. After gauge-fixing, the three-point function involving $\mathcal{V}_{\Delta,\ell}$ appearing in \eqref{eq:scpw_ssform} becomes a bosonic three-point function involving $V_{\Delta,\ell}$ and the three-point function involving the supershadow operator becomes
\begin{equation}\label{eq:3ptfn_gaugefix}
\frac{\expval{\widetilde{V}_{\widetilde{\Delta},\ell;\mu_{1}\ldots\mu_{\ell}}(\infty)\Phi(X_{3})\overline{\Phi}(\overline{X}_{4})}}{\expval{\Phi(X_{3})\overline{\Phi}(\overline{X}_{4})}} 
= (-1)^{\ell} \,
	\frac{\expval{\widetilde{V}_{\widetilde{\Delta},\ell;\mu_{1}\ldots\mu_{\ell}}(\infty)\phi(x_{3})\phib(x_{4})}}{\expval{\phi(x_{3})\phib(x_{4})}}\bigg|_{x_{34} \to z_{43}}
\end{equation}
where on the r.h.s we replace $x_{34}$ with $z_{34}$ in the bosonic three-point function. The desired inner product, including the Berezinian for this choice of gauge-fixing, is thus given by
\begin{equation}\label{eq:zero-rungip_shadow}
\begin{split}
\expval{\Upsilon_{\Delta,\ell},\mathcal{F}_{0}} 
&= 
	-\frac{2}{\pi}
	\int d^{3}x_{3}\,d^{2}\theta_{3}\,d^{3}x_{4}\,d^{2}\thetab_{4}\,
		\frac{|z_{43}|^{\widetilde{\Delta}-\ell-4\Delta_{\Phi}}z_{43,\mu_{1}} \ldots z_{43,\mu_{\ell}}(\mathbf{1}^{\mu_{1}}\ldots\mathbf{1}^{\mu_{\ell}}-\mathrm{traces})}{|z_{13}|^{2\Delta_{\Phi}}|z_{42}|^{2\Delta_{\Phi}}}\bigg|_\mathfrak{X} \,, \\
\mathfrak{X} 
&\equiv 
	\{\thetab_{1}=\theta_{2}=0;x_{1} = \mathbf{1},x_{2}=0\}.		
\end{split}
\end{equation}
The factor $-2/\pi$ comes from the fact that the Berezinian gives $-1/(2^{4}\mathrm{Vol}(SO(2))) = -1/(2^{5}\pi)$ and then we multiply by a factor of $2^{6}$ because this is how we normalized the inner product below \eqref{eq:supergaugefixmeas}. The key observation is that this integral takes the exact same form as the ladder kernel eigenvalue equation \eqref{eq:kernelDl} for an operator with dimension $\widetilde{\Delta}$ and spin $\ell$ after fixing coordinates in the same way as we have done here. Therefore,
\begin{equation}\label{eq:zero-rungip_shadow_final}
\expval{\Upsilon_{\Delta,\ell},\mathcal{F}_{0}} = 
	-\frac{1}{\pi\, J \,b_{\phi}^{3}} \; k(\widetilde{\Delta},\ell) = -\frac{1}{\pi \,J\, b_{\phi}^{3}} \; k(\Delta,\ell),
\end{equation}
where we have used the invariance of the kernel eigenvalue under $\Delta \to \widetilde{\Delta}$. Comparison with \eqref{eq:zero-rung_ip_final} gives
\begin{equation}\label{eq:fnorm}
\mathfrak{f}(\Delta,\ell) = (-1)^{\ell+1} \, \frac{2^{\ell}\, (\Delta+\ell)(\Delta-\ell-1)\,\Gamma\left(\ell+\frac{1}{2}\right)}{\sqrt{\pi}\,\Gamma(\ell+1)}.
\end{equation}
%

\section{Determinants for gauge-fixing}
\label{sec:det}

In this appendix, we explain how to compute the determinant and Berezinian for the gauge-fixing maps used to fix coordinates in the bosonic \eqref{eq:bosip} and $\mathcal{N}=2$ \eqref{eq:ip} inner products, respectively. 

Let us begin with the bosonic case. Given the unfixed measure appearing in \eqref{eq:bosip_unfixed}:
\begin{equation}\label{eq:unfixedmeas}
d\alpha \equiv \frac{d^{3}x_{1}\,d^{3}x_{2}\,d^{3}x_{3}\,d^{3}x_{4}}{x_{12}^{6}x_{34}^{6}},
\end{equation}
we want to fix the coordinates to $x_{1} = \mathbf{0}$, $x_{2} = (\frac{z+\zb}{2},\frac{z-\zb}{2i},0)$, $x_{3} = \mathbf{1} = (1,0,0)$, $x_{4} = \infty$ using the conformal group. We will denote the corresponding gauge-fixing map by $\mathfrak{P}$. The Jacobian $\mathfrak{J}_{\mathfrak{P}}$ of the map $\mathfrak{P}$, i.e., induced by the action of the conformal group, is obtained by contraction of the measure with the generators of the conformal algebra where the latter are represented by vector fields. More precisely, we want to fix the $10$ coordinates of the 4-points 
$\{x_1,x_2,x_3,x_4\}$, which we denote by $w_{j}$ $(j=1,\ldots,10)$ using the $10$ generators of the conformal algebra: translations $P_{\mu}$, special conformal transformations $K_{\mu}$, rotations $M_{\mu\nu}$, and dilatation $D$, which we denote in turn by $V_{j}$ $(j=1,\ldots,10)$. Let $v^{(i)}$ denote a generator acting on the $i^{\rm th}$ spatial coordinate. The generator $V$ acting on the four coordinates is then given by the sum on each of the coordinates, viz.,
\begin{equation}\label{eq:gen4coords}
V = v^{(1)} + v^{(2)} + v^{(3)} + v^{(4)}.
\end{equation}
The Jacobian is now constructed from the contraction of the $V_{i}$ with the $w_{j}$:
\begin{equation}\label{eq:Jac}
\left(\mathfrak{J}_{\mathfrak{P}}\right)_{ij} = dw_{j}(V_{i}).
\end{equation}
It is straightforward to compute the determinant of this Jacobian. Furthermore, there exists a discrete symmetry $z \to \bar{z}$ that must be gauged in the inner product. We therefore need to replace the factor $\Im(z)$ appearing in $\det\left(\mathfrak{J}_{\mathfrak{P}}\right)$ with $\abs{\Im(z)}$ and divide by a factor of $1/2$. Thus we obtain the gauge-fixed measure
\begin{equation}\label{eq:gaugefixmeas}
d\alpha_{\mathrm{fixed}} = dzd\zb\frac{\det\left(\mathfrak{J}_{\mathfrak{P}}\right)}{x_{12}^{6}x_{34}^{6}}\bigg|_{x_{1} = \mathbf{0}, x_{3} = \mathbf{1}, x_{4} = \infty} \to dzd\zb\frac{|z-\zb|}{4\, |z|^{6}}.
\end{equation}
We are free the normalize our inner product such that there is no factor of $\frac{1}{4}$ and hence we obtain the bosonic inner product in the main text \eqref{eq:bosip}.

The supersymmetric case is similar. We start with the unfixed measure
\begin{equation}\label{eq:superunfixedmeas}
d\alpha_{_{\mathcal{N}=2}}  \equiv 
\frac{d^{3}x_{1}d^{3}x_{2}d^{2}\thetab_{1}d^{2}\theta_{2}}{|z_{12}|^{4}} \; \frac{d^{3}x_{3}d^{3}x_{4}d^{2}\theta_{3}d^{2}\thetab_{4}}{|z_{43}|^{4}}.
\end{equation}
The superconformal group $OSp(4|2,2)$ includes four Poincar\'e supercharges $\mathcal{Q}_{\alpha}$ and four conformal supercharges $\mathcal{S}_{\alpha}$ which we use to fix the Grassmann coordinates to $\thetab_{1} = \theta_{2} = \theta_{3} = \thetab_{4} = 0$. The super-Jacobian of the gauge-fixing map $\mathfrak{P}^{\mathcal{N}=2}$ is given by
\begin{equation}\label{eq:superJ}
\mathfrak{s}\mathfrak{J}_{\mathfrak{P}^{\mathcal{N}=2}} = 
				\left(\begin{array}{c|c}
				dw(V) & d\theta(V) \, d\thetab(V) \\
				\hline
				dw(\mathcal{Q},\mathcal{S}) & d\theta(\mathcal{Q},\mathcal{S}) \, d\thetab(\mathcal{Q},\mathcal{S}) \\
				\end{array}\right).
\end{equation}
where we have used the notation $\left( dw(\mathcal{Q},\mathcal{S}) \right) \equiv \left(\begin{matrix} dw(\mathcal{Q}) \\ dw(\mathcal{S}) \\ \end{matrix} \right)$.
Therefore, we obtain the gauge-fixed measure
\begin{equation}\label{eq:supergaugefixmeas}
\begin{split}
d\alpha_{_{\mathcal{N}=2}}^{\mathrm{fixed}} 
&= 
	\frac{d^{3}x_{1}d^{3}x_{2}d^{2}\thetab_{1}d^{2}\theta_{2}}{|z_{12}|^{4}}\, 
	\frac{d^{3}x_{3}d^{3}x_{4}d^{2}\theta_{3}d^{2}\thetab_{4}}{|z_{43}|^{4}} \,
	\mathrm{Ber}(\mathfrak{s}\mathfrak{J}_{\mathfrak{P}^{\mathcal{N}=2}})
	\bigg|_{x_{1} = \mathbf{0}, x_{3} = \mathbf{1}, x_{4} = \infty;\thetab_{1} = \theta_{2} = \theta_{3} = \thetab_{4} = 0} \\
	&\to d^{2}z\,\frac{|z-\zb|}{2^{6}|1-z|^{2}|z|^{4}},
\end{split}
\end{equation}
where there is a factor of $2^{-4}$ coming from the fact that $d^{2}\theta_{j} = \frac{i}{2}d(\theta_{j})_{1}d(\theta_{j})_{2}$. Again, we are free to remove the factor of $2^{-6}$, which gives the inner product in the main text \eqref{eq:ip}.



\input{syk3d-refs}

\end{document}

%% file: syk3d-refs.tex
\providecommand{\href}[2]{#2}\begingroup\raggedright\endgroup